\documentclass[10pt]{book}
\usepackage{amsmath}
\usepackage{amssymb}
\usepackage[hang,flushmargin]{footmisc}
\usepackage{mathcomp}
\usepackage{textcomp}
\usepackage{units}
\usepackage{dsfont}
\usepackage{tikz}
\usepackage{fancyhdr}
\usepackage{emptypage}

\newcommand{\alphap}{\alpha^\prime}
\newcommand{\pp}{{=\!\!\!|}}
\newcommand{\Tr}{\text{Tr}}

\newcommand{\pmu}{\phantom{\mu}}
\newcommand{\pa}{\phantom{a}}

\newcommand{\ab}{\bar{a}}
\newcommand{\bb}{\bar{b}}

\newcommand{\hb}{\bar{h}}

\fancypagestyle{plain}{
\fancyhf{}

\fancyfoot[LE,RO]{\thepage}
}

\usepackage{vubtitlepage}
\author{Dimitri Terryn}
\title{The Generalised K\"{a}hler Geometry of\\ Wess-Zumino-Witten Models}

\promotor{Prof. Dr. Alexander Sevrin}

\faculty{Faculteit Wetenschappen en Bio-ingenieurswetenschappen}
\department{Departement Theoretische Natuurkunde}
\reason{Proefschrift ingediend met het oog op het behalen\\van de graad
van Doctor in de Wetenschappen}
\date{November 2012}

\begin{document}

\frontmatter

\faculty{Faculty of Science and Bio-engineering Sciences}
\department{Department of Physics - Theoretical Physics}
\reason{Dissertation submitted to the Vrije Universiteit Brussel in partial fulfillment\\of the requirements for the degree Doctor of Science}

\maketitlepage

\null
\thispagestyle{empty}
\vfill
\noindent Print: Silhouet, Maldegem\\
\\
\copyright \phantom{} 2012 Dimitri Terryn\\
\\
2012 Uitgeverij VUBPRESS Brussels University Press \\
VUBPRESS is an imprint of ASP nv (Academic and Scientific Publishers nv) \\
Ravensteingalerij 28 \\
B-1000 Brussels \\
Tel. +32 (0)2 289 26 50 \\
Fax +32 (0)2 289 26 59 \\
E-mail: info@vubpress.be \\
www.vubpress.be \\
\\
ISBN 978 90 5718 231 0\\
NUR 910 / 925\\
Legal deposit D/2012/11.161/166\\
\\
All rights reserved. No parts of this book may be reproduced or transmitted in any form or by
any means, electronic, mechanical, photocopying, recording, or otherwise, without the prior
written permission of the author.
\mainmatter
\setcounter{page}{1}
\pagenumbering{roman}
\tableofcontents

\newpage
\thispagestyle{empty}
\mbox{}
\chapter*{Acknowledgements}
First of all, I would like to thank my advisor, Prof. Dr. Alexander Sevrin, for giving me the opportunity to work on this very interesting topic, and for his patience and guidance over the years.\\
\\
I would also like to thank the Vrije Universiteit Brussel for allowing me the honour of assisting with the various courses offered by the Department of Physics. I will never forget the years spent humbly helping the next generation of bright young  students in furthering their understanding of physics.\\
\\
A big thank you for the Profs. Alexander Sevrin, Ben Craps, Jan Danckaert, Johan Schoukens and Thomas Durt for the enjoyable collaboration when assisting them with their various courses of the years. My thanks also to the different colleagues I had the pleasure of encountering and interacting with during my stay at the VUB. You all made the work a lot more enjoyable and I wish you well with your future endeavours.\\
\\
I would like to express my deepest gratitude to the members of my jury, Profs. Verentennicoff, Craps, Sevrin, Danckaert, Kieboom, Boulanger and Lindstr\"{o}m, for generously offering up their time to read and study this work and suggesting numerous improvements.\\
\\
I would also like to offer thanks to the Erasmus Hogeschool Brussel and the Ecole Europ\'{e}enne de Bruxelles III for entrusting me with a number of teaching duties over the years. It was my pleasure to be of assistance.\\
\\
To my friends and other associates whom I had the good fortune of meeting during my time spent at the Vrije Universiteit Brussel, thank you for the numerous adventures we shared together. Debbie, David, Dieter, Fre, Gregory, Koen, Steven, Thibaud, and many, many others, thank you for allowing me to keep my sanity (or indulging in different kinds of insanity, as the case may be).\\
\\
Last but certainly not least, I would like to thank my grandmother for her care and patience during times when she had a stressed thesis writer around the house.

\chapter{Introduction}
\thispagestyle{empty}
\setcounter{page}{1}
\pagenumbering{arabic}
\section{Our mathematical universe}

Theoretical physics, at its most basic level, seeks to identify the fundamental constituents of our observable universe and to formulate a consistent description of how they interact. This aim may seem to be nothing short of hubris, after all, there is no guarantee that such a description even exists, much less that is accessible to the intellectual tools humans have at their disposal. Setting aside the attainability of this ultimate goal one is forced to acknowledge that this program has yielded incredible results over the last centuries. From our limited vantage point as an earth-bound species utilising the limited intellectual frameworks and material tools available to us we have already discovered a great deal as to how our universe operates, even though undoubtedly much greater discoveries and understanding yet await.\\
\\
What is it then, that enables us to formulate confident statements about the inner workings of the various natural phenomena that surround us? AÏngn important turning point came when the collection of intellectual pastimes known as Natural Philosophy gave rise to the new field of Physics by adopting two new frameworks in which to operate : the scientific method and mathematisation of what had previously been a predominantly philosophical endeavour. The \"{u}r-example of this development is no doubt Newton's universal theory of gravitation which sought to not only to describe the movements of the heavens which had enthralled the natural philosophers for centuries, but the movement of terrestrial bodies as well. \\
\\
Although Newton's work is now over three centuries old and has largely been superseded by more accurate theoretical descriptions, it is still indicative of the most important features that guide theoretical physics to this day. First and foremost, it provides a succinct and elegant mathematical model of a large variety of different phenomena. By formulating a limited number of principles which correspond to a precise mathematical expression it is, at least in principle, possible to accurately describe occurrences ranging from falling objects, shooting cannonballs, to planetary orbits. A related point is that in order to accomplish this, the necessary mathematical tools had to be refined or properly applied in the case where they already existed or, more dramatically, invented all-together. This fruitful cross-pollination between mathematical ideas and physical theories was to be a model for many later developments.  \\
\\
Theoretical models, be they elegant, powerful or beautiful, ultimately should be used to make objective statements and accurate predictions about the phenomena they are meant to describe. It is here that we encounter the dramatic shift that the scientific method entailed. Building on the large volume of observational data that his predecessors has amassed, Newton's theory was able to not only reproduce this data but to make predictions as to when astronomical events would occur, with Halley's work on the eponymous comet as the most famous example. Using Newton's laws one is able to theoretical derive previous ``laws'', such as the ones formulated by Johannes Kepler which were, amongst other things, based on the wealth of experimental data amassed by Tycho Brahe. That the phenomenological laws of Kepler can be derived from first principles using Newton's laws of motion and gravitation firmly roots the theory in the realm of physical reality, where measurements and observations have the final say in determining wether or not a particular theoretical model is correct or not. It was the inability of Newton's law to properly account for later observations, such as the perihelion precession of Mercury, more than it's philosophical inconsistencies that ultimately showed the need for a more general framework to be established. \\
\\
Another important theme of theoretical physics is also illustrated by this example, namely the concept of unification. Before Newton's time it was somewhat of an accepted fact that the movements of the heavens were distinct from the movements of bodies on the earth. It was a great triumph of Newton's theory of gravitation that, taken together with his laws of motion, it not only described falling bodies on the surface of the earth but also could account for the sometimes idiosyncratic trajectories of the heavenly bodies without resorting to additional hypotheses. In stating that both earthly and heavenly bodies are subject to the same laws of motion and the same force, in this case gravitation, the theoretical models to describe them are unified by concluding that these a priori distinct physical phenomena are merely different manifestations of the same underlying principle. Many more dramatic examples of this exist, perhaps most famously Maxwell's electromagnetic theory unifying electricity and magnetism or, more recently, the electroweak theory of Sheldon Lee Glashow, Abdus Salam and Steven Weinberg that combines the electromagnetic and weak nuclear forces within a single description. The ultimate aim of this journey would be to formulate a self-consistent set of statements which, translated into the proper mathematical framework, are able to account for all physically observable phenomena, at the very least in principle. Wether this is possible is still very much an open question, but it is undeniable that as our understanding of the physical universe broadens and deepens additional structure becomes apparent and many distinct areas of physics have their theoretical descriptions converging into the same framework.\\
\\
The previous discussion should leave us in awe. On the face of it, there is no compelling reason as to \emph{why} this line of enquiry should yield such spectacular results. The use Eugene Wigner's phrase, this unreasonable effective of mathematics, which some regard as a purely intellectual construct, has the apparent power of not only describing what we observe in an elegant and coherent fashion, but can subsequently be employed to to accurately predict how the world around us will behave. Furthermore, patterns and regularities observed in nature that to the human observe seem to be completely unrelated will take on a new meaning by being shown to be different manifestations of the same underlying principle. An apt phrase to describe this, thinking back to Newton, is \emph{``As above, so below''}, which bereft of its alchemical origins can be taken to mean that there is no inherent difference between bodies moving in the heavens, and those moving on earth. When employed together, the scientific method and the consistent use of mathematics as a way of objectifying physical statements has proven to be a powerful method for describing the world around us.

\section{Gauge theories and the Standard Model}

If we take the reductionist's approach and perhaps somewhat naively state that to understand the universe we must identify its constituent parts and how they interact we are led to that area of theoretical physics currently knows as theoretical particle physics. At the beginning of the 20th century a somewhat optimistic picture emerged that there exist atoms which are made up of electrons, neutrons and protons. These ingredients make up matter, and that this matter interacts via the electromagnetic and gravitational forces, and that was that. As it so often happens this apparent conclusion to the ultimate nature of the universe was somewhat premature. The discovery of the neutrino, originally proposed by Wolfgang Pauli in 1930 to explain the continuous energy spectrum of $\beta$-decays and the muon in 1936, prompting the famous phrase uttered by Isidor Rabi \emph{``Who ordered that?"}, led to the realisation that the microscopic world was not quite yet understood. Further inquiry into the structure of the atomic nucleus and the discovery of the strong and weak nuclear forces made it clear that the story was far from complete and that a new framework was needed.\\
\\
As it stands now it is not possible to put the four different fundamental forces that lead to the various particle interactions on the same footing. The electromagnetic force and both nuclear forces are the main actors when it comes to the subatomic world and are described by a specific class of quantum field theories known known as gauge theories. The simplest example of a gauge theory is Maxwell electrodynamics, where the force-carrying photons are excitations of a gauge field $A_\mu(x)$ which correspond the an electromagnetic field strength
	\begin{equation}
	F_{\mu\nu}=\partial_\mu A_\nu - \partial_\nu A_\mu.
	\end{equation}
The theory is then defined by the action functional
	\begin{equation}
	S=-\frac 1 4 \int d^4 x F^{\mu\nu}F_{\mu\nu}.
	\end{equation}
The important realisation is that the above theory contains a redundancy in its description. Physically, the photon field only has two degrees of freedom by virtue of being massless which are the polarisation states of the photon. The corresponding gauge field however has four degrees of freedom in four dimensions meaning that two of these degrees of freedom are redundant. At the level of gauge field we can see that the theory contains a so called gauge symmetry
	\begin{equation}
	A_\mu(x)\rightarrow A_\mu(x) - \partial_\mu \Lambda(x)
	\end{equation}
where $\Lambda(x)$ is an arbitrary complex number that can depend on the spacetime coordinates, making it a local gauge symmetry as opposed to global one. Performing this gauge transformation leaves the field strength and the action invariant confirming that it is indeed a symmetry. One important realisation was that the possible gauge parameters $\Lambda(x)$ form a group, namely the group of unit complex numbers or $U(1)$. In this case $U(1)$ is called the underlying gauge group of the theory. As the elements of $U(1)$ commute this is the simplest non-trivial example of such a theory known as an abelian gauge theory. When considering more complicated gauge groups we obtain non-abelian gauge theories were the gauge parameters need not be commuting. Collectively, these theories are known as Yang-Mills theories and they are at this time fairly well understood, at least in the perturbative regime where the various couplings are small.\\
\\
This idea, namely that the underlying quantum field theory is a local gauge theory where the gauge parameters form some Lie group turned out to be very powerful. The current \emph{Standard Model of Elementary Particle Physics} is based on the gauge group $SU(3) \times SU(2) \times U(1)$. The $SU(3)$ part contains the eight massless gluon fields who carry the strong nuclear force. It forms the basis for the theory known as Quantum Chromodynamics, which can be used as a model for the strong nuclear force by including the strongly interacting constituents of the Standard Models, called quarks, and the different gluon fields. The other part of the Standard Model corresponds to the $SU(2) \times U(1)$ part of the gauge group and unifies the electromagnetic and the weak nuclear forces. While the electroweak force is unified at mass scales of approximately 100 GeV it appears distinct at lower energies through a process known as electroweak symmetry breaking. The $SU(2) \times U(1)$ part of the gauge group break down to a non-trivial $U(1)$ subgroup giving rise to familiar electromagnetism mediated by the photon field, while the three remaining gauge bosons acquire mass and give rise to the vector bosons $Z,W^\pm$ who carry the weak nuclear force. The process through which this occurs is an example of spontaneous symmetry breaking called the Brout-Englert-Higgs mechanism.\\
\\
The Standard Model has been tested to an incredible degree of accuracy in recent years. The last experimentally unverified piece is very close to being resolved following the discovery of a massive scalar boson at CERN \cite{cmshiggs}\cite{atlashiggs} in 2012. While analysis of this signal is still ongoing it is very likely that this particle is indeed the excitation of the Brout-Englert-Higgs field corresponding to the mechanism of spontaneous symmetry breaking.

\section{Gravity, the odd man out}
The previous section contained a glaring omission in that the Standard Model remains entirely silent on the subject of the fourth fundamental force, namely gravity. The practical reason for this is that on subatomic scales it is completely overshadowed by the three other forces at energy scales that our experimentally accessible to us. From a philosophical standpoint however this omission is unsatisfying, and furthermore there do exist regimes where gravitic interactions will become important at short distances, the very early universe and black holes being two prominent examples.\\
\\
The gauge theories underlying the Standard Model are examples of quantum field theories, meaning that they satisfy the axioms of both quantum mechanics and special relativity. Gravity on the other hand is currently described by Albert Einstein's General Theory of Relativity which is a classical theory and as such does not incorporate quantum effects. Let us first review the basic premise of General Relativity, which states that the gravitational interaction between matter is the result of the curvature of spacetime described by the Riemann curvature tensor $R_{\mu\nu\rho\sigma}$, itself a function of the spacetime metric tensor $g_{\mu\nu}(x)$. The relevant field equations are known as the Einstein equations and can be expressed using contractions of the curvature tensor called the Ricci tensor $R_{\mu\nu}$ and the Ricci scalar $R$,
	\begin{equation}
	R_{\mu\nu}-\frac{1}{2}R g_{\mu\nu}+\Lambda g_{\mu\nu}=8\pi G_N T_{\mu\nu}.
	\end{equation}
Although in practice a highly complicated set of coupled non-linear partial differential equations, the underlying idea giving rise to them has a nice intuitive interpretation. Given a certain distribution of matter and energy given by the lagrangian density $\mathcal{L}_M$ we can calculate the resulting energy-momentum tensor $T_{\mu\nu}$. The distribution of mass and energy will then cause spacetime to bend, with the resulting spacetime then given by a solution of the Einstein equations. Note that we included a term linear in $g_{\mu\nu}$ proportional to the cosmological constant $\Lambda$, which can be roughly regarded as the energy density of the empty spacetime. Although initially not included in the theory this cosmological has been determined to be positive and non-zero,
	\begin{equation}
	\Lambda \approx 10^{-47} \text{GeV}^4,
	\end{equation}
on the basis of astronomical observations. Explaining this highly peculiar value is an important open problem at the present time.\\
\\
Being a field theory the Einstein equations can be derived from the variation of an action functional. This action is known as the Einstein-Hilbert action and has the deceptively simple form
	\begin{equation}
	S= \int d^4 x \sqrt{-g} \left( \frac{R-2\Lambda}{{16 \pi G_N} } + \mathcal{L}_M \right).
	\end{equation}
when including the cosmological constant and matter fields.
Not only does General Relativity reduce to Newton's theory when the gravitational effects involved are small and relativistic effects are negligible, it also has a fantastic track record of explaining various experimental results which could not adequately be explained in the Newtonian theory. Furthermore, it replaced the philosophically disturbing instantaneously mediated gravitational force underpinning Newton's theory with gravitational effects which propagate at no more than the speed of light, as required by special relativity.  \\
\\
Despite these successes and disregarding possible experimental or observational discrepancies from theoretical arguments alone it is clear that General Relativity cannot be the final word with regards to gravity. At best, it is an effective theory which is highly successful at ``low" energies, but as field theories are generally do it announces its own demise at higher energies. Explicitly writing down factors of $\hbar$ and $c$ Newton's constant defines a natural energy scale via the (reduced) Planck mass
	\begin{equation}
	M_P = \sqrt{\frac{\hbar c}{8\pi G_N}} \approx 10^{18} \text{ GeV}
	\end{equation}
which, when compared to accessible energies in particle colliders of around $10^4$ Gev, is far from the energy regime experimentally accessible to us. Nevertheless, at least in principle it should be possible to formulate a quantum mechanical field theory which reproduces General Relativity at large distances. One can use the Planck mass to introduce a dimensionful coupling in a quantum field theory describing a spin-2 graviton coupling to itself by studying small perturbations around a flat Minkowski spacetime\footnote{An pedagogical discussion of this argument can be found in \cite{zeeqft}.}
	\begin{equation}
	g_{\mu\nu}=\eta_{\mu\nu}+\frac{1}{M_P}h_{\mu\nu}
	\end{equation}
and considering the corresponding expansion of the Einstein-Hilbert action which schematically reads
	\begin{equation}
	S=\int d^4 x \left( (\partial h)^2 + \frac{1}{M_P} h  (\partial h)^2 + \frac{1}{M_P^2}h^2  (\partial h)^2 + \dots \right).
	\end{equation}
The above action describes the dynamics of a self-interacting, massless, spin-2 particle called the graviton which has been hypothesised to correspond to the gravitational interaction. The infinite amount of interaction terms correspond to the fundamental fact that anything carrying energy or mass, including the gravitons themselves, couple the gravitational field. When the energy scale $E$ being probed is sufficiently smaller than the Planck mass, in other words as long as $E \ll M_P $, we can trust perturbation theory to make sense. When approaching the Planck scale however it is clear that this effective theory must be enhanced in some way as new physics is expected to appear.\\
\\
It is not surprising that the procedure outlined above does not produce satisfactory results. General Relativity treats the spacetime metric as the fundamental object in a holistic manner, while quantum field theory is largely formulated in the perturbative regime concerned with small perturbations around a flat background. By artificially splitting up the metric in this manner it is perhaps inevitable that important ingredients in theory become lost.As a result, it is unclear how gravity and the different gauge forces fit within a unified framework. This is still very much an open question and perhaps represents one of the most fundamental challenges found within theoretical high energy physics today.

\section{Enter strings}
The most promising candidate for a quantum theory of gravity came about in a roundabout way. While trying to make sense of the various exotic particles produced in collision experiments it was discovered that mesons and baryons, which are now known to correspond to various bound states consisting of quarks but which were not very well understood at the time, could be classified into so-called Regge trajectories. The details need not concern us (for a historical overview, see, for example \cite{divecchia}), but the main feature was that there existed certain linear relationships between the angular momentum of the particles and their scattering amplitudes. Gabriele Veneziano \cite{veneziano} proposed a scattering amplitude that accounted for this strange behaviour but the underlying physical principle remained rather mysterious. It was later found independently by Yoichiro Nambu \cite{nambu}, Holger Bech Nielsen \cite{Nielsen} and Leonard Susskind \cite{susskind} that scattering amplitudes of the Veneziano type could be derived by modelling the various resonances as vibrating, one dimensional strings. This interpretation was at odds with other experimental data and with the development of quantum chromodynamics this line of reasoning was abandoned as a way of describing the strong nuclear force.\\
\\
The string model proposed was studied further and was found to contain several surprising properties. What was originally proposed as a model for the strong interaction turned out to contain additional bosonic degrees of freedom which were not expected. Most notably it contain a massless spin-2 particle, which was especially surprising since the most natural candidate for such a particle was the graviton. The natural appearance of the gravitational interaction in a quantum theory and the subsequent proof that the theory could be formulated in a form that was free from quantum anomalies and ultraviolet divergences, implying that the quantised theory was consistent, led to fluster of activity. String theory quickly became the prime candidate for providing a unified framework to describe not only gravity, but the various gauge forces as well, offering the promise that it would be in principle able to account for all observable phenomena so far. While a comprehensive review of string theory is goes beyond the scope of this work, we offer a review in chapter 2 in order to set the stage.\\
\\
Amongst the attractive properties of these newly discovered string theories was that they seemed to require an additional element which, from a theoretical standpoint at least, was highly desirable. The string theories formulated consider so far only contained bosonic degrees of freedom, prompting the obvious question as to how fermions should be included. Furthermore, the spectrum of the theory contained states which were tachyonic, which implied that its ground state was not stable, which at the very least made perturbative methods questionable. The answer to both these problems was that the theory quite naturally seemed to generalise to what are now known as superstring theories by including an additional symmetry known as supersymmetry. Supersymmetry had been proposed earlier as an additional possible symmetry for quantum field theories.  Roughly speaking, by investigating the possible symmetries of the S-matrix that led to a consistent and non-trivial interacting field theory it was found that the only permissible symmetries were Lorentz invariance and internal gauge symmetries. This property, known as the Coleman-Mandula theorem\cite{colemanmandula}, was later found to be true only when considering bosonic symmetries. By including fermionic symmetries\footnote{Mathematically this can be expressed by considering a graded symmetry algebra.} one could enlarge the allowed symmetry group to include so-called supersymmetries. Although field theories based on a supersymmetric extensions of the Lorentz-algebra had been considered before in the Soviet Union, supersymmetry became well know after the work of Julius Wess and Bruno Zumino \cite{wesszumino}. \\
\\
On a very basic level supersymmetry requires every boson to be part of a supersymmetric multiplet where it is partnered with one or more corresponding fermions, and vice-versa. This offers an intriguing philosophical notion of further unification, as it further blurs the previously clear line between particles making up matter, which are fermions, and force-carrying particles, which are bosons. If supersymmetry is indeed realised in nature it would imply that the observable difference between force carrying particles and matter particles is merely a low-energy occurrence that is not a feature of the underlying high-energy theory. Unfortunately, none of the known particles that comprise the Standard Model can be related to each other by supersymmetry transformations, which forces us to conclude that supersymmetry is broken at some energy scale above those that can be probed by current accelerator technology. As of writing, there is no direct evidence for supersymmetry being an actual feature of our universe. Some indirect indications exist, indeed electroweak precision measurements seem to favour the Minimal Supersymmetric Standard Model (MSSM) over the ordinary Standard Model, but the matter is far from resolved. There are theoretical and experimental reasons that seem to indicate that if supersymmetry as we understand it exists it should be broken around the TeV scale, which is currently being probed by the LHC, where finding evidence for supersymmetry is currently an active area of study.\\
\\
Disregarding the question wether or not supersymmetry is a desirable feature of superstring theory a more glaring phenomenological obstacle exists, namely that the various superstring theories can only be formulated consistently in ten dimensional spacetime. Higher dimensional theories have a long history of being considered as candidates for unifying gravity with the different gauge forces. One way of reconciling this feature with the seemingly obvious observed fact that our universe is four dimensional is the Kaluza-Klein construction where the extra dimensions are compact in some sense and their internal volume is small compared to the length scale implied by our currently accessible energy regime which, taking the electroweak scale for instance, corresponds to a characteristic length of $l_w \approx 10^{-18}$m. Compactifying six of the ten spacetime dimensions in a way that is consistent with the supersymmetry conditions one is led to consider so-called Calabi-Yau manifolds making up the internal space\cite{wittencalabiyau}. While this idea would explain the seeming mismatch between the number of spacetime dimensions required by the superstring theories with our low-energy observations, it introduces new phenomenological problems in that there are many different ways to reduce the various theories to four dimensions, a large subclass of which could give rise to a four dimensional universe consistent with the Standard Model and a positive cosmological constant. With no known natural way to distinguish between the various compactifications this so-called Landscape problem severely impacts the predictive power of superstring theories when utilising them to describe known physics. We offer a brief overview of the various superstring theories in chapter 3.\\
\\

\section{Sigma models and geometry}
Wether or not superstring theories provide a suitable framework to unify our different models for describing the fundamental building blocks of our observable universe remains very much an open question. Nevertheless, it has proven to be an incredibly fertile area of research in terms of developing new mathematical structures that are not only interesting in their own right, but which hold the promise to clarify and expand on our current physical frameworks. We have already alluded to deep connection between geometrical notions and the various symmetries that underly, and in a very real sense define, our current physical understanding. The very usefulness of string theory lies in the intimate connection between the symmetries existing on the string worldsheet and the geometrical implications for the spacetime which the string inhabits. The topic of thesis is an example of this connection and goes by the name of generalised complex geometry. Amongst the bosonic excitations of the string we find not only the spin-2 graviton field giving rise to the spacetime metric, but also a new ingredient that is described by an antisymmetric tensor of rank 2 called the Kalb-Ramond field $b_{\mu\nu}$. Recall that in Einstein's theory of gravity we required spacetime to be a Riemannian manifold. String theory backgrounds contain solutions that are more general in that they are spacetimes that have non-vanishing torsion, which is itself the result of the presence of this $b$-field. Generalised complex geometry provides a natural framework in which these backgrounds can be understood by placing the symmetry relating to the metric, diffeomorphism invariance, and symmetries relating to the $b$-field on the same footing.\\
\\
The mathematical language to describe the interplay between a propagating string and its ambient spacetime is that of non-linear $\sigma$-models. Originally considered as a model for describing $\beta$-decay containing a new scalar called the sigma \cite{gellmann}, $\sigma$-models are a special class of field theories where the field content consists of number of fields $\phi^\mu(x)$ that take on values in a manifold $\mathcal{M}$, called the target space. The $\sigma$-model then consists of a field theory on a submanifold $\Sigma$, which here we will take to be two-dimensional as we are interested in the worldsheet swept out by the propagating string defined by the action\footnote{For conventions and notation, please consult the appendix.}
	\begin{equation}
	S \sim \int_{\Sigma} d^2 \sigma \phantom{.}g_{\mu \nu}(x) \partial_a \phi^\mu(x) \partial^a \phi^\nu(x).
	\end{equation}
Mathematically, the fields $\phi^\mu(x)$ can be regarded a collection of maps
	\begin{equation}
	\phi^\mu : \Sigma \rightarrow \mathcal{M}
	\end{equation}
that are critical points of the Dirichlet energy functional, which is related to the action functional defining the two-dimensional field theory. Such maps are known in the mathematical literature as harmonic maps. The object $g_{\mu\nu}(x)$ now carries a dual interpretation : from the field theory point of view it is an infinite collection of coupling constants between the fields $\phi^\mu(x)$, while from the geometrical point of view it is the metric that describes the geometry of the target manifold.\\
\\
The harmonic maps making up the $\sigma$-model ensure that symmetries of the field theory on the worldsheet constrain the geometry of the target space. An important instance of this appears when one considers supersymmetric non-linear $\sigma$-models where one adds additional Grasmannian coordinates to the worldsheet, turning it into a supermanifold. Invariance under supersymmetry transformations constrains the target space to be a complex manifold \cite{alvarezkahler} \cite{zuminokahler}, in other words it ensures the existence of an almost complex structure $J$ that is integrable. Additional supersymmetry lead to a more complicated geometry. We will review this in some detail in chapter 4, but the main feature is that adding more supersymmetry generators will further constrain the geometry. It is interesting to note that in two dimensions it is possible to add left- and right-handed supersymmetry generators independently, however we will not consider such models here. For a second supersymmetry, the so called $\mathcal{N}=(2,2)$ models, we find that the target manifold must contain a symplectic structure which is compatible with the complex structure, implying that it is a K\"{a}hler manifold. For the maximum amount of supersymmetry, $\mathcal{N}=(4,4)$ the target manifold is required to be hyper-K\"{a}hler, giving us two 2-spheres worth of complex structures, each compatible with the symplectic structure \cite{howetwistor}.\\
\\
The presence of the Kalb-Ramond field $b_{\mu\nu}$ in the string spectrum leads to the possibility of a background geometry which is no longer torsion-free. When including additional supersymmetries this naturally leads us to more involved geometrical structures \cite{howesierra}. In particular, the $\mathcal{N}=(2,2)$ model will give rise to a target space geometry that is bihermitian, as discovered by S. James Gates, Chris Hull and Martin Ro\v{c}ek \cite{gates}. The key feature of bihermitian geometry is that the target space contains two complex structures $J_\pm$, both compatible with the torsionful connection and both hermitian with respect to the metric. The algebraic properties of these complex structures allows for a classification of the necessary constrained superfields that parametrise the target manifold. More specifically, it was long suspected and later proven \cite{lindstrom2007} that one can describe a general $\mathcal{N}=(2,2)$ $\sigma$-model using chiral, twisted-chiral and semi-chiral superfields. Which constrained superfields are required depends on the geometry of the target manifold. When only chiral superfields are present the resulting geometry is the familiar K\"{a}hler geometry, however once other constrained superfields come into play the resulting bihermitian geometry is consists of a broader class of backgrounds. It is within this context that we are naturally led to consider generalised complex geometry.\\
\\
Generalised complex geometry was introduced by Nigel Hitchin in \cite{hitchingcy} and further developed by Marco Gualtieri \cite{gualtieri}. It was first formulated, amongst other reasons, as a natural framework in which one could in some sense unify the previously distinct notions of complex and symplectic geometry in a setting that was broader than the usual K\"{a}hler geometry. One reason to consider such a setup is that the internal Calabi-Yau spaces appearing in string compactification scenarios require the vanishing of so-called form fields (also referred to as internal fluxes) that are present in the spectrum of string  excitations\cite{granacalabi}. Generically, one is left with a number of massless fields called moduli being present in the lower dimensional theory as a result of the compactification. From a phenomenological point of view this is an undesirable feature, so one must find a way to eliminate these moduli from the low-energy theory. One way to achieve this to let the internal fluxes corresponding to the form fields acquire a non-zero expectation value, which will cause the moduli to become very massive, making them unobservable in the low energy regime. By introducing non-zero form fields in the internal space however one is led to geometries which are no longer Calabi-Yau. It is in this sense that Hitchin introduced the notion of a generalised Calabi-Yau manifold. One immediate advantage of this setup was that this new framework provided a novel way of describing a feature of Calabi-Yau compactifications called mirror symmetry, in which two distinct Calabi-Yau manifolds will give rise to the same four-dimensional physics. While a fascinating and extensive area of study, we will not consider flux compactifications further in this work. \\
\\
It was later realised that generalised complex geometry offers a natural setting in which to study $\mathcal{N}=(2,2)$ non-linear $\sigma$-models that have as their target space a manifold which has non-zero torsion. Once a torsion field is present this target space will be given by a bihermitian geometry which is no longer K\"{a}hler. Generalised complex geometry provides a setting in which this bihermitian geometry can be described by what is referred to as a generalised K\"{a}hler geometry, which as the name implies generalises the usual K\"{a}hler is some sense. One of the main features is that any $\mathcal{N}=(2,2)$ $\sigma$-model can be described by a single scalar potential that is a function of the superfields which parametrise the target space. More precisely, for a collection of superfields $\Phi^\mu$ the $\sigma$-model is fully described by the action
	\begin{equation}
	\label{simpleaction}
	S=\int d^2 \sigma d^2 \theta d^2 \hat{\theta} \phantom{.} V(\Phi^\mu).
	\end{equation}
When no torsion is present, this potential for the two-dimensional field theory has a direct geometrical interpretation as being the K\"{a}hler potential that determines the geometrical data of the target space. Since an action of the form (\ref{simpleaction}) is not restricted to this subclass of bihermitian geometries it requires a different interpretation when the target space is no longer K\"{a}hler. It was shown that the potential $V(\Phi^\mu)$ can be thought of us a generalised K\"{a}hler potential for a generalised K\"{a}hler geometry, in the sense that locally fully describes the resulting geometry \cite{lindstrom2007a}. We will review these issues in some detail in chapter 5.\\
\\
In this work we will consider a subclass of supersymmetric non-linear $\sigma$-models called Wess-Zumino-Witten models, which are $\sigma$-models where the target manifold is a reductive Lie group, i.e. a Lie group that is the semi-direct product of semi-simple and abelian factors. By employing the additional structure provided by the Lie algebra underlying the Lie group one can explicitly find expressions for the possible complex structures on the target manifold, as developed in \cite{spindelsevrin}\ and \cite{spindelsevrin1}, which is the topic of chapter 6.  Although the correspondence between generalised complex geometry and and $\mathcal{N}=(2,2)$ $\sigma$-models is now fairly well understood, there are few examples of generalised K\"{a}hler geometry where the generalised K\"{a}hler potential is known explicitly. Of particular interest is a phenomenon known as type-changing, where the geometry of the target space alters drastically and a different superspace description is required. By considering a number of non-trivial examples, which is the topic of chapter 7, we attempt to shed more light on this matter, and hopefully indicate a direction for future study.

\chapter{Bosonic strings}
\thispagestyle{empty}
\noindent \emph{In this chapter we will provide a brief overview of bosonic string theory in order to illustrate how the various properties described in the previous chapter arise. As there are many excellent texts available\footnote{Including, but certainly not limited to, \cite{gsw}, \cite{polchinski} and \cite{beckerbecker}.} on the subject we will only highlight the most important features. We will start with a description of the bosonic string by investigating the classical theory. While unrealistic, many of the the features that are of interest later on will already be present in this description. The quantum theory will then introduce additional subtleties which have important consequences on the geometry of the space in which the string propagates. Of particular interest is the subtle interplay between the symmetries of the worldsheet field theory and the ambient target space, which will be an important theme throughout this work.}

\newpage

\section{Worldlines and worldsheets}

\subsection{The relativistic point particle}

Before we introduce the string we will start by describing the dynamics of the relativistic particle with mass $m$ propagating in $D$-dimensional Minkowski space of signature $\eta_{\mu \nu}=(-1,+1,+1,\dots,+1)$. This ambient space will be referred to as the target space. To make Lorentz-invariance manifest we denote the trajectory of the particle by $x^{\mu}(\tau)$. A natural candidate for an action is to make it proportional to the length of the worldine $\gamma$ of the particle as follows
	\begin{equation}
	S \sim  \int_{\gamma} ds .
	\end{equation}
Since this action is not dimensionless in natural units we have to introduce a proportional factor with units of mass or energy. The only natural parameter with the right units available to us is of course $m$, which leads us to the action\footnote{The minus sign is chosen in order to reproduce the correct relativistic expressions later on and is a consequence of the choice of signature.}
	\begin{equation}
	S =-m  \int_{\gamma} ds .
	\end{equation}
	If we want to relate the line-element $ds$ with the Minkowski-space coordinates we need to consider the pull-back or induced metric on the worldline
	\begin{equation}
	g_{\tau \tau}=\frac{\partial x^{\mu}}{\partial \tau}\frac{\partial x^{\nu}}{\partial \tau}\eta_{\mu \nu}
	\equiv \dot{x}^{\mu} \dot{x}^{\nu}\eta_{\mu \nu},
	\end{equation}
which allows us to rewrite us the action in terms of the target space coordinates
	\begin{equation}
	\label{particlesqrtaction}
	S =-m  \int_{\gamma} d\tau \sqrt{-g_{\tau \tau}}= -m  \int_{\gamma} d\tau \sqrt{-\dot{x}^{\mu}
	\dot{x}^{\nu}\eta_{\mu \nu}}.
	\end{equation}
If we fix a frame $x^{\mu}=(t,\vec{x})$ the action further simplifies to 
	\begin{equation}
	S=-m\int_{\gamma}dt \sqrt{1-\dot{\vec{x}}.\dot{\vec{x}}},
	\end{equation}
after which we can calculate the conjugate momenta $\vec{p}$ conjugate to $\vec{x}$ and the Hamiltonian resulting in
	\begin{eqnarray}
	\vec{p}=\frac{m\dot{\vec{x}}}{1-\dot{\vec{x}}.\dot{\vec{x}}}, \qquad E=\sqrt{m^2+\vec{p}^{\hspace{0.8mm}2}},
	\end{eqnarray}
which we indeed recognize as being the familiar relativistic momenta and energy.\\
\\
An important property of this action is that it is invariant under arbitrary reparametrizations of the wordline $\tilde{\tau}=\tilde{\tau}(\tau)$. As a result, not all $D$ degrees of freedom are physical because of the presence of a redundancy in the description and we are left with $D-1$ independent fields. When looking at the conjugate momenta this translates to them being subject to the following constraint
	\begin{eqnarray}
	p^{\mu}p_{\mu}+m^2=0,
	\end{eqnarray}
which we recognize as being the mass-shell condition of a physical particle of mass $m$.\\
\\
Should one chose to investigate the quantum theory a technical issue arises that will continue to plague us when we move on to extended objects. The action (\ref{particlesqrtaction}) contains a square root making its use in the path integral difficult. Fortunately an elegant workaround exists which can be interpretated in a geometrical sense.\\
\\
One can add an auxiliary field $e(\tau)$ to the theory and considering the following alternate action
	\begin{eqnarray}
	S=\int_{\gamma} d\tau \left( \frac{1}{2e}\dot{x}^2 -\frac{e}{2}m^2 \right).
	\end{eqnarray}
This auxiliary field is not physical. Its equations of motion fix it to be
	\begin{eqnarray}
	e=\frac{1}{m}\sqrt{-\dot{x}^2}.
	\end{eqnarray}
When we plug this back into the action we get
	\begin{eqnarray}
	S=-m \int_{\gamma} d\tau\sqrt{-\eta_{\mu\nu}\dot{x}^{\mu}\dot{x}^{\nu}},
	\end{eqnarray}
which establishes the equivalence with the action (\ref{particlesqrtaction}). The action is now written down in a form that is more easily quantified. An additional upside is that the description can be naturally extended to massless particles.\\
\\
The geometrical interpretation of this procedure is as follows. The auxiliary field $e(\tau)$ can be identified when comparing it to the line element of the wordline
	\begin{eqnarray}
	ds^2=g_{\tau \tau}d\tau^2, \qquad g_{\tau \tau}\equiv e^2.
	\end{eqnarray}
This implies that $e$ is the einbein corresponding to a gravity theory in 0+1 dimensions
	\begin{equation}
	S=\int_{\gamma} d\tau \sqrt{g_{\tau \tau}}\left( \frac{1}{2}g^{\tau \tau}\dot{x}^2 -\frac{1}{2}m^2 		\right).
	\end{equation}
The fact that the einbein is completely determined by the worldine and as such is not physical corresponds to the well-known fact that in $D=1$ gravity is not a dynamical theory.

\subsection{The relativistic string}
Now that we have reviewed how relativistic dynamics naturally arises when we consider the field theory on its wordline we will turn our attention to how this translates when we consider an object that has a certain extention in one dimension, the so-called fundamental string.\\
\\
As before we will take $D$-dimensional Minkowski space as our ambient space with coordinates $X^{\mu}.$\footnote{The change in notation is simply a matter of convention.} The worldsheet of the string is described by the coordinates $\sigma^{\alpha}=(\tau,\sigma)$, $\alpha=1,2$. The target space coordinates $X^{\mu}(\tau,\sigma)$ are taken to be bosonic fields on the string worldsheet $\Sigma$. Our starting point will be an action that is propertional to the surface of the worldsheet
	\begin{equation}
	S\sim \int_{\Sigma}d \mu.
	\end{equation}
As before we rewrite the surface element $d\mu$ in terms of the induced metric on the worldsheet
	\begin{equation}
	\gamma_{\alpha \beta}=\frac{\partial X^{\mu}}{\partial \sigma^{\alpha}} \frac{\partial X^{\nu}}		
	{\partial \sigma^{\beta}}\eta_{\mu \nu}.
	\end{equation}
When written out in terms of this induced metric we get the so-called Nambu-Goto action
	\begin{eqnarray}
	\label{NambuGoto}
	S &=&-T \int_{\Sigma}d^2\sigma \sqrt{-\det\gamma}\nonumber \\
	&=&-T \int_{\Sigma}d^2\sigma \sqrt{-(\dot{X})^2 ( X^{\prime})^2+(\dot{X}.X^{\prime})^2}.
	\end{eqnarray}
The constant of proportionality $T$ is identified with the tension of the string. For historical reasons, the tension is replaced by the parameter $\alphap$ called the Regge-slope as follows
	\begin{equation}
	T=\frac{1}{2\pi \alphap}.
	\end{equation}
In natural units we have $[T]=$ 2 and $[\alphap]=$ -2. Consequently, the only dimensionful parameter in our theory so far defines a length scale, the so-called string length scale $l_s$ via
	\begin{equation}
	l_s^2=\alphap.
	\end{equation}
If we introduce the conjugate momenta as follows
	\begin{eqnarray}
	\Pi_{\mu}^{\tau}=\frac{\partial \mathcal{L}}{\partial \dot{X}^{\mu}}, \qquad \Pi_{\mu}^{\sigma}=\frac{\partial 
	\mathcal{L}}{\partial X^{\prime\mu}},
	\end{eqnarray}
the equations of motion resulting from Nambu-Goto action (\ref{NambuGoto}) take on the simpler form
	\begin{equation}
	\frac{\partial \Pi_{\mu}^{\tau}}{\partial \tau}+\frac{\partial\Pi_{\mu}^{\sigma}}{\partial\sigma}=0.
	\end{equation}
These equations of motion, despite superficially resembling a familiar wave equation, are actually highly non-linear and rather non-trivial due to the explicit form of the momenta $\Pi_{\alpha}^{\mu}$. This can be made somewhat more visible by noting that the equations of motion for the bosonic fields $X^{\mu}$ can be written as
	\begin{equation}
	\partial_{\alpha} \left( \sqrt{-\det \gamma} \gamma^{\alpha \beta}\partial_{\beta} X^{\mu}\right)=0.
	\end{equation}
The Nambu-Goto action posseses reparametrisation invariance as a gauge symmetry, which tells us that not all $D$ degrees of freedom are physical. One can easily show that the action remains unchanged under arbitrary  transformations of the form $\sigma^\alpha \rightarrow \tilde{\sigma}^\alpha(\sigma^\alpha)$.\\
\\
Before we look at the quantisation of this theory we notice that a problem similar to the point particle will make this difficult. As before we will solve this by promoting the induced metric to an independent field on the worldsheet $\gamma_{\alpha \beta}\rightarrow g_{\alpha \beta}(\tau,\sigma)$. The resulting action is called the Polyakov action\footnote{This action was first considered independently in \cite{brinkvecchia} and \cite{deserzumino}, after which it  acquired its name through its use in quantising the string in \cite{polyakov}.}
	\begin{equation}
	\label{Polyakov}
	S=-\frac{1}{2\pi\alphap}\int_{\Sigma}d^2\sigma \sqrt{-\det g}g^{\alpha\beta}\partial_\alpha 
	X^\mu \partial_\beta X^\mu\eta_{\mu\nu}.
	\end{equation}
The equations of motion for the fields $X^\mu$ become
	\begin{equation}
	\label{EOMPolyakov}
	\partial_{\alpha} \left( \sqrt{-\det g} g^{\alpha \beta}\partial_{\beta} X^{\mu}\right)=0.
	\end{equation}
The new ingredient here is that the worldsheet metric $g_{\alpha\beta}$ obeys its own equations of motion. We can find these by varying the action with respect to $g_{\alpha\beta}$, which yields
	\begin{eqnarray}
	\frac 1 2 g_{\alpha \beta}g^{\rho \sigma} \partial_\rho X \cdot \partial_\sigma X = \partial_\alpha X \cdot \partial_
	\beta X.
	\end{eqnarray}
These can be solved by introducing the function
	\begin{eqnarray*}
	f(\sigma^\alpha)=g^{\rho \sigma}\partial_\rho X \cdot \partial_\sigma X,
	\end{eqnarray*}
so that the worldsheet metric can be written in the form
	\begin{eqnarray*}
	g_{\alpha \beta} = 2 f^{-1}(\sigma^\alpha) \partial_\alpha X \cdot \partial_\beta X
	\end{eqnarray*}
Cast in this form the worldsheet metric differs from the induced metric $\gamma_{\alpha \beta}$ by a so-called conformal factor. The fact that this conformal factor does not enter in the equations of motion for $X^\mu$ is a manifestation of another symmetry of the theory, namely Weyl-invariance, which means the worldsheet theory is invariant under arbitrary local rescalings of the worldsheet metric $g(\sigma^\alpha)\rightarrow \Omega^2(\sigma^\alpha)g(\sigma^\alpha)$. Any two worldsheet metrics which are related by such a Weyl-transformation will give rise to the same physical theory.\\
\\
Taken together, the invariance under diffeomorphisms and Weyl transformations will allow us to write (\ref{EOMPolyakov}) in a simpler form. By choosing a suitable parametrisation we can fix two of the three independent components of the worldsheet metric and go to the so called conformal gauge
	\begin{equation}
	g_{\alpha\beta}=e^{2\phi}\eta_{\alpha\beta},
	\end{equation}
where the arbitrary function $\phi(\sigma^\alpha)$ is the remaining degree of freedom, that can be subsequently gauged away by invoking Weyl invariance. Setting $\phi=0$ the equations of motion now describe $D$ free harmonic oscillators
	\begin{equation}
	\label{EOMwave}
	\partial^2 X^\mu=0.
	\end{equation}
After gauging the worldsheet metric still needs to obey its equations of motion. Using this gauge this amounts to demanding that the stress-energy tensor on the worldsheet vanishes $T_{\alpha\beta}=0$. Written out explicitly in component form we get the following constraints
	\begin{eqnarray}
	\label{constraints}
	\dot{X}.X^{\prime}&=&0 \nonumber \\
	\dot{X}^2+X^{\prime 2} &=&0.
	\end{eqnarray}
Roughly speaking, these constraints respectively limit the string to transversal modes of oscillation and relates it's instantaneous velocity to its length.\\
\\
Having put the equations of motion for $X^{\mu}$ in a manageable form we can now write down the general solution to (\ref{EOMwave}) and expand it into Fourier modes\footnote{We assume the worldsheet coordinate $\sigma$ is subject to the periodicity condition $\sigma \sim \sigma +2\pi$. This corresponds to considering closed strings. When taking the boundary term of the variation of the Polyakov action into account one can also consider an open string with either Neumann or Dirichlet boundary conditions. In what follows we will focus on the closed string as the general features of the theory are similar.}. It is convenient to introduce light-cone coordinates $\sigma^{\pp}=\tau+ \sigma$ and $\sigma^{=}=\tau-\sigma$, after which the general solution takes the form $X^\mu (\sigma)=X_L^\mu(\sigma^\pp)+X_R^\mu(\sigma^=)$ with
	\begin{eqnarray}
	X_L^\mu(\sigma^\pp)&=&\frac{x^\mu}{2}+\frac{\alphap p^\mu}{2}\sigma^\pp +i\sqrt{\frac{\alphap}{2}} 
	\sum_{n \neq 0} \frac{\tilde{\alpha}^\mu_n}{n}e^{-in\sigma^\pp} \nonumber \\
	X_R^\mu(\sigma^{=})&=&\frac{x^\mu}{2}+\frac{\alphap p^\mu}{2}\sigma^+ +i\sqrt{\frac{\alphap}{2}}
	\sum_{n \neq 0} \frac{\alpha^\mu_n}{n}e^{-in\sigma^=},
	\end{eqnarray}
where we have introduced the real left- and right moving oscillator modes $\tilde{\alpha}^\mu_n$ and $\alpha^\mu_n$ and string momenta and positions $p^\mu$ and $x^\mu$. The momenta can be used to define zero modes
	\begin{eqnarray}
	\alpha_0^\mu \equiv \sqrt{\frac{\alphap}{2}}p^\mu, \qquad \tilde{\alpha}_0^\mu \equiv \sqrt{\frac{\alphap}{2}}p^
	\mu.
	\end{eqnarray}
In lightcone coordinates the constraints (\ref{constraints}) become
	\begin{equation}
	(\partial_\pp X)^2=(\partial_= X)^2=0.
	\end{equation}
One can easily show these constraints can be solved by introducing Fourier modes $L_n$ and $\tilde{L}_n$ for the constraints and demanding that they vanish
	\begin{eqnarray}
	L_n&=&\frac{1}{2}\sum_m \alpha_{n-m} \cdot \alpha_m =0\nonumber \\
	\tilde{L}_n&=&\frac{1}{2}\sum_m \tilde{\alpha}_{n-m} \cdot \tilde{\alpha}_m =0.
	\end{eqnarray}Just like the point particle these constraints force the excitations of the string to obey a mass shell condition. From the specific form of the $n=0$ constraint
	\begin{equation}
	L_0=\alphap p^2 +\sum_{m=1}^\infty \alpha_{-m} \cdot \alpha_{m}=0,
	\end{equation}
and similarly for $\tilde{L}_0$, we find the relation
	\begin{equation}
	M^2=\frac{4}{\alphap}\sum_{m=1}^\infty \alpha_{-m} \cdot \alpha_{m}=\frac{4}{\alpha^		
	\prime}\sum_{m=1}^\infty \tilde{\alpha}_{-m} \cdot \tilde{\alpha}_{m}.
	\end{equation}
The fact that the right- and left moving oscillations need necessarily have the same mass is known as level matching and is the only constraint linking the left- and right-movers.

\section{The quantum string}
The Polyakov action will be the starting point when we attempt to quantise the string. We will see that the quantisation of the bosonic string gives rise to an unstable vacuum and as such we will not dwell on it too long. We will however construct the physical Fock space and note that demanding that the quantum theory is consistent implies strong restrictions on the dimensionality of the target space.

\subsection{Canonical quantisation of the bosonic string}
Having defined the oscillators $\alpha^\mu_n$ and $\tilde{\alpha}^\mu_n$ one can show that they obey the following relations
	\begin{eqnarray}
	\{\alpha^\mu_n,\alpha^\nu_m\}=\{\tilde{\alpha}^\mu_n,\tilde{\alpha}^\nu_m\}&=&in \eta^{\mu\nu}
	\delta_{n	+m,0} \nonumber \\
	\{\alpha^\mu_n,\tilde{\alpha}^\nu_m\}&=&0 ,
	\end{eqnarray}
where $\{,\}$ denotes the Poisson bracket. Following the usual procedure for canonical quantisation we promote the oscillator modes to operators and replace the Poisson brackets by commutators
	\begin{eqnarray}
	 \left[\alpha^\mu_n,\alpha^\nu_m\right]=\left[\tilde{\alpha}^\mu_n,\tilde{\alpha}^\nu_m\right]&=&- 
	 n \eta^{\mu\nu}\delta_{n	+m,0} \nonumber \\
	\left[\alpha^\mu_n,\tilde{\alpha}^\nu_m\right]&=&0.
	\end{eqnarray}
The appearance of the Minkowski metric will give rise to negative norm states which tells us that the theory contains unphysical states. This is to be expected since we still need to impose the constraints. The quantum equivalent of demanding that the $L_n$'s and $\tilde{L}_n$'s vanish becomes a condition on the physical states $| \phi \rangle$
	\begin{eqnarray}
	L_n | \phi \rangle=\tilde{L}_n | \phi \rangle=0 && n>0.
	\end{eqnarray}
One should be careful to considering operators consisting of products of oscillators due to the non-trivial commutation relations. Specifically, there arises an ordering ambiguity in the definition of the zero modes $L_0$ and $\tilde{L}_0$. One can chose normal ordering for these operators at the price of picking up an undetermined constant $a$ at the level of the zero mode constraints
	\begin{equation}
	(L_0-a) | \phi \rangle=(\tilde{L}_0-a) | \phi \rangle=0.
	\end{equation}
The constant $a$ has as a consequence that the masses of the excitations are shifted depending on its value
	\begin{eqnarray}
	M^2&=&\frac{4}{\alphap}\left( -a +\sum_{m=1}^\infty \alpha_{-m} \cdot \alpha_{m} \right)
	\nonumber \\
	&=&\frac{4}{\alphap}\left( -a +\sum_{m=1}^\infty \tilde{\alpha}_{-m} \cdot \tilde{\alpha}_{m} 
	\right). 
	\end{eqnarray}
We conclude this subsection by arguing that the quantisation scheme described here results in a Lorentz invariant theory. Give the usual definition for the generators of the Lorentz algebra we can calculate
	\begin{equation}
	\label{Lorentzgenerator}
	J^{\mu\nu}=x^\mu p^\nu-x^\nu p^\mu -i \sum \frac{1}{n}(\alpha^\nu_{-n} \alpha^\mu_{n} -\alpha^
	\mu_{-n} \alpha^\nu_{n}) + \cdots
	\end{equation}
where the $\dots$ denote a similar oscillator piece in terms of the $\tilde{\alpha}$'s. The generators now $J^{\mu\nu}$ consist of a familiar piece relating to the string center of mass and additional terms  due to the oscillators. Given this expression a straightforward calculation shows that 
	\begin{equation}
	[L_n, J^{\mu\nu}]=0,
	\end{equation}
guaranteeing that physical states obeying the constraints will consist of Lorentz multiplets, thereby proving that the resulting theory is indeed Lorentz invariant.

\subsection{The bosonic string spectrum}
So far in our study of the quantum string we kept Lorentz covariance manifest. The price to pay for this was that the Weyl symmetry developed an anomaly and that we had to project out part of the states in order to avoid ghosts. To determine the string spectrum explicitly it is convenient to use residual gauge invariance to manifestly make all the states physical, at the cost of losing manifest Lorentz invariance.\\
\\
We proceed by singling out two directions $X^\pm=\frac{1}{\sqrt{2}}\left(X^0 \pm X^{D-1}\right)$ as lightcone coordinates. As described in the previous subsection we have residual Weyl transformations remaining. Invoking this remaining freedom we will work in the so called lightcone gauge where $X^+$ takes a particularly simple form
	\begin{equation}
	\label{lightconegauge}
	X^+=x^+ + \alphap p^+ \tau.
	\end{equation}
Checking that the gauge choice (\ref{lightconegauge}) is consistent with the constraints we find that the constraints themselves fully determine $X^-$ in terms of the transverse directions $X^i$ (barring two constants of integration)
	\begin{eqnarray}
	\partial_+ X_L^- = \frac{1}{\alphap p^+}\sum_{i=1}^{D-2}\partial_+X^i\partial_+X^i \nonumber \\
	\partial_- X_R^- = \frac{1}{\alphap p^+}\sum_{i=1}^{D-2}\partial_-X^i\partial_-X^i.
	\end{eqnarray}
The advantage of this gauge choice is that the operators $X^\pm$ no longer contain oscillators. Effectively we have eliminated the need for longitudinal oscillation modes altogether, where previously these had to be truncuated from the state space by imposing constraints.\\
\\
The price we pay for this convenience is that we have lost explicit Lorentz invariance as the theory will have developed a Lorentz anomaly at the quantum level. For generic choices of $a$ and $D$ we are dealing with a so-called non-critical string theory where, depending on the description, the Weyl or Lorentz anomaly is still present. A number of different ways exist to calculate the precise values of $a$ and $D$ where the quantum theory becomes anomaly-free, at various levels of rigor. \\
\\
One way of determining the allowed values of $a$ en $D$ is to look at the generators defined in (\ref{Lorentzgenerator}). If the theory is to be Lorentz invariant they must obey the Lorentz algebra
	\begin{equation}
	\label{Lorentzalgebra}
	[J^{\mu\nu},J^{\rho\sigma}]=\eta^{\nu\rho}J^{\mu\sigma}-\eta^{\mu\rho}J^{\nu\sigma}-\eta^{\nu
	\sigma}J^{\mu\rho}+\eta^{\mu\sigma}J^{\nu\rho}.
	\end{equation}
Quantum effects give rise to an obstruction when considering the commutators of the elements $J^{i-}$. A detailed calculation shows that
	\begin{equation}
	[J^{i-},J^{i-}] \sim \sum_{n>0}\left( \left[\frac{D-2}{24}-1\right]n+\frac{1}{n}\left[a-\frac{D-2}
	{24}\right]\right).
	\end{equation}
Since (\ref{Lorentzalgebra}) would imply that this commutator must vanish we are force to conclude that quantum consistency of the theory is only possible when $a=1$ and $D=26$.\\
\\
Needless to say this is a very strong constraint on the target space, and one that carries significant phenomenological difficulties. As mentioned in the introduction, various scenarios have been suggested to reconcile this fact with the observation that our visible universe consists of a four-dimensional spacetime.\\
\\
Having determined the critical values we can now investigate the particle spectrum of the closed bosonic string. It is convenient the introduce the so-called levels $N$ and $\tilde{N}$ as follows
	\begin{eqnarray}
	N=\sum_{i=1}^{D-2}\sum_{n>0}\alpha^i_{-n}\alpha^i_n, \qquad \tilde{N}=\sum_{i=1}^{D-2} 
	\sum_{n>0}\tilde{\alpha}^i_{-n}\tilde{\alpha}^i_n.
	\end{eqnarray}
In terms of these levels the masses of the various states are given by
	\begin{equation}
	\label{masscondition}
	M^2=\frac{4}{\alphap}(N-a)=\frac{4}{\alphap}(\tilde{N}-a).
	\end{equation}
The previously derived fact that $a=1$ now has an important consequence on the theory as a whole when we investigate the spectrum. We define the vacuum state $| 0,p^\mu \rangle$ as
	\begin{eqnarray}
	\hat{p}^\mu| 0, p^{\mu} \rangle=p^\mu | 0,p^\mu \rangle, \qquad \alpha^\mu_i | 0,p^\mu \rangle = 
	\tilde{\alpha}^\mu_i| 0,p^\mu \rangle =0
	\end{eqnarray}
Since no oscillators are present we find that the mass of these excitations is given by
	\begin{equation}
	M^2=-\frac{4}{\alphap}, 
	\end{equation}
in other words, the ground state is tachyonic. At the very least, this implies that the chosen ground state does not provide a stable vacuum. We will currently ignore this issue and continue acting once with the oscillators on the vacuum state
	\begin{equation}
	\tilde{\alpha}^j_{-1}\alpha^i_{-1} | 0,p^\mu \rangle,
	\end{equation}
which is massless $M^2=0$. Since we have 24 different values for $i,j$ this state transforms as rank two tensor under $SO(24)$. This state decomposes into irreducible representations as
	\begin{equation}
	\mathbf{24} \otimes \mathbf{24} =\mathbf{299} \oplus \mathbf{276} \oplus \mathbf{1},
	\end{equation}
in other words a traceless symmetric tensor $G_{\mu\nu}$, an asymmetric tensor $B_{\mu\nu}$ and the trace $\Phi$. The conclusion is that the massless sector of the closed string comprises of
\begin{itemize}
\item The graviton $g_{\mu\nu}(X)$
\item The Kalb-Ramond 2-form $b_{\mu\nu}(X)$
\item The dilaton $\Phi(X)$.
\end{itemize}
The natural appearance of massless spin-2 excitations was one of the main surprises that led to the initial enthusiasm for string theory. The Kalb-Ramond field is a new ingredient which will have a profound influence on the geometry of the target space as we will see later.\\
\\
As equation (\ref{masscondition}) describes an infinite tower of massive states there are many more possible excitations beyond the two sectors considered thus far. However, since these masses are proportional to $\frac{1}{\alphap} \sim M_P^2$ there is little need to investigate them further at this point.\\
\\
We could repeat this analysis for the open string spectrum where we would find a similar situation. The ground state is once again tachyonic, the massless sector consists of a $SO(24)$ vector corresponding to a gauge field $A_\mu(X)$, and beyond that there exists an infinite tower of massive states.\\
\\
This concludes our overview of the bosonic string. Although many novel features are present, the tachyon and the critical dimension $D=26$ are cause for concern. In the next chapter we will consider the superstring by adding supersymmetry to the worldsheet.

\chapter{Superstrings}
\thispagestyle{empty}
\noindent \emph{In the previous chapter we reviewed the basic properties of the bosonic string theory. Even though it was possible to formulate a consistent quantum theory we encountered some worrisome elements. Firstly, the entire particle spectrum was bosonic, which obviously makes it unsuitable for a fundamental theory since any such theory must include fermions as well. Secondly, the quantum theory was only consistent for a specific choice of space-time dimension. Lastly, the tachyonic ground state raises questions about the validity of the chosen ground state.\\
\\
In this chapter we will include fermionic degrees of by introducing supersymmetry. There exist four major ways of formulating a theory of superstrings
\begin{itemize}
\item The Ramond-Neveu-Schwarz (RNS) formalism which makes the worldsheet supersymmetric.
\item The Green-Schwarz (GS) formalism which starts with a supersymmetric spacetime.
\item The Berkovits or pure spinor formalism.
\item The superembedding formalism, developed by Bandos, Sorokin, and others.
\end{itemize}
In order to elucidate the superstring spectrum we will focus on the RNS formalism in this chapter. }
\newpage
\section{The RNS formalism}
Our starting point will be the Polyakov action in the conformal gauge
	\begin{equation}
	S_P=-\frac{1}{2 \pi \alphap} \int d^2\sigma \partial_\alpha X^\mu \partial^\alpha X_\mu,
	\end{equation}
where as before the fields $X^\mu(\tau,\sigma)$ are bosonic scalar fields on the worldsheet and the components of a Lorentz vector in the target space. We now add a spacetime Lorentz vector $\psi^\mu(\tau,\sigma)$ whose components are Majorana spinors on the worldsheet
	\begin{equation}
	\Psi^\mu(\tau,\sigma)=\left(
	\begin{array}{c}
 	\psi_-^\mu(\tau,\sigma)\\
	 \psi_+^\mu(\tau,\sigma)
	\end{array}\right).
	\end{equation}
By introducing the 2-dimensional Dirac matrices $\{\rho^\alpha,\rho^\beta\}=2\eta^{\alpha\beta}$ we can add the standard kinetic term form the fermions to the action as follows
	\begin{equation}
	\label{susystring}
	S=-\frac{1}{2 \pi \alphap} \int d^2\sigma \left(\partial_\alpha X^\mu \partial^\alpha X_\mu+\bar{\psi}^\mu\rho^\alpha 
	\partial_\alpha \psi_\mu\right).
	\end{equation}
Expanding this action into the chiral components $\psi^\mu_{\pm}$ of the fermions and by choosing an appropriate basis for the Dirac algebra the equations of motion for the new degrees of freedom become the familiar Dirac equation
	\begin{eqnarray}
	\label{Dirac}
	\partial_+\psi_-^\mu=0, \qquad \partial_-\psi_+^\mu=0.
	\end{eqnarray}
The action (\ref{susystring}) now has a new global symmetry that relates the bosonic and fermionic fields. We call the infinitesimal parameter characterising this symmetry $\epsilon$, itself a constant Majorana spinor. This global supersymmetry takes the following form
	\begin{eqnarray}
	\delta X^\mu &=& \bar{\epsilon} \Psi^\mu \nonumber \\
	\delta \Psi^\mu &=&\rho^\alpha \partial_\alpha X^\mu \epsilon.
	\end{eqnarray}
One easily checks that these transformations leave the action invariant up to a total derivative. \\
\\
In general the supersymmetry algebra defined by the previous transformations does not close off-shell. One can check that the commutator of two consecutive transformations of the fermion fields $\delta_1 \psi^\mu$ and $\delta_2 \psi^\mu$ yields
	\begin{equation}
	[\delta_{\epsilon_1},\delta_{\epsilon_2}] \Psi^\mu = - \bar \epsilon_1 \rho^\alpha \epsilon_2 \partial_\alpha \Psi^\mu
	+ \bar \epsilon_1 \rho_\beta \epsilon_2 \rho^\beta \rho^\alpha \partial_\alpha \Psi^\mu.
	\end{equation}
The first term corresponds to a translation, whereas the second term can be made to vanish by using the equations of motion. If we want the supersymmetry algebra to close off-shell we will have to introduce an auxiliary field that does not carry additional dynamics. This is a generic feature which will prove to be important later on. 

\subsection{Open string RNS boundary conditions}

In our discussion of the bosonic string we glossed over the fact that when varying the action one has to take the resulting boundary term into account when solving the equations of motion. For the bosonic string this amounted to taking the $\sigma$ coordinate to be periodic, leaving us with a closed string, or imposing so-called Dirichlet or Neumann boundary conditions on the endpoints, resulting in an open string with free moving or fixed endpoints, respectively. When considering the superstring we will have to impose similar boundary conditions. We shall now review them in some detail as they will give raise to the different particle sectors of the theory.\\
\\
Consider the fermionic part of the action (\ref{susystring})
	\begin{equation}
	S_F \sim \int d^2 \sigma \left( \psi_-\partial_+ \psi_- + \psi_+\partial_- \psi_+ \right).
	\end{equation}
When varying this part of the action we are left with the following boundary term
	\begin{equation}
	\label{susystringvar}
	\delta S_F \sim \int d\tau \left( \psi_-\delta\psi_- - \psi_+\delta\psi_+\right)\left. \right|_{\sigma=\pi} -  \left( \psi_-\delta
	\psi_- - \psi_+\delta\psi_+\right)\left. \right|_{\sigma=0}.
	\end{equation}
These terms vanish if we impose the following condition
	\begin{equation}
	\psi^\mu_+ = \pm \psi^\mu_-.
	\end{equation}
Since the overall sign can be chosen freely, we can take one endpoint to obey
	\begin{equation}
	\psi^\mu_+ \left. \right|_{\sigma=0}= \psi^\mu_-\left. \right|_{\sigma=0}.
	\end{equation}
This leaves us with the other endpoint where we have two different possibilities. These two choices correspond to distinct sectors of the theory and are called the Ramond and Neveu-Schwartz sectors.
\begin{itemize}
\item If we chose the other endpoint to obey
	\begin{equation}
	\psi^\mu_+ \left. \right|_{\sigma=\pi}= \psi^\mu_-\left. \right|_{\sigma=\pi},
	\end{equation}
the so-called Ramond boundary condition, we can expand the solutions to the Dirac equation (\ref{Dirac}) as follows
	\begin{eqnarray}
	\psi^\mu_-(\tau,\sigma)&=&\frac{1}{\sqrt{2}}\sum_{n \in \mathbb{Z}}d^\mu_n e^{-in(\tau-\sigma)} \nonumber \\
	\psi^\mu_+(\tau,\sigma)&=&\frac{1}{\sqrt{2}}\sum_{n \in \mathbb{Z}}d^\mu_n e^{-in(\tau+\sigma)}.
	\end{eqnarray}
The Fourier components $d^\mu_n$ again represent oscillator modes and are real due to the Majorana condition imposed on $\psi^\mu(\tau,\sigma)$.
\item  If, on the other hand, we chose the other endpoint to obey
	\begin{equation}
	\psi^\mu_+ \left. \right|_{\sigma=\pi}=- \psi^\mu_-\left. \right|_{\sigma=\pi},
	\end{equation}
we are left with Neveu-Schwarz boundary conditions. Imposing these boundary conditions we are left with the following solutions
	\begin{eqnarray}
	\psi^\mu_-(\tau,\sigma)&=&\frac{1}{\sqrt{2}}\sum_{r \in \mathbb{Z}+\frac{1}{2}} b^\mu_r e^{-ir(\tau-\sigma)} 
	\nonumber \\
	\psi^\mu_+(\tau,\sigma)&=&\frac{1}{\sqrt{2}}\sum_{r \in \mathbb{Z}+\frac{1}{2}}b^\mu_r e^{-ir(\tau+\sigma)}.
	\end{eqnarray}
Note that in addition to having different Fourier components $b^\mu_r$, the index $r$ now runs over half integer values.
\end{itemize}

\subsection{Closed string RNS boundary conditions}
As with the bosonic string, the closed string essentially consists of two copies of the open string oscillator modes by considering periodic boundary conditions. The different possibilities are
	\begin{equation}
	\psi^\mu_\pm (\tau,\sigma)= \pm \psi^\mu_\pm(\tau,\sigma+\pi),
	\end{equation}
which gives rise to the aforementioned Ramond (corresponding to the positive sign) and Neveu-Schwartz (corresponding to the negative sign) boundary conditions. The resulting solutions take the following form when imposing Ramond boundary conditions
	\begin{eqnarray}
	\psi^\mu_-(\tau,\sigma)&=&\frac{1}{\sqrt{2}}\sum_{n \in \mathbb{Z}}d^\mu_n e^{-2in(\tau-\sigma)} \nonumber \\
	\psi^\mu_+(\tau,\sigma)&=&\frac{1}{\sqrt{2}}\sum_{n \in \mathbb{Z}}\tilde{d}^\mu_n e^{-2in(\tau+\sigma)},
	\end{eqnarray}
where the main difference is that we have different sets of oscillator modes $d^\mu_n$ and $\tilde{d}^\mu_n$ for the left- and right-movers. The Neveu-Schwartz boundary conditions similarly give rise to the solutions
\begin{eqnarray}
	\psi^\mu_-(\tau,\sigma)&=&\frac{1}{\sqrt{2}}\sum_{r \in \mathbb{Z}+\frac{1}{2}} b^\mu_r e^{-2ir(\tau-\sigma)} 
	\nonumber \\
	\psi^\mu_+(\tau,\sigma)&=&\frac{1}{\sqrt{2}}\sum_{r \in \mathbb{Z}+\frac{1}{2}}\tilde{b}^\mu_r e^{-2ir(\tau+\sigma)}.
	\end{eqnarray}
Since both boundary conditions can be applied independently to the left-moving and right-moving solutions, and that the full solution consists of superpositions of both, we are left with four different sectors, the R-R, NS-R, R-NS and NS-NS sectors.

\section{Light-cone quantisation of the RNS superstring}
Having constructed the classical solutions to the equations of motion obeying the various boundary conditions we can proceed to canonically quantise the RNS superstring. As usual we promote the different oscillator modes to operators. The oscillators corresponding to the $X^\mu$ fields obey the same commutation relations as the ones characterising the bosonic string, while the new element is that the oscillators making up the worldsheet fermions $\Psi^\mu_\pm$ obey anti-commutation relations of the form
	\begin{eqnarray}
	\{b^\mu_r,b^\nu_s\}=\delta_{r,-s}\eta^{\mu\nu}, \qquad \{d^\mu_m,d^\nu_n\}=\delta_{m,-n}\eta^{\mu\nu},
	\end{eqnarray}
in the case of the open string, with identical relations for the left movers in the closed string case.

\subsection{Super-Virasoro conditions}
Due to the appearance the worldsheet metric $\eta^{\mu\nu}$ we once again have ghost states in the spectrum that need to be dealt with. On the classical level, the worldsheet theory is still Weyl-invariant but the addition of supersymmetry on the worldsheet enlarges the this symmetry. This is again reflected in the vanishing of the various components stress-energy tensor $T_{\alpha\beta}$
	\begin{eqnarray}
	T_{\pp =}=T_{= \pp}&=0  \nonumber \\
	T_{\pp \pp}=T_{= =}&=0 .
	\end{eqnarray}
The Fourier components of the stress-energy tensor $L_n$ and $\tilde{L}_n$ now receive additional contributions from the fermionic modes, the explicit form of which need not concern us here. Associated to the supersymmetry parameters $\epsilon_{\pm}$ are two conserved supercurrents
	\begin{eqnarray}
	j^\mu_\pm=\Psi^\mu_\pm \partial_\pm X_\mu,
	\end{eqnarray}
 that can be expanded into Fourier components as well. These expansions take on different forms in the NS and R sectors of the theory and we will denote their components by $(G_r,\tilde{G}_r)$ and $(F_n,\tilde{F}_n$), respectively.\\
 \\
As might be expected the theory acquires an anomaly after quantisation that is a function of the central charge (the value of which is related to the dimensionality of the target space). We are now left with the Fourier components as operators which form the super-Virasoro algebra\footnote{We omit the obvious extension to the closed string for clarity purposes.}
	\begin{eqnarray}
	\left[L_m,L_n\right]&=&(m-n)L_{m+n}+\frac{D}{8}m(m^2-1)\delta_{m,-n} \nonumber \\
	\left[L_m,G_r\right]&=&\left( \frac{m}{2}-r\right) G_{m+r} \nonumber \\
	\{G_r,G_s\}&=&2L_{r+s}+\frac{D}{2}\left( r^2-\frac{1}{4}\right)\delta_{r,-s}
	\end{eqnarray}
for the NS sector and
	\begin{eqnarray}
	\left[L_m,L_n\right]&=&(m-n)L_{m+n}+\frac{D}{8}m^3\delta_{m,-n} \nonumber \\
	\left[L_m,F_n\right]&=&\left( \frac{m}{2}-n\right) F_{m+r} \nonumber \\
	\{F_m,F_n\}&=&2L_{m+n}+\frac{D}{2}m^2\delta_{m,-n}
	\end{eqnarray}
for the R sector.\\
\\
Related to the conformal anomaly the operator $L_0$ acquires an ordering ambiguity. Normal ordering then introduces ordering constants $a_R$ and $a_{NS}$ which will differ between both sectors of the theory. We are now ready to impose the super-Virasoro conditions which will eliminate the ghost states from the theory. The conditions
	\begin{eqnarray}
	L_n| \phi \rangle&=&0 \hspace{3mm} n>0 \nonumber \\
	G_r | \phi \rangle &=&0 \hspace{3mm} r>0 \nonumber \\
	(L_0-a_{NS})| \phi \rangle&=& 0,
	\end{eqnarray}
define the physical states in the NS-sector and 
	\begin{eqnarray}
	L_m| \phi \rangle&=&0 \hspace{3mm} n> 0 \nonumber \\
	F_n | \phi \rangle &=&0 \hspace{3mm} n\geq 0 \nonumber \\
	(L_0-a_{R})| \phi \rangle&=& 0,
	\end{eqnarray}
do likewise in the R-sector. \\
\\
The additional zero mode constraint $F_0| \phi \rangle$ in the NS-sector goes by the name of the Dirac-Ramond equation, which can be regarded as a stringy generalisation of the familiar Dirac equation. The $L_0$ conditions once again imply a mass-shell condition which will now use to investigate the ground state of the theory.

\subsection{The ground states}
As with the bosonic string there exists residual gauge symmetry, now for both the bosonic and fermionic degrees of freedom. This allows us to go to the light-cone gauge which eliminates the longitudinal oscillation modes, at the price of losing manifest Lorentz invariance. Proceeding in this way we will now investigate the ground states of the quantised superstring, more specifically the R- and NS-sector of the open superstring. As usual the ground states will be those states that are eliminated by the creation operators which gives us
	\begin{eqnarray}
	\label{superground}
	\alpha^i_{n} | 0,p^\mu \rangle_{NS}=b^i_r| 0,p^\mu \rangle_{NS}&=&0 \text{  with  } n,r>0\\
	\alpha^i_{n} | 0,p^\mu \rangle_{NS}=d^i_r| 0,p^\mu \rangle_{NS}&=&0 \text{  with  } n>0.
	\end{eqnarray}
Before defining the mass-squared operator the ordering ambiguities still need to be dealt with. One can show that quantum consistency of the super-Virasoro algebra requires $D=10$, or, equivalently, $a_R=0$ and $a_{NS}=\frac{1}{2}$. \\
\\
According to this choice the masses of the NS-sector excitations are given by
	\begin{equation}
	\alphap M^2 = \sum_{n=1}^\infty \alpha_{-n}^i \alpha_n^i +\sum_{r=\frac{1}{2}}^\infty b^i_{-r} b^i_{r}-\frac{1}{2},
	\end{equation}
which unfortunately implies that our chosen ground state is again tachyonic. The first excited state, obtained by acting once with fermionic oscillators, is massless and has 8 degrees of freedom, signifying that this is an $SO(8)$ vector. Subsequent excited states are constructed by acting on them with the various oscillation modes.\\
\\
The R-sector ground state on the other hand is massless as can be seen by considering the mass-shell condition in the R-sector
	\begin{equation}
	\alphap M^2 = \sum_{n=1}^\infty \alpha_{-n}^i \alpha_n^i +\sum_{n=1}^\infty d^i_{-n} d^i_{n}.
	\end{equation}
In addition to being massless, the ground state defined in (\ref{superground}) is degenerate,  as can be shown as follows. In contrast to the NS-sector the solutions $\psi_\pm(\tau,\sigma)$ consistent with the Ramond boundary conditions contain a zero mode $\psi^\mu_0$ satisfying (up to normalization) the Dirac algebra $\{\psi_0^\mu,\psi_0^\nu\}=\eta^{\mu \nu}$. For these zero modes to act on the ground state (and indeed all subsequent states in the R-sector) it must furnish a representation of the ten-dimensional Dirac algebra. In other words, the R-sector states are ten dimensional Dirac spinors.\\
\\
To conclude this section we note that by virtue of the reality conditions imposed on the oscillators the R-sector states are Majorana (real) spinors. We can simultaneously impose a further chirality or Weyl condition on these spinors, something that is only possible when $D=2\text{ mod }8$. This will give rise to two distinct superstring theories, respectively called type IIA and type IIB superstring theory.
	
\subsection{Getting rid of the tachyon : the GSO projection}

After canonical quantisation and fixing the critical dimension the theory still has problems. Firstly, even though supersymmetry was introduced on the worldsheet it is not yet realised in the target space. Secondly, we still have a tachyonic ground state. In contrast with the bosonic string there exists a systematic way to eliminate the tachyonic state from the particle spectrum. As an added bonus, the same projection realises supersymmetry in the target space.\\
\\
First we introduce two operators that count the number of fermionic oscillators in the R and NS sectors
	\begin{eqnarray}
	F_{NS}=\sum_{r}^\infty b^i_{-r}b^i_r , \qquad F_R=\sum_n^\infty d^i_{-n} d^i_n.
	\end{eqnarray}
This allows us to define the following two parity operators
	\begin{eqnarray}
	G_{NS}=(-1)^{F_{NS}+1}, \qquad G_{R}=\Gamma^{11}(-1)^{F_R},
	\end{eqnarray}
where we have introduced the 10 dimensional chirality operator $\Gamma^{11}=\prod_{i=0}^9 \Gamma^i$. The parity operators $G$ define a projection, named after Gliozzi, Scherk, and Olive, dividing the spectrum in states of positive and negative G-parity. In the NS-sector, this condition
	\begin{equation}
	G_{NS} | \Omega \rangle_{NS}=(-1)^{F_{NS}+1}| \Omega \rangle_{NS}= | \Omega \rangle_{NS},
	\end{equation}
forces us to only keep states created out of an odd number of $b^{\mu}_r$ oscillators. More specifically, the tachyonic ground state is projected out and we are left with the massless $SO(8)$ vector as the new ground state.\\
\\
In the R-sector, the appearance of the $\Gamma^{11}$ operator leaves us with a choice of an odd or even number of $d^\mu_n$ oscillators depending on the choice of chirality of the ground state as alluded to in the previous subsection. This choice will correspond to Type IIA or Type IIB superstring theory, respectively.\\

\subsection{Physical superstring spectrum}
After imposing the super-Virasoro conditions and truncating the resulting ghost-free spectrum by preforming the GSO-projection we are left with the final particle spectrum of a variety of a variety of superstring theories. We will provide a brief overview.

\subsubsection{Type I superstring theory}
The starting point for Type I superstring theory are the open superstrings considered thus far. After the GSO-projection the surviving massless ground states. Massless particle states in $D=10$ will transform in an 8-dimensional representation of its little group $SO(8)$, or more correctly its dual cover $Spin(8)$. $Spin(8)$ is unique in that its spinor representations have the same dimensionality as the vector representation $\mathbf{8}_V$\footnote{This feature of the Lie Algebra $D_4$ is known as \emph{triality}, which plays a role in the Green-Schwarz formalism and is related to the fact that in $D=10$ the particle states possess spacetime supersymmetry.}. In addition, $Spin(8)$ possesses two inequivalent spinor representations which we shall call $\mathbf{8}_S$ and $\mathbf{8}_C$.\\
\\
As we have seen, in the NS-sector of the theory the tachyonic ground state (a $Spin(8)$ singlet) gets projected out leaving us with a massless $SO(1,9)$ vector transforming as $\mathbf{8}_V$. The R-sector on the other hand consisted of a massless Majorana spinor that can be further be decomposed in two Majorana-Weyl spinors transforming as $\mathbf{8}_S \oplus \mathbf{8}_C$. The GSO-projection picks out one of these components, say $\mathbf{8}_S$. The massless states of the open Type I superstrings thus consist of a spacetime $\mathcal{N}=1$ supermultiplet containing the vector bosons $A_\mu^a$ and their superpartners $\lambda_\mu^a$. The associated low-energy effective action describing these degrees of freedom in spacetime is $D=10$, $\mathcal{N}=1$ super-Yang-Mills theory.\\
\\
Consistency of the theory requires the presence of closed strings as well as open strings. The closed strings in Type I superstring theory are obtained by tensoring the open string ground states according to
	\begin{eqnarray}
	(\mathbf{8}_V \oplus \mathbf{8}_S) \otimes (\mathbf{8}_V \oplus \mathbf{8}_S) \mod \mathbb{Z}_2 =
	( \mathbf{8}_V \otimes \mathbf{8}_V) \oplus (\mathbf{8}_V \otimes \mathbf{8}_S).
	\end{eqnarray}
The modding by $\mathbb{Z}_2$ ensures that the closed Type I strings have $\mathcal{N}=1$ spacetime supersymmetry, which is required for their consistent coupling to the open string theory\footnote{This requirement can also be obtained from considering the modular invariance of the theory. We will not discuss these issues here.}. Decomposing this into irreducible $SO(1,9)$ representations we obtain the field content 
	\begin{equation}
	(\mathbf{35}_V \oplus \mathbf{28}_V \oplus \mathbf{1}) \oplus (\mathbf{56}_S \oplus \mathbf{8}_C),
	\end{equation}
corresponding to the graviton $g_{\mu\nu}$, the Kalb-Ramond field $b_{\mu\nu}$, the dilaton $\Phi$, the gravitinos $\Psi_\mu$ and the dilatino $\varphi_\mu$. The low energy effective field theory corresponding to this field content is  $D=10$, $\mathcal{N}=1$ supergravity.

\subsubsection{Type IIA and Type IIB superstring theory}
The Type II superstring theories consist of closed strings giving rise to fields exhibiting extended spacetime supersymmetry. In the Type I theory we obtained closed strings from tensoring together two copies of the open string spectrum followed by a truncation by $\mathbb{Z}_2$. In contrast, the closed strings in Type II theory are obtained from tensoring the full spectrum. This allows for two inequivalent choices resulting from the fact that the GSO-projection allowed for two possible R-sector ground states of opposite chirality. The two superstring theories corresponding to this choice are called Type IIA and Type IIB. Their massless particle states are obtained by considering
	\begin{eqnarray}
	\text{Type IIA : } (\mathbf{8}_V \oplus \mathbf{8}_S) \otimes (\mathbf{8}_V \oplus \mathbf{8}_C) \\
	\text{Type IIB : } (\mathbf{8}_V \oplus \mathbf{8}_S) \otimes (\mathbf{8}_V \oplus \mathbf{8}_S).
	\end{eqnarray}
The massless states of the theories consist of four distinct sectors. The first, called the NS-NS sector contains bosonic states and is common to both theories
	\begin{equation}
	\mathbf{8}_V \otimes \mathbf{8}_V = \mathbf{35}_V \oplus \mathbf{28}_V \oplus \mathbf{1}
	\end{equation}
We recover the graviton, Kalb-Ramond 2-form and dilaton fields just like in the Type I theory. The other bosonic sector is called the R-R sector and contains the fields
	\begin{eqnarray}
	\text{Type IIA : } \mathbf{8}_S \otimes \mathbf{8}_C \\
	\text{Type IIB : } \mathbf{8}_S \otimes \mathbf{8}_S,
	\end{eqnarray}
which correspond to bosonic $p$-form fields called RR-fields. The Type IIA theory contains a 1-form $C_\mu^{(1)}$ and a 3-form $C_{\mu\nu\rho}^{(3)}$, while the IIB theory contains a 0-form $C^{(0)}$, a 2-form $C_{\mu\nu}^{(2)}$ and a self-dual 4-form $C^{(4)}_{\mu\nu\rho\sigma}$. As alluded to in the introduction, these RR-fields play an important part in string compactification scenarios where dimensional reduction gives rise to unwanted scalar fields, the so called moduli of the compactification. Non-zero RR-fluxes can be used to stabilise these moduli. We will not consider them further in this work.\\
\\
The NS-R and R-NS sectors of the Type II theories will give rise to spacetime fermions. These sectors are described by the decompositions
	\begin{eqnarray}
	\text{Type IIA : } \mathbf{8}_V \otimes \mathbf{8}_C = \mathbf{8}_S \oplus \mathbf{56}_C\\
	\text{Type IIB : } \mathbf{8}_V \otimes \mathbf{8}_S = \mathbf{8}_S \oplus \mathbf{56}_S.
	\end{eqnarray}
Each sector contains a gravitino and a dilatino. The two gravitini have the same chirality in the Type IIB theory, and opposite chiralities in the Type IIA theory.\\
\\
Taken together, the low energy effective theories corresponding with the massless content of the Type II superstring theories the two unique $\mathcal{N}=2$ $D=10$ supergravity theories, conveniently called Type IIA and Type IIB supergravity. More specifically, the type IIA theory corresponds to $\mathcal{N}=(1,1)$ supergravity while the type IIB theory corresponds to $\mathcal{N}=(2,0)$ supergravity in $D=10$. It is the NS-NS sector of these two theories that will be the focus of this work.
 
\subsubsection{Heterotic string theories}
There are two additional consistent superstring theories that can be formulated in $D=10$. These so-called heterotic theories use the fact that for closed strings the right- and left-moving sectors of the spectrum decouple. The idea is to combine the $D=26$ bosonic string theory for the left movers with a $D=10$ supersymmetric string theory for the right movers. The resulting theory can consistently be formulated and can be shown to have $\mathcal{N}=1$ spacetime supersymmetry. Furthermore, quantum consistency requires that these models contain massless Yang-Mills multiplets corresponding to either a $SO(32)$ or $E_8 \times E_8$ gauge symmetry, giving rise to two types of heterotic string.\\
\\
This concludes our overview of supersymmetric string theories. In the next chapter, we will examine the low-energy effective action of a superstring probing spacetimes that are no longer flat, since it should be expected that more complicated backgrounds can arise by virtue of excitations of the graviton and Kalb-Ramond fields. The resulting theories are known as non-linear $\sigma$-models and will be the main focus of this thesis.

\chapter{Non-linear $\sigma$-models}
\thispagestyle{empty}
\noindent \emph{The previous two chapters have set the stage for the rest of this thesis. So far we have considered (super)strings moving in a flat Minkowski target space. The various oscillation modes of the string then gave rise to various dynamical fields that we then identified with, amongst others, the graviton field $g_{\mu\nu}$ and the Kalb-Ramond field $b_{\mu\nu}$. We should expect the excitations of these fields to determine the various geometrical structures that result when strings interact with each other. In principle, one could be expected to calculate these resulting low-energy background fields given an arbitrary number of interacting superstrings, however as of the time of writing this is an open problem.\\
\\
The approach we need to take then becomes more subtle. We will assume that the string propagates in a spacetime which is no longer flat but contains non-trivial background fields that the string will probe. The framework in which these configurations can be analysed are known as non-linear $\sigma$-models, for various historical reasons. Our starting point will be a non-linear $\sigma$-model describing the various fields on the worldsheet of a propagating string which will take values in an ambient target space. As we will see, quantum consistency of the non-linear $\sigma$-model will then constrain the geometry of the target space in various ways, depending on the amount of symmetry present on the worldsheet.}

\section{A bosonic $\sigma$-model with background fields}
The Polyakov action introduced in chapter 2 was used to describe the movement of a bosonic string in a flat Minkowski target space. We will now investigate this action from a more general $\sigma$-model point of view. Having the string propagate in Minkowski space meant that the action included the metric $\eta_{\mu\nu}$, which can be regarded as a collection of coupling constants between the different bosonic fields of the theory. For this specific choice of metric, which is diagonal, the various kinetic terms in the action do not couple to each other, which allowed us to quantise the theory as a collection of decoupled harmonic oscillators. More generally, we can couple the worldsheet scalars to each other by using the field $g_{\mu\nu}(X)$
	\begin{equation}
	S_G=-\frac{1}{4 \pi \alphap}\int d^2\sigma \sqrt{-\gamma}\gamma^{\alpha \beta}g_{\mu\nu}(X)\partial_\alpha X^\mu
	\partial_\beta X^\nu.
	\end{equation}
This action is not the most general renormalisable $d=2$ non-linear $\sigma$-model. It is possible to add two additional terms
	\begin{eqnarray}
	S_B&=&-\frac{1}{4 \pi \alphap} \int d^2\sigma \epsilon^{\alpha \beta}b_{\mu\nu}(X)\partial_\alpha X^\mu \partial_
	\beta X^\nu \\
	S_D&=&\frac{1}{4 \pi}\int d^2\sigma \sqrt{-\gamma}R^{(2)}\Phi(X).
	\end{eqnarray}
From the perspective of the $\sigma$-model $b_{\mu\nu}$(X) is an antisymmetric rank 2 tensor whose components consist of additional coupling constants and $\Phi(X)$ a scalar coupling. From the perspective of the ambient geometry in which the string propagates these objects correspond to the geometrical data characterising the background geometry. \\
\\
For $\Phi(X)=$ constant the term (4.3) is a total derivative and as such only defined by the global topology of the worldsheet. The topological invariant appearing in this manner is the Euler characteristic 
\begin{equation*}
\chi(\Sigma) = \frac{1}{4 \pi} \int_\Sigma d^2 \sigma \sqrt{-\gamma} R^{(2)}(\gamma)
\end{equation*}
One of the main results from investigating the Polyakov action was that requiring the theory to remain Weyl invariant after quantisation limited the spacetime dimension to a fixed value, which is a strong constraint on the ambient target space. Similarly, we want our general $d=2$ $\sigma$-model to be Weyl invariant as well, and we should expect equally stringent constraints on the background geometry to arise in this fashion. One way of proceeding is calculating the $\beta$-function as a perturbative series expansion, where at each loop we have an additional series in $\alphap$ corresponding to the string-related effects. Requiring it to vanish to first order in the loop expansion and up to first order in $\alphap$ leads to the following equations \cite{callanbackground},\cite{fradkinbackground}
	\begin{eqnarray}
	\label{beta functions}
	\beta^g_{\mu \nu}&=&\alphap R_{\mu\nu}-\frac{\alphap}{4}H_{\mu\sigma\rho}H^{\sigma \rho}_{\phantom{\sigma \rho}\nu} + 
	2 \alphap \nabla_\mu \nabla_\nu \Phi +\mathcal{O}(\alpha^{\prime 2}) \nonumber \\
	\beta^b_{\mu\nu}&=&\frac{\alphap}{2}\nabla^{\sigma}H_{\sigma\mu\nu} + \alphap \nabla^\sigma \Phi H_{\sigma\mu
	\nu} + \mathcal{O}(\alpha^{\prime 2}) \nonumber \\
	\beta^\Phi_{\mu\nu}&=&\frac{D-26}{6}-\frac{\alphap}{2} \Delta \Phi + \alphap \nabla_\sigma \Phi \nabla^\sigma \Phi - 
	\frac{\alphap}{24} H_{\mu\nu\sigma}H^{\mu\nu\sigma} +\mathcal{O}(\alpha^{\prime 2}),
	\end{eqnarray}
where we have introduced the 3-form field $H_{\mu\nu\rho}$ as the exterior derivative of the field $b_{\mu\nu}$
	\begin{equation}
	H_{\mu\nu\rho}=\left( \partial_{\mu}b_{\nu \rho}+\partial_{\nu}b_{\rho\mu}+\partial_{\rho}b_{\mu\nu} \right).
	\end{equation}
Demanding that the theory remains Weyl invariant up to first order in $\alphap$ requires the vanishing of the $\beta$-function, i.e.
	\begin{equation}
	\beta^g_{\mu \nu} =\beta^b_{\mu\nu}=\beta^\Phi_{\mu\nu}=0.
	\end{equation}
Let us discuss the dilaton first since its role is rather unique. Since the Ricci scalar in $D=2$ is a topological quantity we can write the relevant part of the action as
	\begin{equation}
	S_D=\frac{1}{4 \pi}\int d^2\sigma \sqrt{-\gamma}R^{(2)}\Phi(X)= \chi \phantom{.} \Phi(X)
	\end{equation}
where $\chi$ is the Euler characteristic of the string worldsheet. It is completely determined by the genus $g$ of the worldsheet when considered as a closed orientable surface
	\begin{equation}
	\chi= 2-2g.
	\end{equation}
As it turns out when calculating string scattering amplitudes the dilaton controls the string coupling $g_s=e^\Phi(X)$. Restricting ourselves to lowest order in string perturbation theory implies that we can restrict ourselves to a constant dilaton $\Phi=\Phi(X)$. Upon fixing the critical dimension $D=26$ the contributions of the dilaton to the $\beta$-function vanish.\\
\\
The remaining terms in the $\beta$-function can be regarded as the equations of motion of a low-energy effective action
	\begin{equation}
	S=\int d^{26}X \sqrt{-g} e^{-2\Phi} \left( R-\frac{1}{12}H_{\mu\nu\rho}H^{\mu\nu\rho}\right).
	\end{equation}
Indeed, varying this action yields the equations of motion
	\begin{eqnarray}
	\alphap R_{\mu\nu}-\frac{\alphap}{4}H_{\mu\sigma\rho}H^{\sigma \rho}_{\phantom{\sigma \rho}\nu}&=&0 \nonumber \\
	\frac{\alphap}{2}\nabla^{\sigma}H_{\sigma\mu\nu} &=&0.
	\end{eqnarray}
These equations of motion tell us that the background geometry is that of a flat Riemannian manifold with torsion. To show this, we introduce the torsionful connections
	\begin{equation}
	\Gamma^{\mu}_{\pm \nu \sigma}= \Gamma^{\mu}_{\phantom{\mu}\nu\sigma} \pm  \frac{1}{2}g^{\mu \rho}H_{\rho\nu
	\sigma},
	\end{equation}
where $\Gamma^\mu_{\phantom{\mu}\nu \sigma}$ corresponds to the usual Levi-Civita connection. We can explicitly write down the curvature tensors as	
	\begin{equation}
	R^\mu_{(\pm)\nu\rho\sigma} = \partial_{\rho}\Gamma^{\mu}_{(\pm)\nu \sigma} - \partial_{\sigma}
	\Gamma^{\mu}_{(\pm)\nu \rho} + \Gamma^{\mu}_{(\pm)\tau \rho} \Gamma^{\tau}_{(\pm)\nu\sigma} - \Gamma^{\mu}
	_{(\pm)\tau\sigma} \Gamma^{\tau}_{(\pm)\nu \rho}.
	\end{equation}
Both curvature tensors are related to each other as
	\begin{equation}
	R^{(+)}_{\mu\nu\rho\sigma} = R^{(-)}_{\rho\sigma \mu\nu} ,
	\end{equation}
so there is no need to consider them separately. The corresponding Ricci tensor can be written out in terms of the torsionless Ricci tensor and the torsion as follows
	\begin{equation}
	R_{\mu\nu}^{(+)}=R_{\mu\nu}+\frac{1}{2}\nabla^{\sigma}H_{\sigma \mu \nu}-\frac{1}{4}H_{\mu \rho \sigma}H^{\rho
	\sigma}_{\phantom{\rho\sigma}\nu}.
	\end{equation}
This allows us to neatly express the equations of motion for $G_{\mu \nu}$ and $B_{\mu \nu}$, or equivalently the vanishing of the $\beta$-function up to first order in $\alphap$, as
	\begin{equation}
	R_{\mu\nu}^{(+)}=0.
	\end{equation}
This result is rather remarkable if one considers that it is derived from a consistency condition of the two-dimensional field theory on the worldsheet of the propagating string. By demanding that the theory remain Weyl invariant we have put strong constraints on the background geometry. It is this subtle interplay between symmetry considerations on the worldsheet and geometric properties of the target space that allows for an incredibly rich mathematical structure to be investigated. This is even more pronounced when considering supersymmetric non-linear $\sigma$-models, as we will now do.\\

\section{Introducing $\mathcal{N}=(1,1)$ supersymmetry}
Before we move on the the general $\sigma$-model we will return to the superstring moving around in a flat spacetime to introduce the concept of superspace.
\subsection{The superstring revisited}
Our previous discussion of the superstring can be recast using the so-called superspace formalism. While still only considering a flat background spacetime several features that will later be of interest already emerge in this setting. Recall that the superstring action was given by
	\begin{equation}
	\label{superstring}
	S=\int d^2\sigma  \left(\partial_\pp X \partial_= +2i \psi_+ \partial_= \psi_+ + 2i \psi_- \partial_\pp \psi_-\right).
	\end{equation}
While realising supersymmetry on the worldsheet and providing a consistent quantum theory, the superstring action considered in the previous chapter was not manifestly supersymmetric. The easiest way of rewriting the theory so that the supersymmetries become manifest is to recast it using superspace formalism. In order to do this we introduce additional supercoordinates on the worldsheet $\theta^\pm$ that are Grassman numbers and form a Majorana spinor. A general superfield $\Phi$ on the worldsheet can then be written as an expansion in $\theta$ in the following manner\footnote{Lorentz indices have been suppressed for clarity.}
	\begin{equation}
	\Phi(\sigma,\theta^\pm)=X(\sigma) +i \theta^+\psi_+(\sigma^\pp) + i \theta^-\psi_-(\sigma^{=}) + i
	\theta^+ \theta^- F(\sigma) 
	\end{equation}
The first two terms in the expansion are identified with the original bosonic and fermionic fields, while the new field $F$ is auxiliary, as we will shortly discuss. Introducing the supergenerators
	\begin{eqnarray}
	Q_+=\frac{\partial}{\partial \theta_+}+\frac{i}{2}\theta ^+ \partial_\pp, \qquad
	Q_-=\frac{\partial}{\partial \theta_-}+\frac{i}{2}\theta ^- \partial_=
	\end{eqnarray}
and the supercovariant derivates
	\begin{eqnarray}
	D_+=\frac{\partial}{\partial \theta_+}-\frac{i}{2}\theta ^+ \partial_\pp, \qquad 
	D_-=\frac{\partial}{\partial \theta_-}-\frac{i}{2}\theta ^- \partial_=,
	\end{eqnarray}
we obtain the $\mathcal{N}=(1,1)$ supersymmetry algebra
\begin{eqnarray}
D_+^2= - \frac{i}{2}\, \partial _\pp \,,\qquad D_-^2=- \frac{i}{2}\, \partial _= \,,
\qquad \{D_+,D_-\}=0.\label{App2}
\end{eqnarray}
The $\mathcal{N}=(1,1)$ integration measure is explicitly given by
\begin{eqnarray}
\int d^ 2 \sigma \,d^2 \theta =\int d\tau \,d \sigma \,D_+D_-.
\end{eqnarray}
The supercovariant derivatives have been defined such that the fields $\Phi$ and $D_\pm \Phi$ transform in the same manner under supersymmetry transformations
	\begin{equation}
	\delta_{\epsilon}\Phi=\epsilon^+ Q_+ \Phi + \epsilon^- Q_- \Phi.
	\end{equation}
We can expand the superfield into components
	\begin{eqnarray}
	\delta  X &=& i \left( \epsilon^+ \psi_+ +  \epsilon^- \psi_- \right)\nonumber \\
	\delta  \psi_+ &=& - \frac{1}{2} \epsilon^+ \partial_\pm X + \epsilon^- F \nonumber \\
	\delta  \psi_- &=&  - \frac{1}{2} \epsilon^- \partial_= X + \epsilon^+ F  \nonumber \\
	\delta F &=&\frac{i}{2}\left( \epsilon^+ \partial_\pm \psi_- - \epsilon^- \partial_= \psi_+ \right)
	\end{eqnarray}
Up to terms involving the new field $F$ these are the same supersymmetry transformations we have considered in the previous chapter. Whereas the supersymmetry algebra only closed on-shell before, the fact that the transformation $\delta F$ contains terms that are proportional to the equations of motion ensures that this algebra closes off-shell.\\
\\
To reproduce the superstring action (\ref{superstring}) we can consider
	\begin{equation}
	S=4 \int d^2\sigma d^2 \theta D_+ \Phi D_- \Phi.
	\end{equation}
Varying this action gives
	\begin{equation}
	\delta S=4\int d^2  \sigma d^2 \theta \left[\epsilon^+ Q_+ \left( D_+\Phi D _-\Phi \right) +
	\epsilon^- Q_- \left( D_+\Phi D _-\Phi \right) \right]
	\end{equation}
which vanishes up to total derivates, establishing that the action is manifestly invariant under supersymmetry transformations. When written out in component form and performing the integral over the supercoordinates this action reduces to
	\begin{equation}
	S=\int d^2\sigma  \left(\partial_\pp X \partial_= X +2i \psi_+ \partial_= \psi_+ + 2i \psi_- \partial_\pp \psi_- +4F^2\right).
	\end{equation}
We have correctly reproduced the action (\ref{superstring}), with the addition of an extra term involving the additional field $F$.  As this term contains no derivatives the equations of motion for this field are purely algebraic, $F=0$, proving that it is indeed auxiliary and carries no additional dynamics.\\
\\
Having connected the superspace description with our earlier results we will conclude this section by remarking that the advantage of using the superspace formalism is twofold. As remarked before, supersymmetry on the worldsheet is now manifest and does not need to be checked independently. Secondly, the presence of the auxiliary field $F$ ensures the closure of the supersymmetry algebra without having to impose the equations of motion. This is a rather general feature that will encounter again when we consider extended supersymmetry later on.

\subsection{The general $\mathcal{N}=(1,1)$ $\sigma$-model}
Having introduced the superspace formalism to describe a superstring propagating on a flat spacetime we now turn our attention to the more general case where the target space is described by a background metric $g_{\mu\nu}$ and a background Kalb-Ramond field $b_{\mu\nu}$. This setup is described by the $\sigma$-model in $\mathcal{N}=(1,1)$ superspace 
	\begin{equation}
	\label{N1action}
	S=8 \int d^2 \sigma d^2 \theta D_+ \Phi^\mu \left( g_{\mu\nu} + b_{\mu\nu}\right) D_- \Phi^\nu.
	\end{equation}
This model is of course manifestly invariant under the supersymmetry transformation\footnote{When considering open strings as well as closed strings boundary terms will of course be present.}
	\begin{equation}
	\delta_\epsilon \Phi^\mu= \epsilon^+ Q_+ \Phi^\mu + \epsilon^- Q_- \Phi^\mu.
	\end{equation}
Varying the action we are led to the equations of motion for the fields $\Phi^\mu$
	\begin{equation}
	D_+ D_- \Phi^\mu + \Gamma^\mu_{(-)  \nu \rho} D_+ \Phi^\nu D_- \Phi^\rho=0.
	\end{equation}
When restricting the equations of motion to the bosonic components, i.e. $\theta=0$, this is the geodesic equation for a free moving particle in a background geometry which has a torsionful connection. 

\section{$\mathcal{N}=(2,2)$ supersymmetric non-linear $\sigma$-models}

\subsection{Extending the supersymmetry}
A natural question to ask ourselves is, can we introduce additional supersymmetry on the worldsheet, and if so, what does this additional symmetry imply for the target space geometry? This was investigated and determined in \cite{gates}, the results of which we will now review. On dimensional grounds this additional supersymmetry must take the form
	\begin{equation}
	\delta \Phi^\mu = \hat{\epsilon}^+  (J_+)^\mu_{\pmu\nu} D_+ \phi^\nu+\hat{\epsilon}^-(J_-)^\mu_{\pmu\nu}  D_- \phi^
	\nu,
	\end{equation}
where we have introduced two rank $(1,1)$ tensors $J_{\pm}$ that are, at the moment, arbitrary. We have enhanced the supersymmetry of the sigma model to N=(2,2) supersymmetry by introducing two additional chiral supersymmetries $\hat{\epsilon}^\pm$. In order to be consistent, we must have that the supersymmetry algebra closes. We can see that this is not generally true by considering
	\begin{eqnarray}
	\label{N2SUSY}
	[\delta_1,\delta_2]\Phi^\mu &=&-i \hat{\epsilon}_1\hat{\epsilon}_2 (J^2_+)^{\mu}_{\pmu\nu} \partial_\pp \Phi^\nu
	- i \hat{\epsilon} _1\hat{\epsilon}_2 (J^2_-)^{\mu}_{\pmu \nu} \partial_{=} \Phi^\nu \nonumber \\
	&-& \frac{1}{2}\hat{\epsilon}_1\hat{\epsilon}_2  N(J_+)^\mu_{\pmu \nu \sigma} D_+ \Phi^{\nu}D_+\Phi^{\rho}
	\nonumber \\
	&-& \frac{1}{2}\hat{\epsilon}_1\hat{\epsilon}_2  N(J_-)^\mu_{\pmu \nu \sigma} D_- \Phi^{\nu}D_-\Phi^{\rho}
	\nonumber \\
	&+&  \Xi^\mu.
	\end{eqnarray}
We have introduced a (1,2) tensor $N(J_\pm)$ called the Nijenhuis tensor\footnote{For an explicit form and useful relations we will refer the reader to the appendix.} and an auxiliary object $\Xi^{\mu}$ that we will analyse shortly.\\
\\
If we require the supersymmetry algebra to close the first four terms in (\ref{N2SUSY}) must reduce to spacetime translations. This is only possible if one imposes the following conditions on the tensors $J_\pm$
	\begin{eqnarray}
	\label{complexmanifold}
	J^2_\pm = -\mathds{1}, \qquad N(J_\pm)=0.
	\end{eqnarray}
The first condition defines two almost complex structures on the target manifold, while the second condition ensures that the both almost complex structures are integrable\footnote{This definition of integrability is correct, although the underlying structure is more subtle. We will review this in chapter 5}. \\
\\
So far demanding that the $\mathcal{N}=(2,2)$ supersymmetry algebra closes implied that the target space is a complex manifold equipped with two complex structures $J_+$ and $J_-$. There is however additional structure present. Since our starting point was a model that was only manifestly invariant under $\mathcal{N}=(1,1)$ supersymmetry it remains to be checked that the action (\ref{N1action}) is in fact invariant under the additional supersymmetry. One can show that this results in two further conditions on the complex structures $J_\pm$.\\
\\
The first one requires that both complex structures are compatible with the background metric in the following sense
	\begin{eqnarray}
	\label{hermitian}
	(J_\pm)^{\rho}_{\pa \mu}g_{\rho \sigma}(J_\pm)^{\sigma}_{\pa \nu}=g_{\mu \nu}.
	\end{eqnarray}
A complex manifold where the complex structure is compatible with the metric is called a hermitian manifold. As in the non-supersymmetric case we can define two torsionful connections. Invariance of the action requires that the complex structures are covariantly constant with respect to these connections
	\begin{equation}
	\label{covconstant}
	\nabla^{(\pm)}_\rho (J_\pm)^\mu_{\pa \nu}=0.
	\end{equation}
A geometry described by the conditions (\ref{complexmanifold}),(\ref{hermitian}) and (\ref{covconstant}) is called \emph{bihermitian} or, anticipating the more general framework discussed in the next chapter, \emph{generalised K\"ahler}. To establish that the name generalised K\"ahler is warranted we will first consider the special case of where there is no torsion present. The complex structures are then taken to be covariantly constant with respect to the Levi-Civita connection
	\begin{equation}
	\nabla_\rho (J_\pm)^\mu_{\pa \nu}=0.
	\end{equation}
A hermitian complex manifold that satisfies this condition is known as a K\"ahler manifold. As a result of this, the 2-form defined by
	\begin{equation}
	\omega_{\mu\nu}=-g_{\mu\rho}J^\rho_{\pa \nu }
	\end{equation}
which is well defined on any hermitian manifold, is closed
	\begin{equation}
	d\omega=0,
	\end{equation}
in which case this 2-form is known as the K\"ahler-form. The presence of a closed 2-form $\omega$ gives the manifold the structure of a symplectic manifold. K\"{a}hler manifolds can be regarded as complex manifolds that are symplectic manifolds whose symplectic structure is compatible with the complex structure. \\
\\
This condition is rather restrictive in that for a K\"{a}hler manifold the geometry, at least locally, is fully determined in terms of a single scalar function, the K\"ahler potential. Suppose we work in a coordinate patch where we have chosen complex coordinates $(z^a,z^{\ab})$ adapted to the complex structure, the K\"{a}hler potential $K(z^a,z^{\ab})$ fully determines the metric via
	\begin{equation}
	g_{a \bb}=\partial_a \partial_{\bb} K.
	\end{equation}
Since this is a local expression the K\"ahler potential can vary between coordinate patches. Given two different patches $\mathcal{U}$ and $\mathcal{V}$ the K\"ahler potentials defined on these patches $K_{\mathcal{U}}$ and $K_{\mathcal{V}}$ are related by a so-called K\"ahler transformation
	\begin{equation}
	K_{\mathcal{V}}=K_{\mathcal{U}}+f(z)+ g(\bar{z})
	\end{equation}
where $f(z)$ and $g(\bar{z})$ are arbitrary holomorphic and anti-holomorphic functions respectively. It can easily been seen that this does not alter the background geometry.\\
\\
When the target manifold has non-vanishing torsion, the background geometry is no longer K\"ahler as can be determined by the fact that the 2-form $\omega$, which would have to play role of the K\"ahler-form, is no longer closed. However, there are two properties of K\"ahler geometry that we would like to recover.

\begin{itemize}
\item The existence of a closed 2-form $\Omega$ analogous to the K\"ahler form.
\item A single scalar function similar to the K\"{a}hler potential locally determining the background geometry.
\end{itemize}
As implied by the name generalised K\"ahler this is indeed possible. However, one element has so far been overlooked, namely the extra term $\Xi^\mu$ in equation (\ref{N2SUSY}). Before we move on, we will have to analyse the superfields and their relation to the complex structures $J_{(\pm)}$ more carefully.

\subsection{The superfield content}
So far, demanding that the target space of the non-linear $\sigma$-model is bihermitian was a necessary condition for the supersymmetry algebra to close, but we have not yet shown that this is sufficient. We collected the additional obstructing terms to the closure of the algebra in the object $\Xi^\mu$. Written out explicitly this object takes the following form (suppressing spacetime indices)
	\begin{eqnarray}
	\label{xi}
	\Xi&=&\hat{\epsilon}_1^+\hat{\epsilon}_2^- \left[ J_+,J_-\right]D_+ D_- \Phi\nonumber \\
	&+& \hat{\epsilon}_1^-\hat{\epsilon}_2^+ \left[ J_+,J_-\right]D_+ D_- \Phi \nonumber \\
	&-& 2 \hat{\epsilon}_1^-\hat{\epsilon}_2^+ M(J_-,J_+) D_+ \Phi D_- \Phi \nonumber \\
	&-& 2 \hat{\epsilon}_1^+\hat{\epsilon}_2^- M(J_-,J_+) D_+ \Phi D_- \Phi,
	\end{eqnarray}
where we have introduced the quantity $M(A,B)^\mu_{\pa \nu \rho}$ defined by

Closure of the supersymmetry algebra necessitates $\Xi=0$. One can show that this is true on-shell by using the equations of motion and condition (\ref{covconstant}). Ideally, we would require the closure of the algebra without using the equations of motion. One way of accomplishing this is requiring that the complex structures commute
	\begin{equation}
	\left[ J_+,J_-\right]=0,
	\end{equation}
which explicitly sets the first two terms of (\ref{xi}) to zero. Furthermore, it can be shown that the two remaining terms can be rewritten to be proportional to the commutator as well, so having the complex structures commute realises off-shell supersymmetry. While this constitutes an attractive class of models we would like to find a more general description where the two complex structures do not commute.\\
\\
Another way of obtaining the closure of the full algebra, already employed in our discussion of the superstring, is introducing auxiliary fields. To see how this can be accomplished we will analyse the possible superfield content in more detail. A general $\mathcal{N}=(2,2)$ superfield $\mathbb{X}(x,\theta,\bar{\theta},\hat{\theta},\hat{\bar{\theta}})$ does not constitute an irreducible representation of the supersymmetry algebra. To reduce the degrees of freedom of the superfields in a covariant way one can use the supercovariant derivatives
	\begin{eqnarray}
	D_+=\frac{\partial}{\partial \theta_+}-\frac{i}{2}\theta ^+ \partial_\pp, \qquad D_-=\frac{\partial}{\partial \theta_-}-\frac{i}
	{2}\theta ^- \partial_= \nonumber \\
	\hat{D}_+=\frac{\partial}{\partial \hat{\theta}_+}-\frac{i}{2}\hat{\theta} ^+ \partial_\pp, \qquad D_-=\frac{\partial}{\partial 
	\hat{\theta}_-}-\frac{i}{2}\hat{\theta} ^- \partial_=.
	\end{eqnarray}
We can now distinguish between three varieties of superfields that satisfy constraints linear in the superderivates\footnote{Other possibilities exist, for example linear and twisted-linear superfields that satisfy constraints that are quadratic in the super derivatives. We will not consider them here.}
\begin{enumerate}
\item Chiral superfields $(\phi,\bar{\phi})$, defined by
	\begin{eqnarray}
	\label{chiraldef}
	\hat{D}_\pm \phi=i D_\pm \phi, \qquad \hat{D}_\pm \bar{\phi}=-i D_\pm \bar{\phi}.
	\end{eqnarray}
The result is a chiral multiplet, consisting of a pair of fields with the same chirality.
\item Twisted-chiral superfields $(\chi,\bar{\chi})$, defined by
	\begin{eqnarray}
	\label{twisteddef}
	\hat{D}_\pm \chi=\pm i D_\pm \chi, \qquad\hat{D}_\pm \bar{\chi}=\mp i D_\pm \bar{\chi}.
	\end{eqnarray}
These conditions define a twisted chiral multiplet, consisting of a pair of fields with opposite chirality.
\item Semi-chiral superfields$(\mathbb{X},\bar{\mathbb{X}},\mathbb{Y},\bar{\mathbb{Y}})$, defined by
	\begin{eqnarray}
	\label{semidef}
	\hat{D}_+ \mathbb{X}=i D_+ \mathbb{X}\phantom{-},& \qquad \hat{D}_+ \bar{\mathbb{X}}=-i D_+ \bar{\mathbb{X}}, \noindent 	\nonumber \\
	\hat{D}_- \mathbb{Y}=-i D_- \mathbb{Y},& \qquad \hat{D}_- \bar{\mathbb{Y}}= i D_- \bar{\mathbb{Y}}.
	\end{eqnarray}
This semi-chiral multiplet consists of a four-tuple of fields, two of left chirality and two of right chirality. The remaining degrees of freedom $\hat{D}_- \mathbb{X}$, $\hat{D}_- \bar{\mathbb{X}}$, $\hat{D}_+ \mathbb{Y}$ and $\hat{D}_+\bar{\mathbb{Y}}$ are auxiliary from a $\mathcal{N}=(1,1)$ point of view.
\end{enumerate}
It was conjectured \cite{sevrintroost}, and later proven in \cite{lindstrom2007}, that these three varieties of superfields are sufficient to fully characterise any bihermitian geometry. The key point is to consider the following decomposition of the tangent bundle of the target manifold $T(\mathcal{M})$ induced by the operator $\left[J_+,J_-\right]$
	\begin{equation}
	T(\mathcal{M})=\ker \left[J_+,J_-\right] \oplus \left( \mbox{im} \left[J_+,J_-\right] g^{-1} \right),
	\end{equation}
where the kernel can subsequently by decomposed as
	\begin{equation}
	\ker \left[J_+,J_-\right] = \ker(J_+ - J_-) \oplus \ker(J_+ + J_-).
	\end{equation}
This provides a natural way of distinguishing between different parts of the tangent space of target manifold that is consistent with the complex structures, as developed in \cite{gates}, \cite{buscher} and \cite{lindstrom2007a}. Recall that when $[J_+,J_-]=0$ the supersymmetry algebra closes without having to introduce auxiliary degrees of freedom. Therefore it would be convenient if we could identify the subspaces defined by $[J_+,J_-]$ with the various kinds of superfields. This is indeed possible as can be seen in the following manner. Given a set of superfields $\Phi^a$ parametrising the target space a general constraint that is first order in the covariant superderivates consistent with Lorentz invariance and dimensionality takes the form
	\begin{equation}
	\label{generalconstraint}
	\hat{D}_\pm \Phi^\mu =i \, \mathbb{J}^\mu_{\pm \nu} D_\pm \Phi^\nu,
	\end{equation}
where $\mathbb{J}_\pm$ are a priori undetermined (1,1) tensors. One can show that consistency with the supersymmetry algebra requires the operators $\mathbb{J}_\pm$ to be complex structures, leading us to the identification $\mathbb{J}_\pm=J_\pm$. Comparing this general constraint with the superfields defined earlier now allows us to identify these superfields in terms of the complex structures.\\
\\
When $J_+$ and $J_-$ commute we can diagonalise them both simultaneously, with eigenvalues $\pm i$. Depending on whether the constrained superfield under consideration has the same or the opposite eigenvalue under $J_+$ and $J_-$ it will parametrise either $\ker(J_+ - J_-)$ or $\ker(J_+ + J_-)$. Comparing the general constraints (\ref{generalconstraint}) with the definitions (\ref{chiraldef}) and (\ref{twisteddef}) we see that $\ker(J_+ - J_-)$ corresponds to chiral superfields, while $\ker(J_+ + J_-)$ corresponds to twisted chiral superfields. In \cite{ivanovkim} it was shown that it is always possible to parametrise $\ker[J_+,J_-]$ in this manner.\\
\\
In contrast, the parametrisation of $\mbox{im}[J_+,J_-]g^{-1}$ was long an open problem. It was conjectured that this could be accomplished solely by semi-chiral superfields, see \cite{sevrintroost}, \cite{sevrintroost1} and \cite{sevrinmaes}. In \cite{lindstrom2007} it was proven that this is indeed the case. Since this proof relies heavily on the existence of Poisson structures related to the complex structures we will provide an overview of such structures in the next subsection. Suffice it to say for now that the semi-chiral superfields defined by (\ref{semidef}) are indeed the correct kind of superfields to fully parametrise $\mbox{im}[J_+,J_-]g^{-1}$.\\
\\
We are thus led to the full off-shell description of the two dimensional $\mathcal{N}=(2,2)$ supersymmetric $\sigma$-model with $d$-dimensional target space $\mathcal{M}$
\begin{center}
  \begin{tabular}{ c || c || c }
   \bf{Subspace} & \bf{Field content} & \bf{Dimension$_\mathbb{R}$} \\ 
    \hline
    $\ker(J_+-J_-)$ & Chiral $(z,\bar{z})$ & $2n_c $\\ \hline
    $\ker(J_+ + J_-)$ & Twisted-chiral $(w,\bar{w})$ & $2n_t$ \\ \hline
    $ \mbox{im}[J_+,J_-]g^{-1}$ & Semi-chiral $(l,\bar{l},r,\bar{r})$ & $4n_s$ \\
  \end{tabular}
\end{center}
In what follows will use the following convention to denote the different components of the various types of superfields :

\begin{itemize}
\label{superfieldindex}
\item Chiral : $(z^\alpha,z^{\bar{\alpha}})$ with $\alpha,\bar{\alpha}\in \{1, \dots, n_c\}$
\item Twisted-chiral : $(w^\mu,w^{\bar{\mu}})$ with $\mu,\bar{\mu}\in \{1, \dots, n_t\}$
\item Left semi-chiral $(l^{\tilde{\alpha}},l^{\bar{\tilde{\alpha}}})$ with $\tilde{\alpha},\bar{\tilde{\alpha}}\in \{1, \dots, n_s\}$
\item Right semi-chiral $(r^{\tilde{\mu}},r^{\bar{\tilde{\mu}}})$ with $\tilde{\mu},\bar{\tilde{\mu}}\in \{1, \dots, n_s\}$
\end{itemize}

\subsection{Some Poisson geometry}
Our discussion so far has neglected an important aspect of the geometry of the target manifold of our $\mathcal{N}=(2,2)$ sigma model. The existence of coordinates satisfying supersymmetric constraints parametrising the various subspaces defined by the complex structures is closely tied to the existence of another type of geometric structure on the target manifold, namely \emph{Poisson structures}, first considered within the context of $\sigma$-models with extended supersymmetry in \cite{lyakhovic}. In this section we will analyse these structures as they will prove to be useful when we want to move beyond K\"{a}hler geometry.\\
\\
A Poisson structure $p^{\mu \nu}$ is a bivector defining a bracket on differentiable functions $f,g \in C^\infty(\mathcal{M})$ 
	\begin{equation}
	\{f,g \}_p = p^{\mu \nu} \frac{\partial f}{\partial x^\mu} \frac{\partial g}{\partial x^\nu}
	\end{equation}
that is skew-symmetric, satisfies the Jacobi-identity and is a derivation in its first argument. Equivalently, a bivector satisfying these conditions will satisfy the partial differential equation
	\begin{equation}
	p^{\mu \nu}\partial_ {\nu}p^{\rho \sigma}+p^{\rho \nu}\partial_ {\nu}p^{\sigma \mu}+p^{\sigma \nu}\partial_ {\nu}p^{\mu 
	\rho}=0.
	\end{equation}
One can show that the requirement of a closing supersymmetry algebra implies the existence of two Poisson structures defined by
	\begin{equation}
	\pi_\pm=(J_+ \pm J_-)g^{-1}.
	\end{equation}
When the commutator of $J_+$ and $J_-$ does not vanish, one can define a third Poisson structure, introduced in \cite{hitchinpoisson}
	\begin{equation}
	\sigma = [J_+,J_-]g^{-1}.
	\end{equation}
If $\ker[J_+,J_-]=\emptyset$, i.e. the target space is completely parametrised by semi-chiral superfields, $\sigma$ is invertible and be used to define a symplectic structure
	\begin{eqnarray}
	\Omega = \sigma^{-1}, \qquad d\Omega=0.
	\end{eqnarray}
As the two complex structures do not commute they cannot be diagonalised simultaneously. Suppose we chose local coordinates such that in a patch the operator $J_+$ takes its canonical form
	\begin{eqnarray}
	J_+= \left(
	\begin{matrix}
	i\mathds{1} &  0  \\
	0 & -i \mathds{1} 
	\end{matrix} \right).
	\end{eqnarray}
With respect to these coordinates the symplectic structure decomposes as
	\begin{equation}
	\Omega_+ = \Omega^{(2,0)}_+ + \bar{\Omega}^{(0,2)}_+.
	\end{equation}
Since we have that the (2,0) and (0,2) pieces of the symplectic structure are holomorphic and anti-holomorphic respectively
	\begin{equation}
	\bar{\partial} \Omega^{(2,0)}_+ = \partial \bar{\Omega}^{(0,2)}_+ =0,
	\end{equation}
we can find local Darboux coordinates $(p,q)$ and $(\bar{p^a},\bar{q^a})$ and put them in their canonical forms
	\begin{eqnarray}
	\Omega^{(2,0)}_+ = dq^{a} \wedge dp^{a}, \qquad \bar{\Omega}^{(0,2)}_+ = d\bar{q}.
	^{a} \wedge d\bar{p}^{a}
	\end{eqnarray}
We can do the same for $J_-$ and put it in its canonical form. This will yield the coordinates $\{Q^{a'},P^{a'}\}$ and the corresponding decomposition of the symplectic structure
	\begin{eqnarray}
	\Omega^{(2,0)}_- = dQ^{a'} \wedge dP^{a'}, \qquad \bar{\Omega}^{(0,2)}_+ = d\bar{Q}.
	^{a'} \wedge d\bar{P}^{a'}
	\end{eqnarray}
The Poisson structures defined above, and their related symplectic structures, ensure that not only do the constrained superfields parametrise the target space of the $\sigma$-model, they also play an important role in determining the background geometry. This will be the focus of the next subsection.

 \subsection{Finding the geometry}
On dimensional grounds, the superspace lagrangian density needs to be dimensionless. The model is then fully determined by a single scalar function $V$ of the superfields 
	\begin{equation}
	\label{N2action}
	S=4 \int d^2\sigma d^2 \theta d^2 \hat{\theta}\phantom{.} V(z,\bar{z},w,\bar{w},l,\bar{l},r,\bar{r}).
	\end{equation}
We should first note that there is an unavoidable ambiguity in the form of the action (\ref{N2action}). Performing a so-called generalised K\"{a}hler transformation
\begin{eqnarray}
V\rightarrow V+F(l,w,z)+ \bar F(\bar l,\bar w,\bar z)+ G(\bar r,w,\bar z)+ \bar G(r,\bar w,z).\label{genKahltrsf1}
\end{eqnarray}
will result in the same $\sigma$-model. In fact, transformations of this form are a requirement if the model is to be globally consistent \cite{hull2009}. \\
\\
The form of the action (\ref{N2action}) implies that the background geometry must be fully determined by the field content and the form of the potential $V$ alone. To do this, we should reduce (\ref{N2action}) to $\mathcal{N}=(1,1)$ superspace by performing the $\hat{\theta}$ integration
	\begin{equation}
	\label{SIntegration}
	S=\left. \int d^2\sigma d^2 \theta \hat{D}_+ \hat{D}_- V(z,\bar{z},w,\bar{w},l,\bar{l},r,\bar{r})\right|_{\hat{\theta}^+=
	\hat{\theta}^-=0}.
	\end{equation}
We introduce the following notation for convenience to denote the various derivatives of the potential $V$
	\begin{eqnarray}
	V_a = \frac{\partial V}{\partial z^a}, \qquad V_{a b}= \frac{\partial^2 V}{\partial z^a \partial z^b},
	\end{eqnarray}
and complex conjugates, where the indices $a,b$ can be any of the indices defined in (\ref{superfieldindex}), depending on the superfield content.\\
\\
We will now investigate the $\mathcal{N}=(1,1)$ form of the action (\ref{SIntegration}) for the various types of constrained superfields that can parametrise the target manifold. This will yield concrete formulas for calculating the various geometrical structures from the potential $V$. The expressions used in this section were derived in \cite{sevrinwijns}.\\

\subsubsection{First case : No semi-chiral fields}
The simplest case is when $\ker(J_+ + J_-)=0$ and the target space is described by chiral superfields $\{z^\alpha\}$ only. Performing the integration over the second supersymmetry components $\hat{\theta}$ and imposing the chiral constraints yields
	\begin{equation}
	S=4 \int d^2\sigma d^2 \theta\left( 2 V_{\alpha \bar{\beta}} D_+ z^\alpha D_- z^{\bar{\beta}} + 2 V_{\bar{\alpha} 
	\beta} D_+ z^{\bar{\alpha}} D_- z^{\beta} \right).
	\end{equation}
Comparing this to the general form of the $\mathcal{N}=(1,1)$ action we can identify the metric and the $b$-field
	\begin{eqnarray}
	g_{\alpha \bar{\beta}}=V_{\alpha \bar{\beta}} &,& b=0.
	\end{eqnarray}
The target manifold is K\"{a}hler with $V(z,\bar{z}$) taking the role of the familiar K\"{a}hler potential. Note that the $(2,0)$ and $(0,2)$ components of the metric tensor vanish, as they should for a K\"{a}hler manifold. Associated with the two complex structures we have two closed 2-forms, the K\"{a}hler forms
	\begin{eqnarray}
	\omega^{(\pm)}= g J_{\pm}, \qquad d\omega^{(\pm)}=0.
	\end{eqnarray}
The same calculation can be performed when $\ker(J_+ + J_-)=0$ , i.e. only twisted chiral superfields are present. The background geometry is once again K\"{a}hler with 
	\begin{eqnarray}
	g_{\mu \bar{\nu}}=- V_{\mu \bar{\nu}}, \qquad B=0,
	\end{eqnarray}
which corresponds to the chiral case up to a sign. We still have the two K\"{a}hler forms calculated in the usual manner
	\begin{eqnarray}
	\omega^{(+)}= -\omega^{(-)} &,& d\omega^{(\pm)}=0.
	\end{eqnarray}
The situation becomes more interesting when both chiral and twisted chiral superfields are present. The mixed chiral/twisted chiral derivatives in the action can be identified with a Wess-Zumino term, signifying that the background geometry will contain torsion and as a result is no longer K\"{a}hler. The metric is still given by
	\begin{eqnarray}
	g_{\alpha \bar{\beta}}=V_{\alpha \bar{\beta}} &,& g_{\mu \bar{\nu}}=- V_{\mu \bar{\nu}} ,
	\end{eqnarray}
while the new element is the appearance of a non-zero $B$-field
	\begin{eqnarray}
	b_{\alpha \nu}=\frac{1}{2}V_{\alpha \nu} &,& b_{\alpha \bar{\nu}}=-\frac{1}{2}V_{\alpha \bar{\nu}}.
	\end{eqnarray}
As the target space will now have non-zero torsion present this is an example of a bihermitian geometry that is not K\"{a}hler. However, the relevant geometrical structures can still be calculated form a single potential $V(z,\bar{z},w,\bar{w})$. In this sense it is natural to regard the function $V$ as generalising the familiar K\"{a}hler potential. In the next chapter we will present the argument that this is indeed the case by properly defining generalised K\"{a}hler geometry.\\

\subsubsection{Second case : no chiral and twisted-chiral fields}
The next case we will consider is a target space that is fully parametrised by semi-chiral superfields. Recall that the a trivial kernel for the commutator implied the existence of an invertible Poisson structure which in turn defines a symplectic 2-form
	\begin{equation}
	\label{semitwoform}
	\Omega \equiv  2 g ([J_+,J_-])^{-1}.
	\end{equation}
Depending on wether one choses to put $J_+$ or $J_-$ in its canonical form one finds two sets of Darboux coordinates $\{q^a,p^a\}$ and $\{Q^{a'},P^{a'}\}$. As the transformation
	\begin{equation} 
	\{q,p\} \rightarrow \{Q,P\}
	\end{equation}
is a canonical transformation or symplectomorphism one can always find a generating function $\mathcal{K}(q,P)$ so that
	\begin{eqnarray}
	p=\frac{\partial \mathcal{K}}{ \partial q}, \qquad Q=\frac{\partial \mathcal{K}}{ \partial P}.
	\end{eqnarray}
Using this generating function one can express the symplectic 2-form in a way that is linear in derivatives of $\mathcal{K}$ by using the \"{}mixed\"{} coordinates $\{q,P\}$
	\begin{equation}
	\Omega=\mathcal{K}_{a a'} dq^a \wedge dP^{a'}
	\end{equation}
As it turns out the function generating the canonical transformations $\mathcal{K}$ can be identified with the potential $V$ in (\ref{N2action}) \cite{lindstrom2007a}. In what follows we will not continue to distinguish between the two concepts anymore and use the notation $V$ for both. For convenience' sake we can introduce the collective coordinates $l$ and $r$ to compactly collect the second derivatives of the generating function $V$
	\begin{eqnarray}
	V_{l l}=\left(
	\begin{matrix}
	V_{\tilde{\alpha} \tilde{\beta}} & V_{\tilde{\alpha} \bar{\tilde{\beta}}} \\
	V_{\bar{\tilde{\alpha}} \beta} & V_{\bar{\tilde{\alpha}} \bar{\tilde{\beta}}} 
	\end{matrix} \right) 
	V_{l r}=\left(
	\begin{matrix}
	V_{\tilde{\alpha} \tilde{\mu}} & V_{\tilde{\alpha} \bar{\tilde{\mu}}} \\
	V_{\bar{\tilde{\alpha}} \mu} & V_{\bar{\tilde{\alpha}} \bar{\tilde{\mu}}} 
	\end{matrix} \right) &
	V_{r r}=\left(
	\begin{matrix}
	V_{\tilde{\mu} \tilde{\nu}} & V_{\tilde{\mu} \bar{\tilde{\nu}}} \\
	V_{\bar{\tilde{\mu}} \nu} & V_{\bar{\tilde{\mu}} \bar{\tilde{\nu}}} 
	\end{matrix} \right),
	\end{eqnarray}
where by our choice of notation we have emphasised the connection to left- and right-handed semi-chiral superfields that parametrise the manifold. \\
\\
In terms of these matrices the symplectic 2-form becomes
	\begin{equation}
	\Omega = \left(
	\begin{matrix}
	0 & V_{lr} \\
	- V_{rl} & 0
	\end{matrix} \right).
	\end{equation}
Both complex structure are no longer diagonal when expressed in this mixed coordinate system.We can calculate them by applying the appropriate coordinate transformations. First some notation. Introducing a $2n_s \times 2n_s$ projection operator,
	\begin{equation}
	\mathds{P}=\left(
	\begin{matrix}
	\mathds{1} & 0 \\
	0 & -\mathds{1}
	\end{matrix} \right),
	\end{equation}
we can define the convenient combinations
	\begin{eqnarray}
	C_{ll}&=& \mathds{P}V_{ll} - V_{ll}\mathds{P} \\
	&=&\left(
	\begin{matrix}
	0 & 2 V_{\tilde{\alpha} \bar{\tilde{\beta}}} \\
	-2V_{\bar{\tilde{a}} \tilde{\beta}} & 0
	\end{matrix} \right)\\
	A_{ll}&=& \mathds{P}V_{ll} + V_{ll}\mathds{P} \\
	&=&\left(
	\begin{matrix}
	2 V_{\tilde{\alpha} \tilde{\beta}} & 0\\
	0& -2V_{\bar{\tilde{\alpha}} \bar{\tilde{\beta}}} 
	\end{matrix} \right),
	\end{eqnarray}
with $C_{lr},C_{rr},A_{lr}$ and $A_{rr}$ defined analogously.\\
\\
Employing this notation, the complex structure $J_+$ with rows and columns ordered in the order $\{ l, \bar{l}, r , \bar{r}\}$ becomes
	\begin{eqnarray}
	J_+&=& \left( \frac{\partial (q,p)}{\partial (q,P)} \right)^{-1} \left(
	\begin{matrix}
	i\mathds{P} &  0  \\
	0 & i \mathds{P} 
	\end{matrix} \right) \left( \frac{\partial (q,p)}{\partial (q,P)} \right) \\
	&=& \left(
	\begin{matrix}
	i\mathds{P} &  0  \\
	iV_{rl}^{-1} C_{ll} & V_{lr}^{-1}(i\mathds{P})V_{lr} 
	\end{matrix} \right),
	\end{eqnarray}
where by convention $V_{lr}^{-1}=(V_{rl})^{-1}$ . The same can be done for the second complex structure $J_-$
	\begin{eqnarray}
	J_-&=& \left( \frac{\partial (Q,P)}{\partial (q,P)} \right)^{-1} \left(
	\begin{matrix}
	i\mathds{P} &  0  \\
	0 & i \mathds{P} 
	\end{matrix} \right) \left( \frac{\partial (Q,P)}{\partial (q,P)} \right) \\
	&=& \left(
	\begin{matrix}
	V_{lr}^{-1}(i\mathds{P}) V_{rl} &  V_{lr}^{-1}C_{rr}  \\
	0 &i \mathds{P}
	\end{matrix} \right).
	\end{eqnarray}
By using this explicit form for the complex structures we can deduce the expressions for the metric and $b$-field by reducing the $\mathcal{N}=(2,2)$ lagrangian to $\mathcal{N}=(1,1)$ superspace and eliminating the auxiliary components of the semi-chiral superfields by imposing their equations of motion. This calculation yields
	\begin{eqnarray}
	\label{Efield}
	E=g+b&=& \frac{1}{2} J_+^T \left(
	\begin{matrix}
	0 & V_{lr} \\
	-V_{rl} & 0
	\end{matrix} \right) J_- \\
	&=& \frac{1}{2} J_+^T \phantom{.}\Omega\phantom{.} J_-.
	\end{eqnarray}
By extracting the symmetric and anti-symmetric parts of this expression we retrieve the metric, consistent with our initial definition of the symplectic structure $\Omega$, and the $b$-field
	\begin{eqnarray}
	g &=& \frac{1}{4} \phantom{.} \Omega  \phantom{}\ [J_+,J_-], \\
	b &=& \frac{1}{4} \phantom{.} \Omega  \phantom{}\ \{ J_+ , J_- \}.
	\end{eqnarray}
We conclude that the generating function of the symplectomorphisms can be regarded as a generalisation of the more familiar K\"{a}hler potential, since its derivatives fully determine, via the symplectic form $\Omega$ and the complex structures $J_\pm$, the background geometry.
\subsubsection{The general case}
When both chiral, twisted chiral and semi-chiral superfields are present the symplectic 2-form (\ref{semitwoform}) is no longer well-defined. The culprits are the non-zero chiral and twisted chiral directions that parametrise $\ker[J_+,J_-]$ which is no longer trivial. We would like to find another symplectic form that determines the background geometry, ideally given in terms of derivatives of a potential function. Using the fact that
	\begin{equation}
	[J_+,J_-]=(J_+ - J_-)(J_+ + J_-)
	\end{equation}
we can envision two candidates that will solve the problem of invertibility. Let us first consider the case without twisted chiral fields present, or in other words $\ker[J_+,J_-]=\ker(J_+ - J_-)$. This leaves $(J_+ + J_-)$ as the invertible part of the commutator. As a result, the 2-form
	\begin{equation}
	\Omega^{(+)} \equiv 2 g (J_+ + J_-)^{-1}
	\end{equation}
is well defined. Conversely, when only twisted chiral and semi-chiral superfields are present, we have $\ker[J_+,J_-]=\ker(J_+ + J_-)$ and the invertible part of the commutator is $(J_+ - J_-)$. Consequently, we have a well-defined two form
	\begin{equation}
	\Omega^{(-)} \equiv  2 g(J_+ - J_-)^{-1}.
	\end{equation}
To find an explicit expression for these two 2-forms we need to compare them to the $\sigma$-model action by comparing the reduction of the $\mathcal{N}=(2,2)$ superspace action to $N=(1,1)$ and integrating out the auxiliary degrees of freedom as before. When all three types of superfields are present we get the following expression for the combination $E=g+b$
	\begin{equation}
	\label{EFieldfull}
	E= \frac{1}{2} J_+^T M_1 J_- + M_2
	\end{equation}
with the matrices $M_1$ and $M_2$ fully determined by the derivatives of generalised K\"{a}hler potential as follows
	\begin{eqnarray}
	M_1 &=& \left(
	\begin{matrix}
	0& V_{lr} & V_{lw} & V_{lz}\\
	-V_{rl} &0 & 0&0 \\
	0&V_{wr} &V_{ww} & V_{wz}\\
	0& V_{zr}&V_{zw} &V_{zz}
	\end{matrix} \right) \\
	M_2&=& \left(
	\begin{matrix}
	0& 0& -V_{lw}&V_{lz} \\
	0& 0&-V_{rw} &V_{rz} \\
	-V_{wl}&-V_{wr} &-2V_{ww} & 0\\
	V_{zl }& V_{zr}& 0 & 2 V_{zz}
	\end{matrix} \right),
	\end{eqnarray}
where the dimensions and matrix structure of the matrices $V$ are determined by the number of chiral, twisted chiral and semi-chiral superfields present. The symmetric and anti-symmetric pieces of this expression will yield the metric and the $b$-field respectively, as before.\\
\\
By comparing (\ref{EFieldfull}) to the previously derived expressions of the metric in terms of the 2-forms $\Omega^{(\pm)}$ we can calculate them explicitly.\\
\\
For $\ker(J_+ + J_-)=\{ 0 \}$, in other words, no twisted chiral superfields present, we have
	\begin{equation}
	g= \frac{1}{2} \Omega^{(+)} (J_+ +J_-)
	\end{equation}
which implies the following expression for the 2-form in terms of the generalised K\"{a}hler potential
	\begin{equation}
	\Omega^{(+)}=\frac{i}{2}\left(
	\begin{matrix}
	C_{ll} & C_{lr} & C_{lz}\\
	C_{rl} & C_{rr} & C_{rz} \\
	C_{zl} & C_{zr} & C_{zz}
	\end{matrix}\right)
	\end{equation}
The geometrical data is neatly encapsulated by the expressions
	\begin{eqnarray}
	\label{geomnotwisted}
	g&=& +\frac{1}{2} \Omega^{(+)} (J_+ + J_-) \\
	b&=& -\frac{1}{2} \Omega^{(+)} (J_+ - J_-)
	\end{eqnarray}
This fully determines the background geometry in terms of the generalised K\"{a}hler potential and the complex structures.\\
\\
Conversely, when $\ker(J_+-J_-)=\{ 0 \}$ and there are no chiral superfields we get
	\begin{equation}
	g= \frac{1}{2} \Omega^{(-)} (J_+- J_-)
	\end{equation}
and we find the following expression in terms of the generalised K\"{a}hler potential
	\begin{equation}
	\Omega^{(-)}=-\frac{i}{2}\left(
	\begin{matrix}
	C_{ll} & A_{lr} & C_{lw}  \\
	-A_{rl} &-C_{rr} & -A_{rw} \\
	C_{wl} & A_{wr} & C_{ww}
	\end{matrix} \right)
	\end{equation}
which in turn defines the geometrical objects
	\begin{eqnarray}
	\label{geomnochiral}
	g&=& +\frac{1}{2} \Omega^{(-)} (J_+ - J_-) \\
	b&=& -\frac{1}{2} \Omega^{(-)} (J_+ + J_-).
	\end{eqnarray}
The situation becomes somewhat more involved in the most general case, with all types of contained superfields present. Equations (\ref{geomnochiral}) and (\ref{geomnotwisted}) suggest the following expressions for the metric and the $b$-field
	\begin{eqnarray}
	g&=& +\frac{1}{2} \Omega^{(+)} J_+ + \Omega^{(-)} J_- \\
	b&=& -\frac{1}{2}\Omega^{(+)} J_+ +  \Omega^{(-)} J_-.
	\end{eqnarray}
By explicitly comparing these expression with (\ref{EFieldfull}) we can find the explicit form of the two 2-forms $\Omega^{(\pm)}$ in terms of the generalised K\"{a}hler potential. We find however that these expressions are no longer linear, and what is more the 2-forms are no longer closed. One way of dealing with this is that the physical content of the theory is only defined by the $b$-field up to a gauge transformation $b \rightarrow b+d\xi$. By employing a suitable gauge transformation it is possible to put either $\Omega^{(+)}$ or $\Omega^{(-)}$ in the desired form, but not both at the same time.\\
\\
By choosing a gauge for the $b$-field such $\Omega^{(+)}$ is closed we obtain
	\begin{equation}
	\label{oplusfinal}
	\Omega^{(+)}=-\frac{i}{2} \left(
	\begin{matrix}
	C_{ll} & A_{lr} & C_{lw} & A_{lz}\\
	-A_{rl} & -C_{rr} & -A_{rw} & -C_{rz} \\
	C_{wl} & A_{wr} & C_{ww} & A_{wz} \\
	 -A_{zl} & -C_{zr} & -A_{zw} & -C_{zz}
	\end{matrix}\right).
	\end{equation}
As $\Omega^{(+)}$ is closed, it is exact and we can find a 1-form $\Xi$ such that $d\Xi=\Omega^{(+)}$. One can easily show that
	\begin{equation}
	\Xi=i \left( V_{l},-V_{\bar{l}},-V_{r},V_{\bar{'}},V_{w},-V
	_{\bar{w}},-V_{z},V_{\bar{z}} \right)
	\end{equation}
is a suitable 1-form. Having made this gauge choice, the other 2-form $\Omega^{(-)}$ is no longer closed, nor is it completely determined by terms linear in the generalised K\"{a}hler potential.\\
\\
By the same token we could have chosen a gauge in which $\Omega^{(-)}$ is closed but $\Omega^{(+)}$ is not. Expressed linearly in terms of the generalised K\"{a}hler potential the 2-form becomes
	\begin{equation}
	\label{ominfinal}
	\Omega^{(-)}=\frac{i}{2} \left(
	\begin{matrix}
	C_{ll} & C_{lr} & C_{lw} & C_{lz}\\
	C_{rl} & C_{rr} & C_{rw} & C_{rz} \\
	C_{wl} & C_{wr} & C_{ww} & C_{wz} \\
	 C_{zl} & C_{zr} & C_{zw} & C_{zz}
	\end{matrix}\right).
	\end{equation}
Note that expressions (\ref{oplusfinal}) and (\ref{ominfinal}) reduce to our previously found expressions if not all kinds of constrained superfield are present.\\
\\
This concludes our overview of the geometric aspects of the most general $\mathcal{N}=(2,2)$ supersymmetric non-linear $\sigma$-model.

\chapter{Generalised complex geometry}
\thispagestyle{empty}
\emph{In chapter 4 we have provided an overview on how $\sigma$-models with extended supersymmetry resulted in a rich geometrical structure of their target space. Imposing $\mathcal{N}=(2,2)$ supersymmetry implied that the background geometry had to be bihermitian. This bihermitian geometry, at least in a certain sense, generalised some notions of the simpler case of K\"{a}hler geometry in that the geometrical data could be extracted from a single scalar function. In this chapter we will make the correspondence more manifest by describing how this bihermitian geometry fits into the larger framework of generalised complex geometry.}

\newpage
\section{The generalised tangent space}
Consider the case of a $d$-dimensional Riemannian manifold $\mathcal{M}$ equipped with an atlas of charts defining patches with local coordinates $\{x^i\}$. Diffeomorphisms on the manifold are generated by vector fields\footnote{Strictly speaking, complete vector fields.}. The \emph{Lie derivative} along a vector field $X \in \Gamma(T\mathcal{M})$ can be characterised via \emph{Cartan's identity}
	\begin{equation}
	\mathcal{L}_X=\{\iota_X, d\}=\left( \iota_X d + d \iota_X\right),
	\end{equation}
where $\iota_{X}$ is the interior product. Applying this to a differentiable function $f$ on $\mathcal{M}$  we can express this in component form as
	\begin{equation}
	\mathcal{L}_X f= X^i \frac{\partial f}{\partial x^i}.
	\end{equation}
The Lie derivative can be used to define the Lie bracket $[X,Y]$ on the tangent bundle of the manifold by requiring
	\begin{equation}
	\mathcal{L}_{[X,Y]}f=\mathcal{L}_X \mathcal{L}_Y f- \mathcal{L}_Y \mathcal{L}_X f,
	\end{equation}
so that the Lie bracket of two vector fields gives another vector field, given in component form by
 	\begin{equation}
	[X,Y]=\left(X^j \frac{\partial Y^i}{\partial x^j} - Y^j \frac{\partial X^i}{\partial x^j} \right)\frac{\partial}{\partial x^i}.
	\end{equation}
Recall that we defined a complex structure as an almost complex structure that was integrable. Integrability was defined as the vanishing of the Nijenhuis tensor. While for a complex structure these properties are equivalent they do not generalise easily, forcing us to define integrability more carefully.\\
\\
Using Lie derivatives one can more properly define integrability of an almost complex structure as a property of  the complexified tangent bundle $T\mathcal{M} \otimes \mathbb{C}$. It is clear that a rank 2 tensor that squares to minus the identity has $+i$ and $-i$ as non-trivial eigenvalues. By considering the $+i$ and $-i$ eigenbundles of $T\mathcal{M} \otimes \mathbb{C}$ and calling them $L$ and $\bar{L}$ respectively one has defined two  \emph{distributions}
\begin{equation*}
	L,\bar{L} \subset T\mathcal{M} \otimes \mathbb{C}.
\end{equation*}
The distribution $L$ (and equivalently $\bar{L}$) is called \emph{involutive} if it is closed under the Lie bracket if for all vector fields $X,Y \in \Gamma(L)$
	\begin{equation}
	X,Y \in \Gamma(L) \Rightarrow [X,Y] \in \Gamma(L).
	\end{equation}
A distribution $L$ locally spanned by a collection of vectors $\{X_a\}$ is called \emph{integrable} if one can find functions $x^i(\sigma^1, \dots, \sigma^{\text{rank}(L)};x_0)$ in a neighbourhood of  a point $p(x_0) \in \mathcal{M}$ solving the system of partial differential equations
	\begin{equation}
	\frac{\partial x^i}{\partial \sigma^a}=X^i_a.
	\end{equation}
Solving this system of equations is equivalent to finding a coordinate transformation $x'=x'(x)$ such that the subbundle $L$ is spanned by the vectors $\{ \frac{\partial}{\partial x'^i} \}$.These new coordinates are then called \emph{adapted coordinates}. The Frobenius theorem now states that
	\begin{equation}
	J \text{ is integrable} \Leftrightarrow L \text{ is involutive}.
	\end{equation}
The characterisation we have used before, that is to say, defining a complex structure by demanding that the Nijenhuis tensor vanishes is known as the \emph{Newlander-Nirenberg theorem}. This theorem states that the Nijenhuis tensor vanishes if and only if $L$ (and equivalently $\bar{L}$) are involutive under the Lie bracket. Defining integrability as an involutive property of some bracket is the strategy we will use to define a generalised complex structure.\\
\\
Recall that the background geometry of the target space is determined by two objects, the metric tensor $g$ and the Kalb-Ramond 2-form $b$. The metric $g$ is invariant under a diffeomorphism generated by a vector field $X$ if
	\begin{equation}
	\mathcal{L}_X g =0.
	\end{equation}
In addition, there is a gauge symmetry of the $b$-field of the form
	\begin{equation}
	{b} \rightarrow b + d\xi
	\end{equation}
parametrised by 1-forms $\xi$. This leaves the torsion $H$ invariant since
	\begin{equation}
	H=db \rightarrow H =d(b+d\xi)= db + d^2 \xi = db
	\end{equation}
Both transformations can be performed at the same time while leaving the background geometry unchanged. In this sense it would be convenient of we could generalise the notion of the Lie bracket by treating vectors $X$ and 1-forms $\xi$ on equal footing in order to incorporate both diffeomorphisms and transformations of the $b$-field in the same framework.\\ 
\\
As the Lie bracket is defined as acting on the a section of the tangent bundle one introduces the generalised tangent bundle by combining the tangent bundle $T(\mathcal{M})$ and the cotangent bundle $T^*\mathcal{M}$ into one object $T\mathcal{M} \oplus T^*\mathcal{M}$.\\
\\
 If we take two generalised tangent vectors
	\begin{eqnarray}
	\mathbb{X}=X+\xi, \qquad \mathbb{Y}=Y+\eta,
	\end{eqnarray}
with $X,Y \in \Gamma(T\mathcal{M})$ and $\xi,\eta \in \Gamma(T^*\mathcal{M})$, there exists a natural pairing by contracting 1-forms along the vectors in a symmetrical way, which we will call the canonical metric $\mathcal{I}$. This bilinear form on the generalised target bundle is given by
	\begin{equation}
	\mathcal{I}(\mathbb{X},\mathbb{Y})=\frac{1}{2}\left(\xi(Y)+\eta(X) \right),
	\end{equation}
and has signature $(d,d)$. We will call the isometry group of the generalised tangent bundle $O(d,d)$. A generic element of of $O(d,d)$ can be generated by exponentiating
	\begin{equation}
	T=\left(
	\begin{matrix}
	A &  \beta \\
	b & -A^*
	\end{matrix} \right),
	\end{equation}
where $A \in \mathfrak{so}(d)$ gives rise to rotations acting on vectors or 1-forms in the usual way. The off-diagonal components relate vectors to 1-forms and vice versa in a novel way. This includes the $b$-transform
	\begin{equation}
	e^b : X+\xi \rightarrow X + \left( \xi - \iota_X b \right),
	\end{equation}
with $b$ a 2-form and the $\beta$ transform
	\begin{equation}
	e^{\beta} : X+\xi \rightarrow \left( X - \iota_{\xi} \beta \right) + \xi,
	\end{equation}
with $\beta$ a bivector.\\
\\
Having introduced the generalised tangent bundle we now need the analogue of the Lie bracket. One can show that there exists a well-defined bracket called the \emph{Courant bracket} on the generalised tangent bundle defined by
	\begin{equation}
	[\mathbb{X},\mathbb{Y}]_C=[X,Y]+\mathcal{L}_X \eta -\mathcal{L}_Y \eta - \frac{1}{2}d\left(\iota_X \eta - \iota_Y \xi
	\right).
	\end{equation}
This definition of the Courant bracket is compatible with the Lie bracket in the following sense. If we define the projection operator
	\begin{equation}
	\pi_{T\mathcal{M}} : T\mathcal{M} \oplus T^*\mathcal{M} \rightarrow T\mathcal{M} : \pi_{T\mathcal{M}} (X+\xi)=X,
	\end{equation}
the Courant bracket satisfies
	\begin{equation}
	\left[ \pi_{T\mathcal{M}} \mathbb{X}  , \pi_{T\mathcal{M}} \mathbb{Y} \right] =\pi_{T\mathcal{M}} [ \mathbb{X},
	\mathbb{Y}]_C.
	\end{equation}
In other words, the Courant bracket consistently reduces to the Lie bracket  when restricted to the tangent bundle.\\
\\
The Courant bracket as defined above allows for one important generalisation. Given a 3-form $H$ one can define the \emph{twisted Courant bracket}
	\begin{equation}
	[X+\xi,Y+\eta]_H=[X+\xi,Y+\eta]_C+ \iota_X \iota_Y H.
	\end{equation}
This (twisted) Courant bracket has the desirable property that it transforms homogeneously under $b$-transformations
	\begin{equation}
	[e^b(X+\xi),e^b(Y+\eta)]_{H}=e^b[X+\xi,Y+\eta]_{H+db},
	\end{equation}
provided that $db=0$. In order to accommodate $b$-transforms where $db \neq 0$ one can use the twisted Courant bracket instead of the untwisted version.\\
\\ 
Having defined a bracket on the generalised tangent bundle that is both invariant under ordinary diffeomorphisms and $b$-transforms we can move on and use this bracket to generalise the notion of integrability to the generalised tangent bundle. This will lead to a proper definition of generalised complex geometry.

 \section{The generalised complex structure}
Having introduced the concept of a generalised tangent space we can define a generalised almost complex structure in a fairly straightforward way. Consider the map
	\begin{equation}
	\mathcal{J} :T\mathcal{M} \oplus T^*\mathcal{M} \rightarrow T\mathcal{M} \oplus T^*\mathcal{M}
	\end{equation}
which is compatible with the bundle structure $\pi(\mathcal{J} \mathbb{X})=\pi(\mathbb{X})$. If we impose the two conditions
	\begin{eqnarray}
	\mathcal{J}^2&=& - \mathds{1}_{2d \times 2d}\\
	\mathcal{I} \left( \mathcal{J} \mathbb{X} , \mathcal{J} \mathbb{Y}\right) &=& \mathcal{I} \left( \mathbb{X} ,\mathbb{Y}
	\right) ,
	\end{eqnarray}
the operator $\mathcal{J}$ is known as a \emph{generalised almost complex structure}. Given such a generalised complex structure, it is natural to ask ourselves how we should introduce the notion of integrability. One way of doing this is by considering the subbundles
	\begin{equation}
	L_\mathcal{J},\bar{L}_\mathcal{J} \subset ( T\mathcal{M} \oplus T^*\mathcal{M} ) \otimes \mathbb{C}
	\end{equation}
and demanding that they are involutive under the ($H$-twisted) Courant bracket~:
	\begin{eqnarray}
	[\mathbb{X},\mathbb{Y}]_H \in \Gamma(L_\mathcal{J}),& \text{for all } \mathbb{X},\mathbb{Y} \in \Gamma(L_
	\mathcal{J}).
	\end{eqnarray}
Equivalently, one can define integrability with respect to the Courant bracket by considering the generalised Nijenhuis tensor, defined as
	\begin{equation}
	\mathcal{N}_\mathcal{J} (\mathbb{X},\mathbb{Y})=[\mathbb{X},\mathbb{Y}]_C - [\mathcal{J}  \mathbb{X}, \mathcal{J} 
	\mathbb{Y}]_C + \mathcal{J}  \big( [\mathbb{X},\mathcal{J} \mathbb{Y}]_C - [\mathcal{J}  \mathbb{X},\mathbb{Y}]_C
	\big),
	\end{equation}
and demanding that it vanishes $\forall \,\mathbb{X},\mathbb{Y} \in  \Gamma(T\mathcal{M}\oplus T^*\mathcal{M})$.\\
\\
So far this definition of generalised complex geometry remains somewhat abstract, so it is instructive to see how it relates to more familiar geometric structures. If we consider one of its original motivations, one should expect that generalised complex geometry should, in a certain sense, interpolate between ordinary complex and symplectic geometry. More specifically, both structures should be able to be recovered from considering a special case of a generalised complex structure. One can show that this is indeed the case. Starting from an ordinary complex structure $J$ it is easy to see that
	\begin{equation}
	\mathcal{J}_J=\left(
		\begin{matrix}
		J & 0\\
		0&- J^T
		\end{matrix}
		\right)
	\end{equation}
is a generalised complex structure. Conversely, starting from a symplectic structure $\omega$,
	\begin{equation}
	\mathcal{J}_\omega=\left(
		\begin{matrix}
		0 & -\omega^{-1}\\
		\omega& 0
		\end{matrix}
		\right)
	\end{equation}
also defines a generalised complex structure. \\
\\
The fact that a generalised complex structure interpolates between these two cases is particularly clear in the case of hyper-K\"{a}hler manifolds, as was pointed out in \cite{gualtieri} and \cite{interpolate}\footnote{This was also considered in \cite{hitchingcy} using the spinor point of view, which will encounter later in this chapter}. Hyper-K\"{a}hler manifolds (or more generally, holomorphic symplectic manifolds) come equipped with a complex structure $I$ and a closed (2,0)-form $\sigma=\omega_J + i \omega_K$ called the holomorphic symplectic structure. From these objects we can construct two generalised complex structures, which we can call $\mathcal{J}_I$ and $\mathcal{J}_{\omega_J}$, who will be of complex and symplectic type respectively. It can be shown that these can be used to define a one-parameter family of integrable complex structures
	\begin{equation}
	\mathcal{J}_t =(\sin t) \mathcal{J}_I + (\cos t) \mathcal{J}_{\omega_J} , \quad t \in [0,\pi/2],
	\end{equation}
where, for $t \neq \pi/2$, $\mathcal{J}_t$ will be the $b$-transform of a symplectic type generalised complex structure, while for $t=\pi/2$ it will be a complex type generalised complex structure.\\
\\
Recall that a K\"{a}hler manifold is a manifold where the symplectic and complex structures are compatible with each other. A natural question to ask is then, does a similar structure exist in the context of generalised complex geometry?

\section{Generalised K\"{a}hler structures}
In the previous chapter we foreshadowed the link between bihermitian geometry and generalised complex geometry by claiming that bihermitian geometry is equivalent to generalised K\"{a}hler geometry. We are now in a position to argue that this name is indeed warranted. To do this however we first need to clarify what we mean by a generalised K\"{a}hler structure.\\
\\
First, we will focus in the case with vanishing Kalb-Ramond field $b=0$, leaving us with a background geometry that is torsionless. Without torsion the background geometry is that of a K\"{a}hler manifold, consisting of a metric tensor $g$, a complex structure $J$ and a symplectic 2-form $\omega$. As noted above, both the complex and the symplectic structure define generalised complex structures $\mathcal{J}_J$ and $\mathcal{J}_\omega$ respectively. These generalised complex structures commute and their product
	\begin{equation}
	\mathcal{G}=-\mathcal{J}_J \mathcal{J}_\omega = \left(
	\begin{matrix}
		0& g^{-1}\\
		g& 0
		\end{matrix}
		\right)
	\end{equation}
defines a positive definite metric on the generalised tangent bundle. Following this example we call a pair of generalised complex structures $(\mathcal{J}_1,\mathcal{J}_2)$ a generalised K\"{a}hler structure if they obey the following conditions :
\begin{itemize}
\item $\mathcal{J}_1$ and $\mathcal{J}_2$ commute.
\item $\mathcal{G}=-\mathcal{J}_1 \mathcal{J}_2$ is a positive definite metric on $ T\mathcal{M} \oplus T^*\mathcal{M}$.
\item $\mathcal{G}^2=1$
\end{itemize}
Using this definition as a starting point, let us now see what happens when a non-vanishing $b$-field is present. Given a generalised K\"{a}hler structure $(\mathcal{J}_1,\mathcal{J}_2)$ one can easily show that performing a $b$-transform yields another generalised complex structure $(\mathcal{J}^b_1,\mathcal{J}^b_2)=(b \mathcal{J}_1 b^{-1}, b \mathcal{J}_2 b^{-1})$. Applying this transformation explicitly to $\mathcal{J}_J$ and $\mathcal{J}_\omega$ we get
	\begin{eqnarray}
	\mathcal{J}_J^B=\left(
		\begin{matrix}
		-J & 0\\
		bJ+J^Tb& J^T
		\end{matrix}
		\right),
		&
	\mathcal{J}_\omega^b=\left(
		\begin{matrix}
		\omega^{-1} &-\omega^{-1}\\
		\omega+b\omega^{-1}b& -b \omega^{-1}
		\end{matrix}
		\right).
	\end{eqnarray}
The resulting metric on the generalised tangent bundle is
	\begin{equation}
	\mathcal{G}^b=\left(
	\begin{matrix}
		-g^{-1} b& g^{-1}\\
		g-bg^{-1}b& bg^{-1}
	\end{matrix} \right)
	\end{equation}
which has the interesting property that its restriction to the tangent bundle, $g-bg^{-1}b$, is a Riemannian metric. Gualtieri furthermore showed that the most general form of a generalised K\"{a}hler metric is fully determined by a Riemannian metric $g$ and a $b$-field via
	\begin{equation}
	\mathcal{G}=\left(
	\begin{matrix}
		-g^{-1} b& g^{-1}\\
		g-bg^{-1}b& bg^{-1}
	\end{matrix}
	\right)=\left(
	\begin{matrix}
		1 & 0\\
		b &1
	\end{matrix}
		\right) \left(
	\begin{matrix}
		0 & g^{-1}\\
		g &0
	\end{matrix}
	\right) \left(
	\begin{matrix}
		1 & 0\\
		-b &1
	\end{matrix}
	\right).
	\end{equation}
Note that this generalised K\"{a}hler metric $\mathcal{G}$ is not obtained from the $b$-transform of a Riemannian metric $g$ since $b$ is not closed when torsion is present.\\
\\
We are now in a position to show how a generalised K\"{a}hler is equivalent to bihermitian geometry. Recall that a bihermitian geometry was determined by the data $(g,b,J_+,J_-)$. Associated with the hermitian complex structures we have two 2-forms $\omega_\pm=gJ_\pm$. One can show that given these elements we can construct $\mathcal{J}_{1/2}$ explicitly as follows
	\begin{equation}
	\label{GaultieriUntwisted}
	\mathcal{J}_{1/2}=\frac{1}{2}\left(
	\begin{matrix}
		1 & 0\\
		b &1
	\end{matrix}
		\right) \left(
	\begin{matrix}
		J_+ \pm J_- & -\left( \omega^{-1}_+ \mp \omega^{-1}_- \right)\\
		\omega_+ \mp \omega_- &-\left(J_+^T \pm J_-^T\right)
	\end{matrix}
	\right) \left(
	\begin{matrix}
		1 & 0\\
		-b &1
	\end{matrix}
	\right).
	\end{equation}
with respect to the untwisted Courant bracket. One can also consider the generalised complex structures with respect to the twisted Courant bracket $[\cdot,\cdot]_H$ using $H=db$ in which case the $b$-transform is no longer needed. In this case, the generalised complex structures corresponding to the bihermitian geometry become
	\begin{equation}
	\label{GaultiereTwisted}
	\mathcal{J}_{1/2}=\frac{1}{2}\left(
	\begin{matrix}
	J_+ \pm J_- & -\left( \omega^{-1}_+ \mp \omega^{-1}_- \right)\\
	\omega_+ \mp \omega_- &-\left(J_+^T \pm J_-^T\right)
	\end{matrix}\right)
	\end{equation}
In the simplest case, i.e. $J_+=J_-$ and $B=0$ this description should reduce to ordinary K\"{a}hler geometry. This is indeed the case, as can be seen from equations (\ref{GaultieriUntwisted}) or (\ref{GaultiereTwisted}). The generalised complex structures defined above reduce to $\mathcal{J}_1 = \mathcal{J}_J$ and $\mathcal{J}_2=\mathcal{J}_\omega$ as should be expected for K\"{a}hler geometry. It is now clear that generalised K\"{a}hler structures contain ordinary K\"{a}hler structures as a subset in addition to being able to describe more general backgrounds, specifically bihermitian geometries with non-commuting complex structures and non-vanishing torsion. The conclusion we are led to is that generalised K\"{a}hler structures provide a natural framework in which to study supersymmetric $\mathcal{N} ~=~(2,2)$ $\sigma$-models.

\section{Pure spinors and type changing}
So far we have determined that generalised complex geometry neatly encapsulates the notions of both complex and symplectic geometry in a natural way by interpolating between the two, yielding a broader class of geometric structures. The existence of a complex or symplectic structure ensures that, locally at least, there exists a preferred set of coordinates that puts them in a canonical form.\\
\\
As we alluded to earlier, when an almost complex structure is integrable, the Newlander-Nirenberg theorem ensures that one can always find coordinates $\{z^a,z^{\bar{a}}\}$ in an open neighbourhood of any point $p\in \mathcal{M}$ that put the complex structure $J$ in its canonical form
	\begin{equation}
	J_0=\left(
	\begin{matrix}
		i \mathds{1} & 0\\
		0 & -i \mathds{1}
	\end{matrix}
		\right).
	\end{equation}
In other words, a complex manifold $(\mathcal{M},J)$ is locally diffeomorphic to the standard complex space ($\mathbb{C}^n,J_0)$. \\
\\
Conversely, given a symplectic manifold $(\mathcal{M},\omega)$ the \emph{Darboux theorem} states that there exist local coordinates $\{x^1,\dots,x^{2d} \}$ called \emph{Darboux coordinates} such that the symplectic form $\omega$ is put in its canonical form
	\begin{equation}
	\omega_0=\left(
	\begin{matrix}
		0 & \mathds{1}\\
		-\mathds{1} & 0 
	\end{matrix}
		\right).
	\end{equation}
The Darboux theorem tells us that locally we can always find symplectomorphisms between the symplectic manifold $(\mathcal{M},\omega)$ and the standard symplectic space $(\mathbb{R}^{2d},\omega_0)$.\\
\\
We would like to do something similar for a generalised complex manifold so that locally 
	\begin{equation}
	(\mathcal{M},\mathcal{J}) \sim (\mathbb{C}^k,J_0) \times (\mathbb{R}^{2d-2k},\omega_0).
	\end{equation}
This is indeed possible as stated by the \emph{generalised Darboux theorem}. The integer $k$ is called the \emph{type} of the generalised complex structure. As we will soon see, $k$ can vary along the manifold, a phenomenon known as \emph{type changing}. In order to make these statements somewhat more concrete we must first introduce spinor representations.\\
\\
Spinor representations correspond to a realisation of the Clifford module characterised by
	\begin{equation}
	\mathbb{X}^2=\mathcal{I} ( \mathbb{X},\mathbb{X} ).
	\end{equation}
We can realise the Clifford algebra by considering polyforms on $\mathcal{M}$, i.e. elements of the exterior algebra $\Lambda^\bullet(T^*\mathcal{M})$. A section of the generalised tangent bundle can be made to act on a polyform $\phi$ as
	\begin{equation}
	\Gamma_{\mathbb{X}} \cdot \phi = \iota_X \phi + \xi \wedge \phi.
	\end{equation}
An easy computation shows that the operators $\Gamma_\mathbb{X}$ furnish a representation of the Clifford module
	\begin{equation}
	\Gamma_\mathbb{X}^2 \cdot \phi = \mathcal{I}( \mathbb{X},\mathbb{X} ) \cdot \phi.
	\end{equation}
Since the Clifford algebra is contained in the dual cover of $SO(d,d)$, $Spin(d,d)$, we have shown that one can regard polyforms as spinors on a generalised complex manifold. One can define a bilinear form on the exterior algebra called the \emph{Mukai pairing}. For any two $\phi_1, \phi_2 \in \Lambda^\bullet(T^*\mathcal{M})$ this bilinear form is given by
	\begin{equation}
	\langle \phi_1,\phi_2 \rangle = \phi_1 \wedge \sigma(\phi_2) \vert_{\text{top}},
	\end{equation}
where the operator $\sigma$ acts on a $p$-form by reversing the order of its components with respect to the chosen basis and $ \vert_{\text{top}}$ projects out the top component of the polyform, which is a $d$-form for a manifold of dimension $d$. The Mukai pairing as defined above has the useful property of being invariant under $O(d,d)$-transformations, and $b$-transforms in particular
	\begin{equation}
	\langle e^b \phi_1,e^b \phi_2 \rangle =\langle \phi_1,\phi_2 \rangle .
	\end{equation}
Given a spinor $\phi$ one can consider its annihilator
	\begin{equation}
	L_\phi = \{ \mathbb{X} \in \Gamma(T\mathcal{M} \oplus T^*\mathcal{M}) \vert \Gamma_\mathbb{X} \cdot \phi =0  \}.
	\end{equation}
It follows immediately from the Clifford relation that such a subspace is isotropic, i.e.	
	\begin{equation}
	\forall \phantom{\gamma} \mathbb{X},\mathbb{Y} \in L_\phi : \mathcal{I}(\mathbb{X},\mathbb{Y})=0.
	\end{equation}
If $L_\phi$ has dimension $d$ it is called \emph{maximally isotropic} and the spinor $\phi$ is called a \emph{pure spinor}. Given a pure spinor we can associate a maximally isotropic subbundle to it, and vice versa. Given two pure spinors $\phi_1$ and $\phi_2$ we can determine if their associated maximally isotropics are distinct by calculating the Mukai pairing
	\begin{equation}
	L_{\phi_1} \cap L_{\phi_2}  = \{ 0 \} \Leftrightarrow \langle \phi_1,\phi_2\rangle \neq 0.
	\end{equation}
Since pure spinors define maximally isotropic subspaces, we would like to know what the general form of these objects is and to which subspace they correspond. Consider a section of an arbitrary subbundle $E \subset T\mathcal{M}$. Its annihilator
	\begin{equation}
	\text{Ann}(E)=\{ \xi \in \Gamma(T^*\mathcal{M}) : \xi(X)=0 ,X \in E\}
	\end{equation}
defines a subspace $L(E,0)=E$ $\oplus$ Ann$(E)$ that is maximally isotropic by construction. In particular, any section $T_p\mathcal{M}$ of the tangent bundle defines a maximally isotropic subspace corresponding to the unit spinor $1 \in \Lambda^\bullet (T^*\mathcal{M})$ since
	\begin{equation}
	\{ X+ \xi \in (T_p \mathcal{M} \oplus  T^*_p \mathcal{M}) : (\iota_X + \xi \wedge)\cdot 1=0 \} = T_p 
	\mathcal{M}.
	\end{equation}
Invariance under $b$-transforms implies that if $\phi$ is a pure spinor, $e^b \phi$ is pure as well. Applying this to the unit spinor we get that any spinor of the form $\phi =1 \wedge e^b = e^b$ is pure for an arbitrary 2-form $b$. We call the associated maximal isotropic
	\begin{equation}
	L(E,B) = \{ (X+(\xi - \iota_X B) : ( X+\xi) \in L(E,0) \}.
	\end{equation}
Similarly, one can take a 1-form $\theta$ and use it to construct a maximally isotropic subspace
	\begin{equation}
	\{ X+ \xi \in (T_p \mathcal{M} \oplus  T^*_p \mathcal{M}) : (\iota_X + \xi \wedge)\cdot \theta =0 \} =\ker (\theta) \oplus 
	\text{span}(\theta),
	\end{equation}
showing us that $\theta$ is pure, as is its $b$-transform $\theta \wedge e^b$. \\
\\
This construction can be generalised as follows. As stated above, any subbundle $E \subset T\mathcal{M}$ defines a maximal isotropic  $L(E,0)=E$ $\oplus$ Ann$(E)$. We call the \emph{type} $k$ of the maximal isotropic the codimension of $E$. The annihilator Ann$(E)$ is then spanned by the 1-forms $(\theta_1, \dots, \theta_k)$. The pure spinor associated with this maximal isotropic can then be written as
	\begin{equation}
	\Omega_k = \theta_1 \wedge \cdots \wedge \theta_k
	\end{equation}
since $(X + \xi) \cdot \Omega_k=0 \Leftrightarrow (X+ \xi) \in L(E,0)$ by construction. We are then free to perform an additional $b$-transform. The only freedom left is multiplying the pure spinor with a constant $c$, which tells us that we are actually looking for sections of a spinor line bundle\footnote{This multiplicative constant need not be the same in every coordinate patch.}. Putting this together we have found the most general form for a pure spinor $\phi$ of type $k$ :
	\begin{equation}
	\phi = c  \phantom{.}\Omega_k \wedge e^b.
	\end{equation}
Pure spinors are useful in that they can be used to characterise generalised complex structures, as we will now show. Recall that the $\pm i$ eigenbundles of an almost complexified generalised complex tangent bundle define two subbundles
	\begin{equation}
	L,\bar{L} \subset \left( T\mathcal{M} \oplus T^* \mathcal{M} \right) \otimes \mathbb{C}
	\end{equation}
We can easily show that $L$ (and $\bar{L}$) are isotropic. For every $x,y$ $\in$ $L$ we have that
	\begin{eqnarray}
	\mathcal{I}(x,y)=\mathcal{I}(\mathcal{J}x,\mathcal{J}y)=\mathcal{I}(ix,iy)=-\mathcal{I}(x,y),
	\end{eqnarray}
so $\mathcal{I}(x,y)=0$. Since both subbundles have maximal dimensionality, they are maximally isotropic subbundles. From the requirement that $L \cap \bar{L} = \{ 0\}$ we can conclude that a generalised complex structure is characterised by a pure spinor
	\begin{equation}
	\phi = c \phantom{.}\Omega_k \wedge e^{b+ i \omega}
	\end{equation}
satisfying $\langle \phi,\bar{\phi} \rangle \neq 0$. Note that we have allowed for a complex $b$-transform defined by the complex 2-form $(b+i \omega)$ due to us working with the complexified generalised tangent bundle.\\
\\
To get a feeling for them we will determine the pure spinors characterising the generalised complex structure associated to a symplectic and an ordinary complex structure. A symplectic type generalised structure
	\begin{equation}
	\mathcal{J}_\omega=\left(
		\begin{matrix}
		0 & -\omega^{-1}\\
		\omega& 0
		\end{matrix}
		\right)
	\end{equation}
is determined by the maximal isotropic 
	\begin{equation}
	L_\omega = \{ X-i\omega(X) : X \in \Gamma(T\mathcal{M} \otimes \mathbb{C})\},
	\end{equation}
which is generated by the pure spinor
	\begin{equation}
	\phi_\omega=e^{i \omega}.
	\end{equation}
We see that a generalised complex structure corresponding to a symplectic structure has type $k=0$. Since the type is determined by the projection of $L_\omega$ to the tangent bundle, a $b$-transform does not change the type of the generalised complex structure. This implies that a generic type 0 pure spinor has the form
	\begin{equation}
	\phi_\omega =c  \phantom{.} e^{b+i\omega}.
	\end{equation}
A generalised complex structure derived from an ordinary complex structure
	\begin{equation}
	\mathcal{J}_J=\left(
		\begin{matrix}
		J & 0\\
		0&- J^T
		\end{matrix}
		\right)
	\end{equation}
has the direct sum of the anti-holomorphic part of the tangent bundle and the holomorphic part of the cotangent bundle as a maximal isotropic. The anti-holomorphic component of the tangent bundle is annihilated by the $(d,0)$-form $\Omega^{d,0}$, in other words the pure spinor has type $d$ and is given by
	\begin{equation}
	\phi_J = \Omega^{n,0}.
	\end{equation}
Just like the previous examples considered, performing a $b$-transform will yield another generalised complex structure of type $n$. \\
\\
The previous discussion makes the generalised Darboux theorem mentioned above concrete. A generalised complex manifold of type $k=0$ corresponds to a symplectic manifold, while the maximal type $k=d$ corresponds to a complex structure. In this sense a generalised complex structure of type $k$ can (locally) be described by $d-k$ symplectic directions and $k$ transverse directions. Using somewhat more precise mathematical language, the complex structure exists transversely to the symplectic foliation associated with the Poisson structure induced by the generalised complex structure. Note that this is a purely local statement. For global considerations, see \cite{bailey}.\\
\\
Since the form of the pure spinors discussed so far is usually defined on a single coordinate patch, the type of a generalised complex structure need not be constant across the entire manifold. As an example, consider the manifold $\mathbb{R}^4$ equipped with the differential form \cite{gualtieri}
	\begin{equation}
	\rho = z_1 + dz_1 \wedge dz_2
	\end{equation}
with $z_1$ and $z_2$ are the standard complex coordinates on $\mathbb{C}^2$. When $z_1 \neq 0$ this defines a symplectic structure
	\begin{equation}
	\rho = z_1 e^{\frac{dz_1 \wedge dz_2}{z1}},
	\end{equation}
which has type $k=0$. This is of course ill-defined on the hypersurface $z_1=0$, where the type changes to $k=0$ and the manifold is complex equipped with standard complex structure corresponding to
	\begin{equation}
	\rho = dz_1 \wedge dz_2.
	\end{equation}
From this discussion we see that the generalised Darboux theorem is only valid for so-called \emph{regular points}, i.e. points where the type is constant in some neighbourhood.\\
\\
Moving on to a generalised K\"{a}hler structure where we have two generalised complex structures $\mathcal{J}_{1,2}$, we have two associated pure spinors $(\phi_1,\phi_2)$ of type
	\begin{eqnarray}
	k_\pm &=& \frac 1 2 \text{corank}_{\mathbb{R}}\left( \omega_+^{-1}\mp \omega_-^{-1}\right) \nonumber \\
	&=& \frac 1 2 \left( 2d - \text{rank}_{\mathbb{R}}  \left( \omega_+^{-1}\mp \omega_-^{-1}\right) \right).
	\end{eqnarray}
We will call $(k_+,k_-)$ the type of the generalised complex structure. In terms of the complex structures, we can write this as
	\begin{equation}
	(k_+,k_-)= \frac 1 2 \left(\dim\ker(J_+ - J_-),\dim\ker(J_+ + J_-) \right).
	\end{equation}
Since the generalised complex structures commute $\mathcal{J}_1$ and $\mathcal{J}_2$ they define distributions on the complexified generalised tangent bundle
	\begin{equation}
	(T\mathcal{M} \oplus T*\mathcal{M}) \otimes \mathbb{C}=L_{++} \oplus L_{+-} \oplus L_{-+} \oplus L_{--},
	\end{equation}
where the subscripts denote the obvious $+i$ and $-i$ eigenbundles of the respective generalised complex structures. We can check that the distributions are involutive in the following manner
\begin{eqnarray}
&&\mathbb{X}_+\in L_{++} \Leftrightarrow \mathbb{X}_+=X_++(g-b)X_+{\mbox{ and }}
\frac 1 2 (1-iJ_+)X_+=X_+\,, \nonumber\\
&&\mathbb{X}_-\in L_{+-} \Leftrightarrow \mathbb{X}_-=X_--(g+b)X_-{\mbox{ and }}
\frac 1 2 (1-iJ_-)X_-=X_-\,, \nonumber\\
&&\bar{\mathbb{X}}_+\in L_{--} \Leftrightarrow \bar{\mathbb{X}}_+=\bar X_++(g-b)\bar X_+{\mbox{ and }}
\frac 1 2 (1+iJ_+)\bar X_+=\bar X_+\,, \nonumber\\
&&\bar{\mathbb{X}}_-\in L_{-+} \Leftrightarrow \bar{\mathbb{X}}_-=\bar X_--(g+b)\bar X_-{\mbox{ and }}
\frac 1 2 (1+iJ_-)\bar X_-=\bar X_-\,.\label{eigenbundle}
\end{eqnarray}
Using this notation, the two associated pure spinors are given by\footnote{Strictly speaking, the pure spinors are only defined up to normalisation, and the associated objects are spinor line bundles. We will not make the distinction here.}
	\begin{eqnarray}
	\Gamma_{\mathbb{X}_+} \cdot \phi_1 = \Gamma_{\mathbb{X}_-} \cdot \phi_1 = 0, &
	\Gamma_{\mathbb{X}_+} \cdot \phi_2 = \Gamma_{\mathbb{X}_-} \cdot \phi_2 = 0.
	\end{eqnarray}
This concludes our overview of generalised complex geometry. Its characterisation in terms of pure spinors will be extremely useful when we will apply this machinery to concrete examples described by $\sigma$-models with $\mathcal{N}=(2,2)$ supersymmetry. In particular we will see that type-changing is a generic feature in these models, which can be made manifest by calculating the various pure spinors underlying the generalised K\"{a}hler geometry.

\section{Generalised K\"{a}hler geometry and $\sigma$-models}
We have now developed the necessary machinery to conclude our discussion of $\mathcal{N}=(2,2)$ supersymmetric $\sigma$-models that we began in chapter 4 in a satisfactory manner. Recall that one of the key properties of a usual K\"{a}hler manifold was that it was equipped with a globally defined closed 2-form $\omega$ called the K\"{a}hler form
	\begin{eqnarray}
	d\omega =0,
	\end{eqnarray}
which could be employed to fully determine the geometry in the absence of torsion. The closure of the $\mathcal{N}=(2,2)$ super algebra required a general background to be bihermitian, which will be no longer K\"{a}hler except when both complex structures are either equal to each other or each other's opposite.\\ 
\\
Both complex structures making up the bihermitian geometry can be used to define a 2-form via
	\begin{eqnarray}
	\omega^{(\pm)}_{\mu\nu}=g_{\nu\rho}J^{\rho}_{\pm\nu},
	\end{eqnarray}
which are not closed in the usual sense due to the presence of a torsion field $H$. Introducing the (real) Dolbeault operators, which when expressed in coordinates adapted to the corresponding complex structure take on the form
	\begin{eqnarray}
	d^c_\pm=i(\bar \partial-\partial),
	\end{eqnarray}
it can be show that the 2-forms $\omega^{(\pm)}$ satisfy\footnote{For a proof, see \cite{gualtieri}.}
\begin{eqnarray}
\pm d^c_\pm \omega^{(\pm)}=\mp H.
\end{eqnarray}
We would now like to make this structure more apparent in the context of the $\sigma$-model. Recall that locally, the choice of $b$-field was only defined up to a gauge transformation $b \rightarrow b+d\xi$. As such, we can introduce two different choices $b^{\pm}$ and, using the locally defined 2-forms  $\Omega^\pm$ introduced before we can write down
\begin{eqnarray}
&&\big(g-b_+\big)(X,Y)=\Omega ^+(X,J_+Y) \nonumber\\
&&\big(g+b_-\big)(X,Y)=\Omega ^-(X,J_-Y),\label{gminb}
\end{eqnarray}
with the torsion given by $db_+ = db_- =H$. Recall that $\Omega^+$ and $\Omega^-$ are not globally defined when both chiral and twisted-chiral fields are present. In order to make contact with the Gualtieri map we can relate $b_+$ and $b_-$ to (\ref{GaultieriUntwisted}) via
\begin{eqnarray}
b&=&b_++\frac 1 2 \,d\big(-V_l\,dl-V_{\bar l}\,d\bar l +V_z\,dz+V_{\bar z}d\bar z\big)
\nonumber\\
&=&b_-+\frac 1 2 \,d\big(-V_r\,dr-V_{\bar r}\,d\bar r +V_w\,dw+V_{\bar w}d\bar w\big).
\label{bgauge}
\end{eqnarray}
The pure spinors can now be explicitly calculated in the form \cite{halmagyi} \cite{hullampere},
\begin{eqnarray}
\phi_+&=&d\bar z^1\wedge d\bar z^2\wedge \cdots\wedge  d\bar z^{n_c}\wedge
e^{i\, \Omega^++\Xi^+} \nonumber\\
\phi_-&=&d\bar w^1\wedge d\bar w^2\wedge \cdots\wedge  d\bar w^{n_t}\wedge
e^{i\, \Omega^-+\Xi^-}\,,\label{psexpl}
\end{eqnarray} 
with
\begin{eqnarray}
\Xi^+&=&\frac 1 2 \,d\big(V_l\,dl+V_{\bar l}\,d\bar l
-V_z\,dz-V_{\bar z}\,d\bar z\big)\, \nonumber\\
\Xi^-&=&\frac 1 2 \,d\big(V_r\,dr+V_{\bar r}\,d\bar r
-V_w\,dw-V_{\bar w}\,d\bar w\big)\, .\label{xidef}
\end{eqnarray} 
In the special case where we only have twisted chiral superfields parametrising the manifold we find that $i\, \Omega^+ + \Xi^+ = - (i\, \Omega^- + \Xi^-)$, while conversely when there are only chiral superfields present we have that $i\, \Omega^+ + \Xi^+ =  i\, \Omega^- + \Xi^-$.\\
\\
Once these expressions have been obtained we can calculate the Mukai pairings
\begin{eqnarray}
\big(\phi_+,\bar \phi_+\big)&=&(-1)^{n_c(n_c+1)/2+n_t+n_s}\,2^{n_t+2n_s}\,\det N_+\, \nonumber\\
\big(\phi_-,\bar \phi_-\big)&=&(-1)^{n_t(n_t+1)/2}\,2^{n_c+2n_s}\,\det N_-\, ,
\label{mukaip}
\end{eqnarray}
where we have employed the notation
\begin{eqnarray}
N_+=\left(
\begin{array}{ccc}
V_{l\bar l}&V_{lr}&V_{l\bar w}\\
V_{\bar r\bar l}&V_{\bar r r}&V_{\bar r \bar w}\\
V_{w\bar l}&V_{wr}&V_{w\bar w}\end{array}
\right)
\end{eqnarray}
and
\begin{eqnarray}
N_-=\left(
\begin{array}{ccc}
V_{l\bar l}&V_{l\bar r}&V_{l\bar z}\\
V_{r\bar l}&V_{r\bar r }&V_{ r \bar z}\\
V_{z\bar l}&V_{z\bar r}&V_{z\bar z}\end{array}
\right).
\end{eqnarray}
These relations are related to the previously performed calculation in superspace of the one-loop for a general $\mathcal{N}=(2,2)$ $\sigma$-model \cite{grisarusevrin1}, where it was found that in order to be superconformally invariant the expression
\begin{eqnarray}
{\cal S}_{1-\mbox{loop}}\,\propto\frac{1}{\varepsilon}\,\int d^2 \sigma \,d^2 \theta \, d^2 \hat \theta \,\ln\,\frac{\det\big(N_+\big)}{\det\big(N_-\big)}\,,
\end{eqnarray}
must vanish. This is indeed the case when 
\begin{eqnarray}
\frac{\det\big(N_+\big)}{\det\big(N_-\big)}\,=\pm|f_+(l,w,z)|^2
|f_-(r, \bar{w},z)|^2
,\label{CYcd2}
\end{eqnarray}
for some arbitrary functions $f_+$ and $f_-$. While condition (\ref{CYcd2}) is required for the model to be superconformally invariant at one-loop it does not guarantee that it is a valid supergravity background. To ensure that this is so, we will need a stronger condition.\\
\\
Using the obtained expressions we are able to check if the generalised K\"{a}hler geometry under consideration obeys a stronger condition, namely the \emph{generalised Calabi-Yau condition}, which requires the pure spinors to be globally defined and closed with respect to the untwisted Courant bracket
\begin{eqnarray}
d\phi_+=0, \qquad d\phi_-=0,
\end{eqnarray}
and the Mukai pairings to satisfy
\begin{eqnarray}
(\phi_+, \bar \phi_+) = c (\phi_- , \bar \phi_-) \neq 0,
\end{eqnarray}
with $c$ a non-zero constant. This is clearly a stronger condition than (\ref{CYcd2}) Employing the notation introduced before it was shown in \cite{hullampere} that this can equivalently stated as 
\begin{eqnarray}
\frac{\det\big(N_+\big)}{\det\big(N_-\big)}=c \neq 0.
\end{eqnarray}
Note that we could have defined this condition with respect to the $H$-twisted Courant bracket, resulting in the condition that the pure spinors are no longer closed but $H$-closed
\begin{eqnarray}
d\phi_\pm= H \wedge \phi_\pm.
\end{eqnarray}
While the condition of the background geometry being generalised Calabi-Yau is a strong one, there is a subtlety that makes it definition ambiguous. Under a coordinate transformation $x \rightarrow x^\prime(x)$ a pure spinor transforms as
\begin{eqnarray}
\phi(x) \rightarrow \phi^\prime(x^\prime)=\sqrt{\frac{\partial x^\prime}{\partial x}}\phi(x),
\end{eqnarray}
in other words, as a density. This is a direct result of the requirement that the Mukai pairing should be invariant under coordinate transformations. Consequently, the generalised Calabi-Yau condition is a coordinate-dependent statement. We will encounter this ambiguity when we consider concrete examples in the later.

\chapter{Wess-Zumino-Witten models}
\thispagestyle{empty}
\emph{In the previous two chapters we have seen that a general $\mathcal{N}=(2,2)$ supersymmetric $\sigma$-model restricts its target space to be a bihermitian manifold and that bihermitian geometry admits a natural interpretation within the framework of generalised complex geometry. Once twisted and semi-chiral superfields are present the target space is no longer K\"{a}hler due to a non-zero background torsion field, resulting in a geometry that interpolates between complex and symplectic geometry. We would like to find specific models where this behaviour is apparent. In this chapter we will focus on a large class of manifolds that provide examples of this kind of background geometry.}

\newpage
\section{$\sigma$-models on group manifolds}
Recall that, at tree level, the vanishing of the $\beta$-functions of the most general bosonic two-dimensional $\sigma$-model restricted the background geometry to be that of a Ricci-flat manifold with a torsionful connection. One example of such manifolds are the so-called \emph{parallelisable manifolds}. A $d$-dimensional parallelisable manifold $\mathcal{M}$ is defined by the existence of $d$ linearly independent vector fields, in other words, as a manifold where the  tangent bundle $T\mathcal{M}$ is a trivial bundle. Such manifolds admit a flat connection, or employing the language we used in chapter 4,
	\begin{equation}
	R_{\mu\nu}^{(+)}=0.
	\end{equation}
At tree level and to first order in $\alphap$ such manifolds are consistent superstring backgrounds. We can distinguish between two kinds of parallelisable manifolds~: Lie groups and the spheres $S^1, S^3$ and $S^7. $\footnote{$S^1$ and $S^3$ being diffeomorphic to the Lie groups $U(1)$ and $SU(2)$, respectively.} In what follows we will consider only Lie groups. We will first show how to connect a $\sigma$-model with a Lie manifold as its target space to our previous formulation of the string action.\\
\\
Recall that we could choose the conformal gauge in which the Polyakov action takes the following form
	\begin{equation}
	S_P=-\frac{1}{4 \pi \alphap}\int_\Sigma d^2 \sigma \, \eta_{\mu\nu} \partial_\alpha X^{\mu}\partial^{\alpha}X^{\nu}.
	\end{equation}
Introducing the complex coordinates
	\begin{eqnarray}
	z=e^{i(\tau + \sigma)},&\bar{z}=e^{i(\tau - \sigma)},
	\end{eqnarray}
we can re-write this more compactly as 
	\begin{equation}
	S_P=-\frac{1}{2 \pi \alphap}\int_\Sigma dz d\bar{z} \left( \partial X \cdot \bar{\partial}X \right).
	\end{equation}
We will now reformulate the Polyakov action as a $\sigma$-model defined on an abelian group manifold. Introducing $d$  Lie algebra generators $\{T_A\}$ satisfying
	\begin{eqnarray}
	[T_A,T_B]=0, & \Tr(T_A T_B)=2 \eta_{AB},
	\end{eqnarray}
we can define the following group element by exponentiation
	\begin{equation}
	U \equiv \exp(i \phantom{.}T_A X^A).
	\end{equation}
Note that the generators ${T_A}$ have to be in the adjoint representation of the algebra, since it is the only faithful $d$-dimensional representation available. The Lie group defined in this fashion is the abelian group $U(1)^d$. Expressed in these terms the equivalent form of the Polyakov action is
	\begin{equation}
	\label{abelaction}
	S=- \frac{1}{\pi \alphap}\int dzd\bar{z} \phantom{.}   \Tr \left( \partial \ln U  \cdot  \bar{\partial} \ln U^{-1} \right).
	\end{equation}
A useful property, and an important historical motivation for considering this class of models, is that the theory now has a global $U(1)^d \otimes U(1)^d$ symmetry,\\
	\begin{equation}
	U(z,\bar{z}) \rightarrow h_- U(z,\bar{z}) h_+,
	\end{equation}
with $h_\pm \in U(1)^d$. Since this symmetry allows for different group elements multiplying the fields $U(z,\bar{z})$ from the left and the right it is known as a \emph{chiral symmetry}.\\
\\
The abelian Lie group considered thus far corresponded to a string propagating in a flat Minkowski space. In order to obtain more involved geometrical backgrounds we take the action (\ref{abelaction}) as a starting point and consider the non-abelian Lie algebra $\mathfrak{g}$ generated by $\{T_A\}$, $A \in \{1, \dots , $dim$(\mathfrak{g})\}$ in the adjoint representation obeying
	\begin{eqnarray}
	[T_A,T_B] &=& i f_{AB}^{\phantom{AB}C} T_C \nonumber \\
	\Tr (T_A , T_B ) &=& 2 \eta_{AB}
	\end{eqnarray}
The non-linear $\sigma$-model with the group manifold $G$ defined by $Lie(G)=\mathfrak{g}$ as the target space now corresponds to the action
	\begin{eqnarray}
	S_0&\equiv&\int_\Sigma dzd\bar{z}  \phantom{.}  \Tr \left( \partial g \cdot\bar{\partial} g^{-1} 
	\right) \nonumber \\
	&=&\int_\Sigma  dzd\bar{z}  \ \phantom{.}  \Tr \left( g^{-1} \partial g \cdot g^{-1} \bar \partial g 
	\right) 
	\end{eqnarray}
By considering non-abelian group manifolds the $\sigma$-model defined above is no longer a permissible string background. In general group manifolds are not Ricci-flat with respect to the usual connection. This can be solved by adding the Wess-Zumino term (in terms of the real coordinates)
	\begin{equation}
	S_1 \equiv \int_\Xi d^2 \sigma dt \phantom{.}\epsilon^{\mu \nu \rho} \Tr \left( g^{-1} \partial_\mu g \cdot  g^{-1} 
	\partial_\nu g \cdot g^{-1} \partial_\rho g  \right).
	\end{equation}
The space $\Xi$ is chosen so that it has the worldsheet of the string as a boundary $\Sigma = \partial \Xi$. The integrand of the Wess-Zumino term is a 3-form defined on a three-dimensional manifold and is as such closed. By the Poincarr\'{e} lemma one can locally find a 2-form so that this 3-form is exact
	\begin{equation}
	 \Tr \left( g^{-1} \partial_\mu g \cdot  g^{-1} \partial_\nu g \cdot  g^{-1} \partial_\rho g \right)=dB.
	\end{equation}
According to Stokes' theorem, integrating an exact 3-form over the manifold $\Xi$ amounts to integrating the 2-form $B$ over its boundary, making its identification with the 2-form $B$ in the non-linear $\sigma$-model natural. It has been shown that by choosing the appropriate combination of the terms $S_0$ and $S_1$ the $\beta$-functions can be made to vanish \cite{wittenwzw}. This combination defines the Wess-Zumino-Witten model 
	\begin{equation}
	S_{WZW}=-\frac{k}{2 \pi} \int dz d\bar{z} \phantom{.}\Tr \left( (\partial g g^{-1})^2 + \frac{2}{3} d^{-1} (dg g^{-1})^3  \right),
	\end{equation}
where $k$ is an integer called the \emph{level} of the theory. The requirement of $k$ being an integer comes from demanding quantum conformal invariance. As in the abelian case, the WZW-models possesses a chiral symmetry corresponding to the currents
\begin{eqnarray}
j=-\partial_+ gg^{-1}, & \bar j = g^{-1}\partial_- g.
\end{eqnarray}
Finally, we note that having chosen a parametrisation for the group element $g$ and by comparing the WZW-action with the general second order action for the $\sigma$-model in real coordinates we can express the geometrical data using the left- and right-invariant vielbeins, defined as
\begin{eqnarray}
 g^{-1}dg=i\,L_\mu^B\,T_B\,dx^\mu,\qquad
 dgg^{-1}=i\,R_\mu^B\,T_B\,dx^\mu\,.
\end{eqnarray}
By using these expressions we can write down the metric and the torsion 3-form as 
\begin{eqnarray}
 g_{\mu\nu}&=&-\frac{k}{8\pi  x}\,\mbox{Tr}\,\partial _\mu g g^{-1}\partial _\nu gg^{-1} \nonumber \\
 &=&
 \frac{k}{8\pi }R_\mu^C\,R_\nu^D\,\eta_{CD}\, \nonumber \\
  H_{\mu\nu\rho}&=&\frac{k}{24\pi x}\,\mbox{Tr}\,dgg^{-1}\wedge dgg^{-1}\wedge dgg^{-1} \nonumber \\
  &=&\frac{k}{48\pi }R_\mu^DR_\nu^ER_\rho^Ff_{DEF}\,dx^\mu\wedge dx^\nu\wedge dx^\rho.
\end{eqnarray}
In order to make contact with the previously considered $\mathcal{N}=(2,2)$ $\sigma$-models we will next investigate under which conditions we can enhance the symmetry.

\section{Supersymmetric WZW models}
Using the machinery reviewed in chapters 3 and 4 we will now consider the supersymmetric WZW-model by using superspace formalism. The obvious generalisation is to replace the ordinary derivatives in the WZW-action by supercovariant $\mathcal{N}=(1,1)$ derivatives and to extend the integration over an additional superspace coordinate $\theta$. This leads us to the action
	\begin{eqnarray}
	S&=&-\frac{k}{\pi x} \int_\Sigma d^2 \sigma d^2 \theta \Tr \left( D_+ g g^{-1} D_- gg^{-1}\right) \nonumber \\
	&+&  \frac{k}{\pi x} \int_\Xi dt  d^2 \sigma d^2 \theta \Tr \left( \partial_t g g^{-1} \{ D_+ g g^{-1},D_- gg^{-1}\} \right).
	\end{eqnarray}
The equations of motion for this model are
	\begin{equation}
	D_+ \left(D_- g g^{-1} \right)=D_- \left( g^{-1} D_+ g\right)=0.
	\end{equation}
As with the non-supersymmetric case, the supersymmetric WZW-model is also invariant under affine transformations
\begin{eqnarray}
 g\rightarrow h_-\,g\,h_+,\label{wzwiso}
\end{eqnarray}
with $h_\pm\in{G}$ and satisfying the constraints
\begin{eqnarray}
 D_+h_-=D_-h_+=0.\label{wzwiso1}.
\end{eqnarray}

\subsection{Complex structures on the Lie algebra}
Having introduced the $\mathcal{N}=(1,1)$ superspace formulation of the WZW-model we want to determine what enhanced supersymmetry the model can possess. When $G$ is a reductive Lie group\footnote{A reductive Lie group is a Lie group whose Lie algebra can be written as the direct sum of a semi-simple Lie algebra and an additional abelian piece.} and even dimensional the model has an additional $\mathcal{N}=(2,2)$ supersymmetry. Analogously to our approach in chapter 4 we add an additional non-manifest supersymmetry, after which we will investigate under which conditions the supersymmetry conditions are satisfied. Expanding the group elements appearing in the lagrangian this symmetry takes the following form
	\begin{eqnarray}
	\left(g^{-1} \delta g \right)^A &=& \epsilon^+ \mathbb{J}^A_{+B}(g^{-1}D_+ g)^B \nonumber \\
	&+& \epsilon^- (LR^{-1})^A_{\phantom{A}
	C} \mathbb{J}^C_{-D}(RL^{-1})^D_{\phantom{D}B}(g^{-1}D_-g)^B \\
	(\delta g g^{-1})^A&=& \epsilon^+ (RL^{-1})^A_{\phantom{A}C} \mathbb{J}^C_{+D} (LR^{-1})^D_{\phantom{D}B} 
	(D_+ g g^{-1})^B \nonumber \\
	&+& \epsilon^- \mathbb{J}^A_{-B} (D_- g g^{-1})^B
	\end{eqnarray}
The transformations defined above satisfy the $\mathcal{N}=(2,2)$ algebra if the operators $\mathbb{J}_\pm$ satisfy the following conditions \cite{spindelsevrin} :
\begin{enumerate}
\item $\mathbb{J}^A_{\pm B}$ are constant and satisfy  $\mathbb{J}^A_{\pm C} \mathbb{J}^C_{\pm B} =-\delta^A_B$.
\item $\mathbb{J}^C_{\pm A} \mathbb{J}^D_{\pm B} \eta_{CD}= \eta_{AB}$. 
\item $f_{DEA} \mathbb{J}^D_{\pm B}\mathbb{J}^E_{\pm C} + f_{DEB} \mathbb{J}^D_{\pm C}\mathbb{J}^E_{\pm A} + f_{DEC} \mathbb{J}^D_{\pm A}\mathbb{J}^E_{\pm B}=f_{ABC}$
\end{enumerate}
Let us investigate these three conditions in turn. The first condition, when applied to the complexified Lie algebra, tells us that $\mathbb{J}^A_{\pm B}$ are almost complex structures and that their possible eigenvalues are $\pm i$. We can always choose the basis of the Lie algebra such that the almost complex structures are diagonal so that their action on the basis elements becomes straightforward
	\begin{eqnarray}
	\mathbb{J}_\pm T_A = +i T_A, & \mathbb{J}_\pm T_{\bar{A}} = -i T_{\bar{A}}.
	\end{eqnarray}
The second condition requires the Cartan-Killing metric on the Lie algebra, defined by
	\begin{equation}
	\eta_{AB}=-\frac{1}{\tilde{h}} f_{AC}^{\phantom{AC}D}f_{BD}^{\phantom{BD}C},
	\end{equation}
with $\tilde{h}$ is the dual Coxeter number of the Lie algebra, to be hermitian with respect to the almost complex structures $\mathbb{J}$. This condition sets the $(2,0)$ and $(0,2)$ pieces of the Cartan-Killing metric to zero	\begin{equation}
	\eta_{AB}=\eta_{\bar{A}\bar{B}}=0.
	\end{equation}
The third and last condition amounts to an integrability condition and introduces the following constraint on the structure constants
	\begin{equation}
	f_{ABC}=f_{\bar{A}\bar{B}\bar{C}}=0.
	\end{equation}
The constraints considered above can be satisfied by choosing a Cartan decomposition
	\begin{equation}
	\mathfrak{g}=\mathfrak{h} \oplus \mathfrak{e}_+ \oplus \mathfrak{e}_-,
	\end{equation}
with $\mathfrak{h}$ the Cartan subalgebra, $\mathfrak{e}_+$ spanned by the positive roots and $\mathfrak{e}_-$, spanned by the negative roots, and have the complex structures act on the Lie algebra in the following way
	\begin{eqnarray}
	\mathbb{J}_\pm \mathfrak{h} &=& \mathfrak{h} \nonumber \\
	\mathbb{J}_\pm e_+&=&+i e_+ \nonumber \\ 
	\mathbb{J}_\pm e_- &=&-i e_+,
	\end{eqnarray}
with $e_+ \in \mathfrak{e}_+$ and $e_- \in \mathfrak{e}_- $. The positive roots have eigenvalue $+i$, the negative roots have eigenvalue $-i$ and the Cartan subalgebra is mapped to itself. We conclude that the only freedom lies in the way the Cartan subalgebra is mapped to itself.

\subsection{Complex structures on the group manifold}
Having chosen the complex structures on the Lie algebra the next step is to determine how they relate to the possible complex structures on the Lie group when considered as a manifold. Any well-defined operator on the Lie algebra can be related to an operator on the group manifold by use of the left- and right- invariant vielbeins $L_a^B$ and $R_a^B$. These can be calculated by expanding the left- and right invariant 1-forms over the Lie algebra generators
	\begin{eqnarray}
 	g^{-1}dg=i\,L_\mu^B\,T_B\,dx^\mu,\qquad
 	dgg^{-1}=i\,R_\mu^B\,T_B\,dx^\mu\,.
	\end{eqnarray}
The complex structures on the group manifold $J_\pm$ are then related to the complex structures on the Lie algebra $\mathbb{J}_\pm$ via
	\begin{eqnarray}
 	\mathbb{J}_+^A{}_B=L_\mu^A\,J^\mu_{+\nu}L^\nu_B,\qquad
	\mathbb{J}_-^A{}_B=R_\mu^A\,J^\mu_{-\nu}R^\nu_B.
	\end{eqnarray}
Note that in order to calculate the left- and right-invariant vielbeins it is necessary to pick an explicit parametrisation of the group manifold. A particularly convenient way of doing so is by considering the following decomposition of the Lie algebra $\mathfrak{g}$ 
	\begin{eqnarray}
	\mathfrak{g}= \mathfrak{b} \oplus \bigoplus^n_{k=1} \mathfrak{d}_k \oplus \bigoplus^n_{k=1} \mathfrak{f}_k,
	\end{eqnarray}
where $\mathfrak{b}$ is an abelian subalgebra, $\mathfrak{d}_k$ are subalgebras isomorphic to $\mathfrak{su}(2)$ and the remaining piece $\mathfrak{f}_k$ commutes with $\mathfrak{d}_j$ for $j > k$ and is closed under the action of an element of $\mathfrak{d}_k$. Once this is done the Lie algebra elements obtained from this decomposition can be exponentiated to yield the group element $G$\footnote{From now on, we will use $G$ instead of $g$ as the group element to avoid confusion with the metric $g$.}.\\
\\
Once the complex structures on the Lie algebra, and by extension on the Lie group, are determined, the field content of the $\sigma$-model will depend on the algebraic properties of the complex structures $J_\pm$ as described in chapter 4. Since a WZW-model requires the Wess-Zumino term in order to be consistent, the target space is necessarily non-K\"{a}hler since background torsion will always be present. Consequently, a description in terms of chiral superfields alone is not possible. In order for no semi-chiral superfields to present, given a choice of complex structures on the Lie algebra, the following equation must hold
	\begin{equation}
	[\mathbb{J}_+,G\mathbb{J}_- G^{-1}]=0
	\end{equation}
for arbitrary group elements $G$ in the adjoint representation. Setting $G=e^{i \alpha}$ and analysing this condition to first order in $\alpha$ tells us that the only non-abelian Lie group that admits such a description is $SU(2) \times U(1)$. All other choices will necessitate at least one semi-chiral superfield parametrising the manifold.\\
\\
Different choices of complex structures on the Lie algebra will correspond to different types of field content of the $\sigma$-model. A systematic investigation of the relation between the two has not yet been performed, however some systematics seem to be present when considering the different possibilities on Lie groups of rank 2 where the Cartan subalgebra is two-dimensional. Taking $\mathbb{J}_+=\mathbb{J}_-$ on the Cartan subalgebra one finds that $\ker(\mathbb{J}_+ + G\mathbb{J}_- G^{-1})$ is always trivial. The contents of $\ker(\mathbb{J}_+ - G\mathbb{J}_- G^{-1})$ can then be explicitly checked, the results of which can be found in table 1, where $n_c$ denotes the number of chiral superfields, $n_t$ the number of twisted-chiral superfields and $n_s$ the number of semi-chiral superfield pairs\footnote{Since left and right semi-chiral always appear together, $n_s$ denotes the number of full semi-chiral multiplets.}. We see that, for rank 2 algebras at least, this choice maximises the number of semi-chiral directions.\\
\\
Another possibility is to take $\mathbb{J}_+=\mathbb{J}_-$ on the roots but opposite in the Cartan subalgebra. A similar analysis can then be performed on $\ker(\mathbb{J}_+ + G\mathbb{J}_- G^{-1})$ and $\ker(\mathbb{J}_+ - G\mathbb{J}_- G^{-1})$, the results of which can be found in table 2. We see that this choice corresponds to the minimal amount of semi-chiral superfields, resulting in a parametrisation of $SU(2) \times U(1)$ in terms of chiral and twisted-chiral superfields alone, as discussed above.

\newpage

\begin{table}
\begin{center}
\begin{tabular}{|c|c|c|}
\hline 
Group& $n_s$& $n_c$\\
\hline\hline
$SU(2)\times U(1)$&1&0 \\ 
$SU(2)\times SU(2)$&1&1 \\ 
$SU(3)$&2&0 \\ 
$SO(5)$&2&1 \\ 
$G_2$&3&1 \\ 
\hline
\end{tabular}
\caption{The superfield content for the rank 2 non-abelian reductive Lie groups when
taking $\mathbb{J}_-=\mathbb{J}_+$ on the Lie algebra. }
\end{center}
\end{table}

\begin{table}[h]
\begin{center}
\begin{tabular}{|c|c|c|c|}
\hline 
Group& $n_s$& $n_c$&$n_t$\\
\hline\hline
$SU(2)\times U(1)$&0&1&1 \\ 
$SU(2)\times SU(2)$&1&0 &1\\ 
$SU(3)$&1&1& 1\\ 
$SO(5)$&1&1 &2\\ 
$G_2$&2&1 &2\\ 
\hline
\end{tabular}
\caption{The superfield content for the rank 2 non-abelian reductive Lie groups when
taking $\mathbb{J}_-|_{\mbox{\small roots}}=\mathbb{J}_+|_{\mbox{\small roots}}$ and $\mathbb{J}_-|_{\mbox{\small CSA}}=-\mathbb{J}_+|_{\mbox{\small CSA}}$ on the Lie algebra.}
\end{center}
\end{table}

\chapter{Some concrete examples}
\thispagestyle{empty}
\emph{We now turn to the main content of this thesis, namely three specific examples of Wess-Zumino-Witten models admitting a description in terms of generalised K\"{a}hler geometry. Starting from the Lie algebra we will investigate the possible complex structures on the target space and the corresponding parametrisation in terms of constrained $\mathcal{N}=(2,2)$ superfields. By providing an explicit form for the generalised K\"{a}hler potential we will investigate the affine isometries that are compatible with the extended supersymmetry and calculate the pure spinors corresponding to the generalised K\"{a}hler structures. We will find that type changing is a generic occurrence and discuss its implications for mirror symmetry. The bulk of this chapter is based on the work in \cite{sevrinterryn}.}

\newpage
\section{$SU(2) \times U(1)$}
The simplest non-trivial example of a $\mathcal{N}=(2,2)$ WZW model has $SU(2) \times U(1)$ as its target space. It was first formulated in terms of one chiral and one twisted-chiral superfield \cite{sevrinblack},\cite{Rocek:1991vk}. Later an alternative description in terms of two semi-chiral superfields was considered \cite{ivanovkim},\cite{sevrintroost}. We will systematically treat the model by considering the possible choices of complex structure on the Lie algebra.

\subsection{Parametrisation and topology}
As is well known the Lie algebra $\mathfrak{su}(2) \oplus \mathfrak{u}(1)$ is generated by the Pauli matrices $\sigma_j$ with $j \in {1,2,3}$ and the $2\times2$ unit matrix $\sigma_0$. We will consider the complexified Lie algebra generated by the elements
	\begin{eqnarray}
	h=\frac{1}{2}\left( \sigma_3 + i \sigma_0 \right), & \bar{h}=\frac{1}{2}\left( \sigma_3 - i \sigma_0 \right), \\
	e=\frac{1}{2}\left( \sigma_1 + i \sigma_2 \right), & \bar{e}=\frac{1}{2}\left( \sigma_1 - i \sigma_2 \right).
	\end{eqnarray}
We choose the following parametrisation for a $SU(2) \times U(1)$ group element
	\begin{eqnarray}
	g= e^{i\rho }\,\left(
	\begin{array}{cc} \cos \psi \, e^{i\varphi _1} & \sin \psi \,e^{i\varphi _2}\\
	-\sin\psi\, e^{-i\varphi _2} & \cos\psi\, e^{-i\varphi _1}\end{array}\right),
	\label{su2u1coor}
	\end{eqnarray}
with $\varphi_1,\varphi_2,\rho \in \mathbb{R}$ mod $2\pi$ and $\psi \in [0,\frac{\pi}{2}[$. Locally the topology of the group is that of a 3-torus $T^3$,  parametrised by $\varphi_1,\varphi_2$ and $\rho$, fibered over a line segment parametrised by $\psi$. The manifold is locally isomorphic to the product $S^3 \times S^1$ and can be identified with the rational Hopf surface $(\mathbb{C}^2/(0,0))/\Gamma$ by introducing
	\begin{eqnarray}
	w&=&e^{-\rho-i \varphi_1} \cos \psi \\
	z&=& e^{-\rho+i \varphi_2} \sin \psi ,
	\end{eqnarray}
with elements of $\Gamma$ acting on $(w,z)$ as
	\begin{equation}
	\Gamma \cdot (w,z) \rightarrow e^{2\pi n}(w,z),n\in \mathbb{Z}.
	\end{equation}
The geometrical data for this manifold is well known. The metric is diagonal with the line element given by
	\begin{equation}
	ds^2 = \frac{k}{2\pi} \left( d\rho^2 + d\psi^2+\cos^2\psi d\varphi_1^2 + \sin^2\psi d\varphi_2^2 \right),
	\end{equation}
while the torsion 3-form is
	\begin{equation}
	H=\frac{k}{2\pi} \sin 2\psi \phantom{.} d\varphi_1 \wedge d\varphi_2 \wedge d\psi,
	\end{equation}
which by choosing an appropriate gauge can be obtained from the 2-form $B$ via $H=dB$ whose only non-zero element 
is $B_{\varphi_1 \varphi_2}=-k\cos2\psi/4\pi$.

\subsection{The complex structures}
The Cartan subalgebra is spanned by the elements $h,\bar{h}$, while $e,\bar{e}$ are the positive and negative roots, respectively. We will order the basis for the complexified Lie algebra as $(h,e,\bar{h},\bar{e})$. The only freedom we have in defining the complex structures on the Lie algebra lies in their action on the elements  $h,\bar{h}$
	\begin{eqnarray}
	\mathbb{J}_1 h = +i h, & \mathbb{J}_1 \hb = -i \hb \\
	\mathbb{J}_2 h = -i h, & \mathbb{J}_2 \hb = +i \hb,
	\end{eqnarray}
resulting in the complex structures
	\begin{eqnarray}
	\mathbb{J}_1 = \left(
	\begin{matrix}
	i \sigma_0 & 0 \\
	0 & -i \sigma_0
	\end{matrix}\right), &
	\mathbb{J}_2 = \left(
	\begin{matrix}
	-i \sigma_3 & 0 \\
	0 & i \sigma_3
	\end{matrix}\right).
	\end{eqnarray}
Our chosen Cartan decomposition is not unique, since we still have the freedom to choose a different decomposition which corresponds to conjugating our chosen complex structures with a group element $G \in SU(2)$ in the adjoint representation. Writing this group element as
	\begin{equation}
	G=\left( \begin{matrix}
	\cos\theta & \sin\theta \\
	-\sin\theta &  \cos\theta
	\end{matrix}\right)
	\end{equation}
we obtain the most general form for the complex structures on the Lie algebra
	\begin{eqnarray}
	\mathbb{J}_1(\theta ,\phi ) &=& G\mathbb{J}_1G^{-1} \nonumber \\
	&= &\left(
	\begin{matrix}
	i \cos\theta & 0 & 0 & e^{-i\phi} \sin\theta \\
	0 & i \cos\theta & -e^{-i \phi} & 0\\
	0 & e^{i \phi}\sin\theta & -i\cos\theta & 0 \\
	-e^{i\phi}& 0 & 0 & -i \cos\theta 
	\end{matrix}
	\right), \nonumber\\
	\mathbb{J}_2(\theta ,\phi ) &=&G\mathbb{J}_2G^{-1} \nonumber \\
	&=& \left(
	\begin{matrix}
	-i\cos\theta & e^{i \phi} \sin\theta & 0 & 0\\
	-e^{-i \phi} \sin\theta & i\cos\theta & 0 & 0\\
	0 & 0 & i\cos\theta & e^{-i \phi} \sin\theta\\
	0 & 0 & e^{i \phi} \sin\theta & -i\cos\theta  
	\end{matrix}
	\right).\nonumber\\
	\end{eqnarray}
This two-parameter family of complex structures is consistent with the fact that $SU(2) \times U(1)$ allows for an enhanced $\mathcal{N}=(4,4)$ supersymmetry, where we have a two-sphere's worth of different complex structures \cite{ivanovkim}.\\
\\ 	
By using the left- and right-invariant vielbeins $L^\mu_B$ and $R^\mu_B$ we can calculate the complex structures on the manifold $J^\mu_{\pm \nu}$. Both these structures can be calculated independently, leaving us with four possible combinations :

\begin{enumerate}
  \item $J_+^\mu{}_\nu=L^\mu_C\,\mathbb{J}_1(\theta ,\phi )^C{}_D\,L^D_\nu$,\quad
  $J_-^\mu{}_\nu=R^\mu_C\,\mathbb{J}_1(\theta ,\phi )^C{}_D\,R^D_\nu\,$.
   \item $J_+^\mu{}_\nu=L^\mu_C\,\mathbb{J}_1(\theta ,\phi )^C{}_D\,L^D_\nu$,\quad
  $J_-^\mu{}_\nu=R^\mu_C\,\mathbb{J}_2(\theta ,\phi )^C{}_D\,R^D_\nu\,$.
  \item $J_+^\mu{}_\nu=L^\mu_C\,\mathbb{J}_2(\theta ,\phi )^C{}_D\,L^D_\nu$,\quad
  $J_-^\mu{}_\nu=R^\mu_C\,\mathbb{J}_1(\theta ,\phi )^C{}_D\,R^D_\nu\,$.
  \item $J_+^\mu{}_\nu=L^\mu_C\,\mathbb{J}_2(\theta ,\phi )^C{}_D\,L^D_\nu$,\quad
  $J_-^\mu{}_\nu=R^\mu_C\,\mathbb{J}_2(\theta ,\phi )^C{}_D\,R^D_\nu\,$.
\end{enumerate}
In order to identify the different kinds of superfields present for any given choice, we need to analyse the kernel of the commutator $[J_+,J_-]$. We need to be careful though, since at certain points of the manifold type changing can occur. We will analyse these loci more carefully when we calculate the pure spinors corresponding with the associated generalised complex structures.

\begin{center}
  \begin{tabular}{ c | c | c | c}
   Choice & $\dim_{\mathbb{C}}\ker(J_+ - J_-)$ & $\dim_{\mathbb{C}}\ker(J_+ + J_-)$ & $\dim_{\mathbb{C}} \ker[J_+,J_-]^\perp$ \\ 
    \hline
    \hline
    1 & 1 & 1 & 0 \\ \hline
    2 & 1 & 1 & 0 \\ \hline
    3 & 0 & 0 & 2 \\ \hline
    4 & 0 & 0 & 2 
  \end{tabular}
\end{center}
We find that choices 1 and 2 correspond to one chiral superfield $z$ and one twisted-chiral superfield $w$, while choices 3 and 4 are described by a pair of semi-chiral superfields $(l,r)$.\\
\\
\underline{\em i. The first choice}\\
When $\theta=\phi=\theta '=\phi '=0$ the complex structures $J_+$ and $J_-$ commute so they can be diagonalised simultaneously. This is accomplished by the coordinates 
\begin{eqnarray}
 \tilde w=\cos\psi \,e^{-\rho -i\,\varphi _1},\qquad
 \tilde z=\sin\psi \,e^{-\rho +i\,\varphi _2}.\label{aux1}
\end{eqnarray}
When considering the more general expressions for the complex structures this is no longer possible and we need to consider them separately. By making the coordinate transformation,
\begin{eqnarray}
\begin{array}{ll}
x_1 = - e^{i \phi} \cos \frac{ \theta}{2}\, \tilde{ w} +
i\, \sin \frac{\theta}{2}\, \bar{\tilde{z}}, & \bar x_1 =
- e^{-i \phi} \cos \frac{ \theta}{2}\, \bar {\tilde{ w}}
- i\, \sin \frac{\theta}{2}\, \tilde{z}, \\
& \\
x_2 =  e^{i \phi} \cos \frac{ \theta}{2}\, \tilde{ z} +
i\, \sin \frac{\theta}{2}\, \bar{\tilde{w}}, & \bar x_2 =
e^{-i \phi} \cos \frac{ \theta}{2}\, \bar {\tilde{ z} }
- i\, \sin \frac{\theta}{2}\, \tilde{w}\, ,
\end{array}
\end{eqnarray}
we can diagonalise $J_+$ for generic values of $\theta $ and $\phi $. If we want to diagonalise the general form of the second complex structure $J_-$ we have to perform a holomorphic coordinate transformation (with respect to $J_+$)\begin{eqnarray}
\begin{array}{ll}
w = - \cos \frac{\theta'}{2} x_1 - i \sin \frac{\theta'}{2}\,
e^{i \phi'}  x_2, & \bar w = -\cos \frac{\theta'}{2} \bar{ x_1} + i \sin \frac{\theta'}{2}\, e^{-i \phi'} \bar{x_2}\, ,\\&\\
z =  i \sin \frac{ \theta'}{2}\,  e^{-i \phi'} x_1+ \cos
\frac{\theta'}{2} x_2,
& \bar z =  - i \sin \frac{ \theta'}{2}\,  e^{i \phi'} \bar{x_1} + \cos \frac{\theta'}{2} \bar {x_2}.
\end{array}
\label{diagco}
\end{eqnarray}
Having put the two families of complex structures in the required form we then find the generalized K{\"a}hler potential
\begin{eqnarray} V(z,
w, \bar z, \bar w) = \frac{k}{4\pi }\,\Big(\int^{\frac{z \bar z}{ w \bar w}} \frac{dq}{q}
\ln\left(1+q\right) - \frac{1}{2} \left( \ln w \bar w \right)^2 \Big)
,\label{su2u1ctc4moduli}
\end{eqnarray}
which correctly reproduces the metric
\begin{eqnarray}
 ds^2=\frac{k}{2\pi }\,\frac{1}{z\bar z+w\bar w}\,\big(dz\,d\bar z+dw\,d\bar w\big),
\label{gchitwi}
\end{eqnarray}
and torsion 3-form,
\begin{eqnarray}
H&=&\frac{k}{4\pi }\,\Big(\frac{\bar w}{(z\bar z+w\bar w)^2}\,dz\wedge d\bar z\wedge dw-
\frac{w}{(z\bar z+w\bar w)^2}\,dz\wedge d\bar z\wedge d\bar w+\nonumber\\
&&\frac{\bar z}{(z\bar z+w\bar w)^2}\,dz\wedge dw\wedge d\bar w-
\frac{ z}{(z\bar z+w\bar w)^2}\,d\bar z\wedge dw\wedge d\bar w\Big)\,,\label{Hchtwch}
\end{eqnarray}
in these coordinates.\\
\\
\underline{\em ii. The second choice}\\
The second choice of complex structures also yields a description in terms of one chiral and one twisted-chiral super field. The relevant coordinates are related to those of the first choice by
\begin{eqnarray}
 \tilde w=\cos\psi \,e^{+\rho -i\,\varphi _1},\qquad
 \tilde z=\sin\psi \,e^{+\rho +i\,\varphi _2},\label{aux2}
\end{eqnarray}
instead of (\ref{aux1}), so the only difference is that we replace $\rho$ by $-\rho $. It can be checked that the resulting
generalized K{\"a}hler potential is still given by (\ref{su2u1ctc4moduli}), and that as a consequence the resulting geometrical data is the same.\\
\\
\underline{\em iii. The third choice}\\
The third choice corresponds to a parametrisation of the target space by two semi-chiral superfields. The starting point is now
\begin{eqnarray}
 l=w,\quad \bar l=\bar w,\quad r=\frac{\bar w}{z},\quad \bar r=\frac{w}{\bar z},\label{semigroup}
\end{eqnarray}
where $z$ and $w$ are the expressions given in eq.~(\ref{diagco}). The
generalised K{\"a}hler potential is now
\begin{eqnarray}
 V(l,\bar l,r,\bar r)=\frac{k}{4\pi }\Big(\ln \frac{l}{\bar r}\,\ln \frac{\bar l}{r}\,-\int^{r\bar r}\frac{dq}{q}\,
 \ln\big(1+q\big)\Big).\label{Vsemi}
\end{eqnarray}
Using this we calculate the metric by considering the $\mathcal{N}=(1,1)$ reduction of the semi-chiral superfields, i.e. $l\equiv l|_{(1,1)}$ and $r\equiv r|_{(1,1)}$,
\begin{eqnarray}
ds^2=\frac{k}{2\pi }\,\Big(\frac{1}{l\bar l}\,dl\,d\bar l+\frac{1}{r\bar r}\,\frac{1}{1+r\bar r}\,drd\bar r-
\frac{1}{lr}\,\frac{1}{1+r\bar r}\,dl\,dr-\frac{1}{\bar l\bar r}\,\frac{1}{1+r\bar r}\,
d\bar l\,d\bar r\Big),
\end{eqnarray}
and the torsion 3-form
\begin{eqnarray}
H=\frac{k}{4\pi }\Big(\frac{1}{l}\,\frac{1}{(1+r\bar r)^2}\,dl\wedge dr\wedge d\bar r-
\frac{1}{\bar l}\,\frac{1}{(1+r\bar r)^2}\,  d\bar l\wedge dr\wedge d\bar r \Big).
\end{eqnarray}
The complex structures are
\begin{eqnarray}
 J_+&=&\left(\begin{array}{cccc}
   +i & 0 & 0 & 0 \\
   0 &-i & 0 & 0 \\
   0 & -2i\,\frac{r}{\bar l} & +i&0\\
   +2i\,\frac{\bar r}{l}&0&0&-i
 \end{array}   \right),\nonumber\\
J_-&=&\left(\begin{array}{cccc}
   i&0&0&-2i\,\frac{l}{\bar r}\,\frac{1}{1+r\bar r} \\
   0& -i& +2i\,\frac{\bar l}{ r}\,\frac{1}{1+r\bar r}&0\\
   0 & 0 & +i & 0 \\
   0 & 0 & 0 & -i
 \end{array}   \right),\label{thecomstr}
\end{eqnarray}
where we labelled the rows and columns in the order $l\bar l r\bar r$.\\
\underline{\em iv. The fourth choice}\\
The second parametrisation in terms of semi-chiral coordinates can be obtained by a similar manipulation as
(\ref{semigroup}) but now with $z$ and $w$ as in the second choice.
The generalised K{\"a}hler potential and all other expressions are then obviously the same
as for the third choice.

\subsection{Duality relations}
The $\sigma$-model description of $SU(2) \times U(1)$ in terms of a chiral/twisted-chiral multiplet is T-dual to the one in terms of a semi-chiral multiplet, or more specifically, one can T-dualise along the $U(1)$ direction provided that sufficient isometries are present \cite{grisarusevrin}. The underlying gauge structure was uncovered in \cite{lindstromryb1},\cite{gatesmerrell}, \cite{lindstromryb2}, and \cite{merrell}. We shall argue here that the lagrangians derived above show that this is indeed the case. Our starting point is the semi-chiral potential (\ref{Vsemi}). One can consider an isometry of the form
\begin{eqnarray}
\delta l=\epsilon \,l,\qquad \delta\bar  l=\epsilon \,\bar l,\qquad
\delta r=0,\qquad \delta \bar r=0,\label{isob1}
\end{eqnarray}
with $\epsilon \in \mathbb{R}$ an arbitrary constant. One easily checks that the potential (\ref{Vsemi}) is indeed invariant under this isometry modulo a generalised K\"{a}hler transformation, so both descriptions should be T-dual to each other. \\
\\
To establish the T-duality more explicitly, we introduce an 
auxiliary semi-chiral multiplet $\{ \hat l$, $\bar{ \hat l}$, $\hat r$, $\bar{ \hat r} \}$ which under the isometry transforms as
\begin{eqnarray}
 \delta \hat l=\epsilon \,\ln l,\qquad \delta\bar{ \hat l}=\epsilon \,\ln\bar  l,
 \qquad\delta \hat r = \epsilon \,\ln r,\qquad \delta\bar{ \hat r} = \epsilon \,\ln \bar r.\label{isob2}
\end{eqnarray}
We can add these auxiliary degrees of freedom to the original lagrangian without modifying the theory since they are total derivative terms from a superfield point of view. This gives us the new potential
\begin{eqnarray}
 V_0=\frac{k}{4\pi }\,\Big(\ln \frac{l}{\bar r}\,\ln \frac{\bar l}{r}\,-\int^{r\bar r}\frac{dq}{q}\,
 \ln\big(1+q\big)-\hat l-\bar{ \hat l}+\hat r+\bar{ \hat r}\Big),
\end{eqnarray}
which is manifestly invariant under the isometry. We now gauge the isometry by introducing a semi-chiral multiplet of gauge parameters $\{ \epsilon_l , \bar \epsilon_l , \epsilon_r, \bar \epsilon_r \}$
\begin{eqnarray}
\bar{\mathbb{D}}_+ \epsilon_l= \mathbb{D}_+ \bar \epsilon_{\bar{l}} = \bar{\mathbb{D}}_- \epsilon_r= \mathbb{D}_- \bar \epsilon_{\bar{r}}=0.
\end{eqnarray}
We next add terms involving an unconstrained real gauge field $X$ and an unconstrained complex gauge field $Y$ ( with $\bar Y= Y^\dagger$) to the potential as follows
\begin{eqnarray}
 V_1=V_0+\frac{k}{4\pi }\,\Big(i\,X\,\ln \frac{l}{\bar l}+\frac 1 2 \, X^2+Y\,\ln r+\bar Y\ln \bar r\Big),
\end{eqnarray}
which exhibits the gauge invariance
\begin{eqnarray}
 &&\delta l=\epsilon _l\,l\,,\qquad \delta \bar l=\bar \epsilon _{\bar l}\,\bar l\,,
 \qquad \delta r=0\,,\qquad \delta \bar r=0\,,\nonumber\\
 &&\delta \hat l=\epsilon _l\,\ln l\,,\qquad \delta \bar{\hat l}=\bar \epsilon _{\bar l}\,\ln \bar l\,,
 \qquad \delta r=\epsilon _r\,\ln  r\,,\qquad \delta \bar{\hat r}=\bar\epsilon _{\bar r}\,\ln \bar r\,,\nonumber\\
 &&\delta X=-i\big(\epsilon _l-\bar \epsilon _{\bar l}\big)\,,
 \qquad \delta Y=\epsilon _l-\epsilon _r\,, \qquad \delta \bar Y=\bar \epsilon _{\bar l}-\bar \epsilon _{\bar r }\,.
\end{eqnarray}
We construct gauge invariant $\mathcal{N}=(2,2)$ field strengths for new gauge fields $X,Y,\bar Y$ via
\begin{eqnarray}
&&  F=i\mathbb{D}_+\bar{\mathbb{D}}_-(X+iY),\qquad \bar F=i \bar{\mathbb{D}}_+\mathbb{D}_-(X-i \bar Y) \nonumber \\
&& G=i\mathbb{D}_+\mathbb{D}_-\bar Y,\qquad \bar G=i \bar{\mathbb{D}}_+\bar{\mathbb{D}}_-Y.
\end{eqnarray}
Using these field strengths we add the lagrange multipliers $u, \bar u, v$ and $\bar v$ (which are unconstrained complex superfields) to yield the first order potential which is equivalent to the original ungauged model
\begin{eqnarray}
 V_2=V_1+ \frac{k}{4\pi }\,\Big(u\,\bar F+\bar u\,F+ v\, \bar G+\bar v\,G\Big).\label{VV2}
\end{eqnarray}
By integrating (\ref{VV2}) by parts and introducing chiral superfield $z$ and twisted-chiral superfield $w$
\begin{eqnarray}
 w=e^{-\bar{\mathbb{D}}_+\mathbb{D}_-u}& \bar w=e^{+ \mathbb{D}_+\bar{\mathbb{D}}_-\bar u} \nonumber \\
 z=e^{i\,\bar{\mathbb{D}}_+\bar{\mathbb{D}}_-v}&  \bar z=e^{i\, \mathbb{D}_+\mathbb{D}_-\bar v}\,.
\end{eqnarray}
we get a new potential
\begin{eqnarray}
 V_2=V_1-\frac{k}{4\pi }\,\Big(i\,X\,\ln\frac{w}{\bar w}\,+Y\,\ln\frac{\bar w}{z}\,+
 \bar Y\,\ln\frac{w }{\bar z}\,\Big). \label{VV3}
\end{eqnarray}
Integrating over the gauge fields $X, Y, \bar Y$ results in the equations of motion
\begin{eqnarray}
 X=i\,\ln\,\frac{w\,\bar l}{\bar w\,l},\qquad
 r=\frac{\bar w}{z},\qquad \bar r=\frac{w}{\bar z},
\end{eqnarray}
which can be plugged into (\ref{VV3}) to yield the second order potential
\begin{eqnarray}
 V_3=\frac{k}{4\pi }\,\Big(\int^{\frac{z \bar z}{ w \bar w}} \frac{dq}{q}
\ln\left(1+q\right) - \frac{1}{2} \left( \ln w \bar w \right)^2+\cdots\,\Big),
\end{eqnarray}
where the omitted terms correspond to terms that are negligible up to a generalised K\"{a}hler transformation. In this manner we have indeed reproduced the T-dual potential obtained before (\ref{su2u1ctc4moduli}).\\
\\
We conclude this section by considering the isometries $g\rightarrow h_-gh_+$ with $h_\pm$ elements of the maximal torus parametrised as 
\begin{eqnarray}
h_-=e^{i\epsilon_-\sigma_0+i\eta_-\sigma_3},\qquad h_+=e^{i\epsilon_+\sigma_0+i\eta_+\sigma_3},
\label{S3S1aff}
\end{eqnarray}
Under this transformation the chiral and twisted-chiral coordinates change as 
\begin{eqnarray}
w\rightarrow e^{-\phi_--\phi_+}\,w,\qquad z\rightarrow e^{-\bar\phi_--\phi_+}\,z.
\end{eqnarray}
where $\phi_\pm=\epsilon_\pm+i\eta_\pm$. In order to keep these transformation consistent with the superconstraints we require that 
\begin{eqnarray}
&&\mathbb{D}_+\phi_-=\bar{\mathbb{D}}_+\phi_-=\mathbb{D}_-\phi_-=0\nonumber\\
&&\mathbb{D}_-\phi_+=\bar{\mathbb{D}}_-\phi_+=\bar{\mathbb{D}}_+\phi_+=0,
\end{eqnarray}
which implies that $\partial_\pp\, \phi_-=\partial_=\,\phi_+=0$. This clearly shows that these transformations correspond to (abelian) affine symmetries. The potential (\ref{su2u1ctc4moduli}) is indeed invariant under these affine symmetries modulo superspace total derivative terms.\\
\\
Looking at the semi-chiral coordinates $l$ and $r$ we find that they transform as 
\begin{eqnarray}
l\rightarrow e^{-\phi_--\phi_+}\,l,\qquad r\rightarrow e^{\phi_+-\bar\phi_+}\,r,
\end{eqnarray} 
which requires 
\begin{eqnarray}
\bar{\mathbb{D}}_+\phi_-=\bar{\mathbb{D}}_+\phi_+=0,\qquad 
\bar{\mathbb{D}}_-\phi_+=\bar{\mathbb{D}}_-\bar\phi_+,\qquad
\mathbb{D}_-\phi_+=\mathbb{D}_-\bar\phi_+,
\end{eqnarray} 
in order to be consistent with the constraints. This is not sufficient to establish invariance (again moduli derivative terms) of the potential. In order to obtain said invariance we will additionally require
\begin{eqnarray}
&&\mathbb{D}_+\phi_-=\bar{\mathbb{D}}_+\phi_-=\bar{\mathbb{D}}_-\phi_-=0,\nonumber\\
&&\mathbb{D}_-\phi_+=\bar{\mathbb{D}}_-\phi_+=\bar{\mathbb{D}}_+\phi_+=0,
\end{eqnarray}
which once more implies  $\partial_\pp\, \phi_-=\partial_=\,\phi_+=0$.

\subsection{Generalised K\"{a}hler geometry and type changing}
We will now explicitly calculate the pure spinors corresponding the generalised complex structures. In what follows we will restrict ourselves to a specific point ($\theta=\theta^\prime=\phi=\phi^\prime=0$) of the $S^3 \times S^1$ moduli space.

\subsubsection{Chiral/twisted-chiral parametrisation}
So far we have disregarded the fact that fully covering the target manifold requires some care in choosing local expressions in different coordinate patches. When calculating the pure spinors however this will become important, especially when we will encounter type-changing later on. As can be seen from explicitly considering the complex coordinates defined on the manifold $SU(2) \times U(1)$
\begin{eqnarray}
w&=&\cos\psi\,e^{-\rho-i \varphi_1} \nonumber\\
z&=&\sin\psi\,e^{-\rho+i \varphi_2}\,,
\end{eqnarray}
the generalised K\"{a}hler potential
\begin{eqnarray} V_{w\neq 0}(
w, \bar w,z, \bar z) = \frac{k}{4\pi }\,\Big(\int^{\frac{z \bar z}{ w \bar w}} \frac{dq}{q}
\ln\left(1+q\right) - \frac{1}{2} \left( \ln w \bar w \right)^2 \Big)\,,
\label{su2u1ctc4modulibb}
\end{eqnarray}
is only valid on a coordinate patch that does include the points where $w=0$, or in terms of the real coordinates, $\psi=\pi/2$. Performing the so-called ``mirror" transform
	\begin{equation}
	V(l, \bar l, r, \bar r, w, \bar w, z, \bar z ) \rightarrow - V(l, \bar l,\bar r, r, z, \bar z ,w, \bar w ),
	\end{equation}
which amounts to replacing the geometric data $\{g,H,J_+,J_- \}$ with $\{g,H,J_+,-J_- \}$, as is explained in  \cite{hull2009} and \cite{sevrinwijns}, will transform the potential to the form
\begin{eqnarray} V_{z\neq 0}(w,
\bar w, z,\bar z) = \frac{k}{4\pi }\,\Big(-\int^{\frac{w \bar w}{ z \bar z}} \frac{dq}{q}
\ln\left(1+q\right) + \frac{1}{2} \left( \ln z \bar z \right)^2 \Big).\label{WZWpot2b}
\end{eqnarray}
This potential is well defined in a any patch not containing points with $z=0$, or $\psi=0$ in real coordinates. Obviously these descriptions should agree on any overlapping patches. This indeed the case as their difference can be calculated to yield a generalised K\"{a}hler transformation
\begin{eqnarray}
V_{w\neq 0}-V_{z\neq 0}=-\frac{k}{4\pi }\,\ln\big( z\bar z\big)\ln\big( w\bar w\big)\,.
\end{eqnarray}
We will start by employing the techniques described in chapter 5 to calculate the pure spinors in the coordinate patch where $w \neq 0$. Starting from the potential in (\ref{su2u1ctc4modulibb}) we obtain the two pure spinors $\phi_+$ 
and $\phi_-$
\begin{eqnarray}
\phi_+ = d\bar z \wedge e^{i\, \Omega^+ + \Xi^+}=
d\bar z \wedge e^{-b\wedge}\, e^{\Lambda^+ }, \nonumber\\
\phi_- = d\bar w \wedge e^{i\, \Omega^- + \Xi^-}=
d\bar w \wedge e^{-b\wedge}\, e^{\Lambda^-}.\label{psctc}
\end{eqnarray}
An explicit calculation of the relevant 2-forms yields
\begin{eqnarray}
i\, \Omega^+ + \Xi^+ =  \frac{k}{ 8 \pi( w \bar w + z \bar z)} \left( - 2 dw \wedge d \bar w - 
\frac{\bar z}{ w}  dw \wedge dz + \frac{z}{w} dw \wedge d \bar z \right. \nonumber \\
 \left. + \frac{\bar z}{\bar w} d \bar w \wedge dz +3 \frac{z}{\bar w} d \bar w \wedge d \bar z - 2 dz 
\wedge  d \bar z  \right),
\end{eqnarray}
\begin{eqnarray}
i\, \Omega^- + \Xi^- = \frac{k}{ 8 \pi( w \bar w + z \bar z)} \left(  2 dw \wedge d \bar w - 
\frac{\bar z}{ w}  dw \wedge dz + \frac{z}{w} dw \wedge d \bar z \right. \nonumber \\
 \left. -3 \frac{\bar z}{\bar w} d \bar w \wedge dz - \frac{z}{\bar w} d \bar w \wedge d \bar z - 2 dz 
\wedge  d \bar z  \right).
\end{eqnarray}
We can also undo the $b$-transform and obtain in this way the $H$-closed pure spinors $\phi_+ = d\bar z \wedge  e^{\Lambda_+}$, $\phi_- = d\bar w \wedge  e^{\Lambda_-}$, with
\begin{eqnarray}
\Lambda_+ &=& -\frac{k}{4\pi} \frac{1}{w \bar w + z \bar z} \left( dw \wedge d\bar w + d z \wedge d\bar z  - 2 \frac{z}{\bar w}d \bar w \wedge d\bar z  \right), \label{Lampsimctcp1}\\
\Lambda_- &=& \frac{k}{4\pi} \frac{1}{w \bar w + z \bar z} \left( dw \wedge d\bar w - d z \wedge d\bar z  - 2 \frac{\bar z}{\bar w}d \bar w \wedge d z  \right), \label{Lammsimctcp1}
\end{eqnarray}
and
\begin{eqnarray}
b = \frac{k}{8 \pi} \frac{1}{z \bar z + w \bar w}\left(  \frac{\bar z}{w} dw \wedge dz - \frac{z}{w} dw \wedge d \bar z - \frac{ z}{\bar w} d \bar w \wedge d \bar z + \frac{z}{\bar w} d \bar w \wedge d \bar z  \right).
\end{eqnarray}
We see that both pure spinors are indeed well defined as long we remain in a coordinate patch where $w \neq 0$. As expected for a parametrisation corresponding to one chiral and one twisted-chiral superfield the generalised K\"{a}hler structure is of type $(1,1)$. As long as we restrict ourselves to this coordinate patch, the type of the generalised K\"{a}hler structure does not change.\\
\\
We now check if the generalised Calabi-Yau conditions are satisfied by calculating the Mukai pairings. This  yields
\begin{eqnarray}
\big( \phi_+, \bar\phi_+\big)=\big( \phi_-, \bar \phi_-\big)=- \frac{k}{2 \pi}\, \frac{1}{z \bar{z}
+w \bar{w}}\,=- \frac{k}{2 \pi}\, e^{2 \rho}\,,
\end{eqnarray}
which is nowhere vanishing and shows that the generalised K\"{a}hler structure satisfies the generalised Calabi-Yau condition.\\
\\
In the second coordinate patch where $z \neq 0$ we can perform the same calculations, this time starting from the potential (\ref{WZWpot2b}). The pure spinors are of the same form as (\ref{psctc}), but with the 2-forms given by
\begin{eqnarray}
i\, \Omega^+ + \Xi^+ =  \frac{k}{ 8 \pi( w \bar w + z \bar z)} \left( - 2 dw \wedge d \bar w + \frac{\bar 
w}{ z}  dw \wedge dz - \frac{\bar w}{\bar z} dw \wedge d \bar z \right. \nonumber \\
 \left. - \frac{w}{z} d \bar w \wedge dz -3 \frac{w}{\bar z} d \bar w \wedge d \bar z - 2 dz \wedge  d 
\bar z  \right), \nonumber\\
i\, \Omega^- + \Xi^- =  \frac{k}{ 8 \pi( w \bar w + z \bar z)} \left(  2 dw \wedge d \bar w + \frac{\bar 
w}{ z}  dw \wedge dz - \frac{\bar w}{\bar z} dw \wedge d \bar z \right. \nonumber \\
 \left. + 3 \frac{w}{z} d \bar w \wedge dz + \frac{w}{\bar z} d \bar w \wedge d \bar z - 2 dz \wedge  d 
\bar z  \right).
\end{eqnarray} 
Similarly, we can undo the $b$-transform to obtain the $H$-closed pure spinors $\phi_+ = d\bar z \wedge  e^{\Lambda_+}$, $\phi_- = d\bar w \wedge  e^{\Lambda_-}$, with
\begin{eqnarray}
\Lambda_+ &=& -\frac{k}{4\pi} \frac{1}{w \bar w + z \bar z} \left( dw \wedge d\bar w + d z \wedge d\bar z  + 2 \frac{w}{\bar z}d \bar w \wedge d\bar z  \right), \label{Lampsimctcp2}\\
\Lambda_- &=& \frac{k}{4\pi} \frac{1}{w \bar w + z \bar z} \left( dw \wedge d\bar w - d z \wedge d\bar z  + 2 \frac{w}{z}d \bar w \wedge d z  \right).\label{Lammsimctcp2}
\end{eqnarray} 
This confirms that the above expressions are indeed well defined for $z \neq 0$. When comparing the expressions of the two pure spinors we now see that, while they both obey the generalised Calabi-Yau equations and are as such superconformally invariant, they do not provide a consistent supergravity background as they are not globally defined.

\subsubsection{Semi-chiral parametrisation}
We now turn to the semi-chiral parametrisation. As described in chapter 4, when there are no chiral or twisted chiral superfields present we can introduce the symplectic 2-form
	\begin{eqnarray}
	\Omega &=& 2 g ([J_+,J_-])^{-1} \\
	& =& \left(
	\begin{matrix}
	0 & V_{lr} \\
	- V_{rl} & 0
	\end{matrix} \right).
	\end{eqnarray}
When only semi-chiral superfields are present the commutator $[J_+,J_-]$ is expected to be everywhere invertible, implying that $\Omega$ is a globally defined symplectic 2-form. This would imply that the torsion 3-form is exact, which is not the case on a group manifold, as remarked upon in \cite{bogaertssevrin} in the context of hyper-K\"{a}hler manifolds. This apparent contradiction is now understood by realising that not every point of the target manifold will be a regular point in this description. As we will now show there will be loci where the type changes and the description in terms of solely semi-chiral fields is no longer valid.\\
\\
We will start by making the following coordinate transformation of the original coordinates
\begin{eqnarray}
l=\ln\left(\cos \psi\,e^{-\rho-i\,\varphi_1}\right),\qquad r=\ln\left(\cot \psi\,e^{i \,\varphi_1-i\, \varphi_2}\right).
\label{cos1}
\end{eqnarray}  
When expressed in these new coordinates the semi-chiral model is determined by the potential
\begin{eqnarray}
 V_{\psi\neq 0}(l,\bar l,r,\bar r)=\frac{k}{4\pi }\Big(\big(l-\bar r\big)\big(\bar l -r\big)-\int^{r+\bar r}dq\,
 \ln\big(1+e^q\big)\Big),\label{Vsemilog}
\end{eqnarray}
as long as we restrict ourselves to a coordinate patch where $\psi \neq 0$. The complex structures are given by 
\begin{eqnarray}
 J_+=\left(\begin{array}{cccc}
   +i & 0 & 0 & 0 \\
   0 &-i & 0 & 0 \\
   0 & -2i & +i&0\\
   +2i&0&0&-i
 \end{array}   \right),\qquad
J_-=\left(\begin{array}{cccc}
   i&0&0&-\frac{2i}{1+e^{r+\bar r}} \\
   0& -i& \frac{2i}{1+e^{r+\bar r}}&0\\
   0 & 0 & +i & 0 \\
   0 & 0 & 0 & -i
 \end{array}   \right).\label{thecomstrlog}
\end{eqnarray}
Starting from the generalised potential we can calculate the geometrical data, such as the line element
\begin{eqnarray}
ds^2= \frac{k}{2 \pi}\left(dl\,d \bar{l}+ \frac{1}{1+e^{r+ \bar{r}}}\left(dr\,d \bar{r}
-dl\,dr-d \bar{l}\,d \bar{r}
\right)\right),
\end{eqnarray}
and the torsion 3-form
\begin{eqnarray}
H=\frac{k}{4\pi }\Big(\frac{e^{r+\bar r}}{(1+e^{r+\bar r})^2}\,dl\wedge dr\wedge d\bar r-
\frac{e^{r+\bar r}}{(1+e^{r+\bar r})^2}\,  d\bar l\wedge dr\wedge d\bar r \Big).
\end{eqnarray}
Since the symplectic 2-form defined by the inverse of the commutator of the complex structures should not be globally exact there must be loci where the type of the generalised complex structure changes. To make this explicit we can calculate the loci where we will have fields parametrising $\ker [J_+,J_-]$
\begin{eqnarray}
\det \big(J_++J_-\big)&=& \frac{16\,e^{2( r+ \bar{r})}}{(1+e^{r +\bar{r}})^2}=
16\cos^4\psi \nonumber\\
\det \big(J_+-J_-\big)&=&\frac{16}{(1+e^{r +\bar{r}})^2}=
16\sin^4\psi,
\end{eqnarray}
so we anticipate type changing to occur at $\psi=\pi/2$ where the locus will be parameterized by a twisted chiral field. Note that since we are working in a coordinate patch where $\psi \neq 0$ the inverse of $\big(J_+-J_-\big)$ is well defined, signalling that no chiral superfields will be required.\\
\\
As we did with the chiral/twisted-chiral parametrisation we can consider the ``mirror" potential of (\ref{Vsemilog}) by introducing the coordinates
\begin{eqnarray}
l=\ln\left(\sin \psi\,e^{-\rho+i\,\varphi_2}\right),\qquad r=\ln\left(\tan \psi\,e^{-i \,\varphi_1+i\, 
\varphi_2}\right).
\label{cos2}
\end{eqnarray}  
By performing the Legendre transformation of (\ref{Vsemilog}) with respect to $l$ and $\bar l$ we obtain the potential
\begin{eqnarray}
 V_{\psi\neq \frac \pi 2}(l,\bar l,r,\bar r)=\frac{k}{4\pi }\Big(-\big(l- r\big)\big(\bar l -\bar r\big)+\int^{r+\bar r}dq\,
 \ln\big(1+e^q\big)\Big),\label{Vsemilog2}
\end{eqnarray}
which is well defined for $\psi \neq \pi/2$. On overlapping patches both potentials are related via a generalised K\"{a}hler transformation in addition to a Legendre transformation, consistent with the fact that when semi-chiral superfields are present we have a larger freedom for reformulating the potential.\\
\\
In terms of the new coordinates and starting from the potential  (\ref{Vsemilog2}) we calculate the complex structures on the manifold  
\begin{eqnarray}
 J_+=\left(\begin{array}{cccc}
   +i & 0 & 0 & 0 \\
   0 &-i & 0 & 0 \\
   2i &0 & -i&0\\
   0&-2i&0&i
 \end{array}   \right),\qquad
J_-=\left(\begin{array}{cccc}
  - i&0&\frac{2i}{1+e^{r+\bar r}}&0 \\
   0& i&0&- \frac{2i}{1+e^{r+\bar r}}\\
   0 & 0 & +i & 0 \\
   0 & 0 & 0 & -i
 \end{array}   \right),\label{thecomstrlog2}
\end{eqnarray}
as well as the line element and the torsion 3-form
\begin{eqnarray}
ds^2&=& \frac{k}{2 \pi}\left(dl\,d \bar{l}+ \frac{1}{1+e^{r+ \bar{r}}}\left(dr\,d \bar{r}
-dl\,d\bar{r}-d \bar{l}\,d r
\right)\right),\nonumber\\
H&=&-\frac{k}{4\pi }\Big(\frac{e^{r+\bar r}}{(1+e^{r+\bar r})^2}\,dl\wedge dr\wedge d\bar r-
\frac{e^{r+\bar r}}{(1+e^{r+\bar r})^2}\,  d\bar l\wedge dr\wedge d\bar r \Big).
\end{eqnarray}
Using the complex structures one finds
\begin{eqnarray}
\det \big(J_++J_-\big)&=& \frac{16}{(1+e^{r +\bar{r}})^2}=
16\cos^4\psi, \nonumber\\
\det \big(J_+-J_-\big)&=&\frac{16\,e^{2( r+ \bar{r})}}{(1+e^{r +\bar{r}})^2}=
16\sin^4\psi.
\end{eqnarray}
We see that points where $\psi =0$ are not regular points and that this defines a locus where the type changes and we will have a description in terms of a chiral superfield.\\
\\
We now turn out attention to the pure spinors. In the first coordinate patch where $\psi \neq 0$ we find that
\begin{eqnarray}
\phi_+ =e^{i\, \Omega^+ + \Xi^+}, \nonumber\\
\phi_- = e^{i\, \Omega^- + \Xi^-},\label{pspss3}
\end{eqnarray}
with the various 2-forms given by
\begin{eqnarray}
 i\, \Omega^+ + \Xi^+  = \frac{k}{8 \pi} \left( 2 dl\wedge d \bar l - dl \wedge dr + 3 d \bar l \wedge 
d\bar r - \frac{2}{1+ e^{r+ \bar r}} dr \wedge d\bar r  \right),
\end{eqnarray}
 and
 \begin{eqnarray}
i\,  \Omega^- + \Xi^-  = \frac{k}{8 \pi} \left( - 2 dl\wedge d \bar l -  dl \wedge dr - d \bar l \wedge 
d\bar r - \frac{2}{1+ e^{r+ \bar r}} dr \wedge d\bar r  \right).
 \end{eqnarray}
Having calculated the pure spinors we compute the Mukai pairings
\begin{eqnarray}
\big( \phi_+,\bar \phi_+\big)&=& \frac{k^2}{4 \pi^2}\, \frac{e^{r+\bar r}}{1+e^{r+\bar r}}\,=
\frac{k^2}{4 \pi^2}\,\cos^2\psi\,, \nonumber\\
\big( \phi_-,\bar \phi_-\big)&=& \frac{k^2}{4 \pi^2}\, \frac{1}{1+e^{r+\bar r}}\,=
\frac{k^2}{4 \pi^2}\,\sin^2\psi\,.\label{Muksc}
\end{eqnarray}
The generalised Calabi-Yau conditions are clearly not satisfied, although the weaker condition
	\begin{equation}
	\frac{\det(N_+)}{\det(N_-)}=\pm |f_+(l,w,z)|^2 |f_-(r, \bar w, z)|^2
	\end{equation}
is, ensuring that the solution is $\mathcal{N}=(2,2)$ superconformally invariant at one loop, even though it is not supergravity solution to the equations of motion. We can repeat this analysis in the second coordinate patch, yielding a similar result as an explicit calculation of the Mukai pairings shows that $(\phi_+,\bar{\phi}_+)=0$ when $\psi=\pi/2$.\\
\\
Since the vanishing of the relevant Mukai pairings occurs at the loci where we expect the type of the generalised complex structure to change it is interesting to investigate this further. One would expect this type changing to be made manifest on the level of the complex structures and the pure spinors when looked at in an appropriate coordinate system. In order to do this we introduce new coordinates
\begin{eqnarray}
w=e^l,\qquad z=e^{-r+\bar l}.
\end{eqnarray}
These coordinates put $J_+$ in its canonical form with respect to $dz$ and $dw$, while $J_-$ becomes
\begin{eqnarray}
\label{jminchange}
J_-&=&\frac{1}{w\bar w+ z\bar z} \left(\begin{array}{cccc}
   +i(w\bar w- z\bar z )& 0 & 0 & +2iwz \\
   0 &-i(w\bar w- z\bar z) & -2i\bar w\bar z & 0 \\
   0 & -2i w z& +i(w\bar w- z\bar z)&0\\
   +2i\bar w\bar z&0&0&-i(w\bar w- z\bar z)
 \end{array}   \right), \nonumber \\
&\phantom{az}&
\end{eqnarray}
with rows and columns labeled in the order $(w, \bar w, z, \bar z)$. The metric $g$ in these coordinates is the one obtained before in (\ref{gchitwi}) and the Kalb-Ramond form $b$, using a suitable gauge choice, is given by
\begin{eqnarray}
b= -\frac{k}{8\pi}\, \frac{w\bar w- z\bar z}{w\bar w+ z\bar z}
\left(\frac{dw\wedge dz}{wz}+\frac{d\bar w\wedge d\bar z}{\bar w\bar z}\right)\,,
\end{eqnarray} 
which results in the torsion 3-form $H=db$ obtained before (\ref{Hchtwch}). 
Performing the coordinate transformation on the pure spinors (\ref{pspss3}) we obtain
\begin{eqnarray}
\phi_\pm= \sqrt{w\bar w z \bar z}\,e^{-b\wedge }\,e^{\Lambda_\pm},\label{psps1}
\end{eqnarray}
where
\begin{eqnarray}
\Lambda_+&=&- \frac{k}{4\pi}\, \frac{1}{w\bar w+ z\bar z}
\left(dw\wedge d\bar w+dz\wedge d\bar z+2 \frac{w}{\bar z}\,d\bar w\wedge d\bar z\right)\,, 
\nonumber\\
\Lambda_-&=&- \frac{k}{4\pi}\, \frac{1}{w\bar w+ z\bar z}
\left(dw\wedge d\bar w+dz\wedge d\bar z-2 \frac{z}{\bar w}\,d\bar w\wedge d\bar z\right)\,,
\end{eqnarray}
One can check that the pure spinors obtained after undoing the $b$-transform are indeed $H$-closed
\begin{eqnarray}
de^{ \Lambda_\pm}=H\wedge e^{ \Lambda_\pm}.
\end{eqnarray}
These expressions are consistent with the fact that the Mukai pairing is invariant under a $b$-transform and a change of coordinates, provided that the term $\sqrt{w\bar w z \bar z}$ is included as it transforms as a density.\\
\\
When looking at the type changing locus $w=0$ we see that the complex structure (\ref{jminchange}) now becomes $J_-=-J_+$. We encountered this situation in our overview of generalised complex geometry in chapter 5, where it was shown that this corresponds to generalised complex structures $\mathcal{J}_+$ and $\mathcal{J}_-$ of the symplectic and complex type, respectively. This is reflected in the pure spinors as well. By rescaling them and performing a $b$-transform the can be put in their canonical form.
\begin{eqnarray}
\phi_+&=&e^{\Lambda_+}, \nonumber\\
\phi_-&=&\bar w\, e^{\Lambda_-}.
\end{eqnarray}
For the regular points where $w \neq 0$ these pure spinors are of symplectic type and correspond to a generalised K\"{a}hler geometry of type (0,0). If we let $w \rightarrow 0$ the spinor $\phi_+$ remains symplectic but $\phi_-$ becomes complex and the type changes to $(0,2)$. An important consequence of this is that while $\phi_+$ remains $H$-closed, this is not true anymore for $\phi_-$. When calculating the Mukai pairings of the spinors in their canonical form we find that they are now non-degenerate even if $w=0$
\begin{eqnarray}
(\phi_+,\bar \phi_+)&=& \frac{k^2}{4 \pi^2}\,\frac{1}{w\bar w+ z\bar z}\, \frac{1}{z\bar z} 
\nonumber\\
(\phi_-,\bar \phi_-)&=& \frac{k^2}{4 \pi^2}\,\frac{1}{w\bar w+ z\bar z}.
\end{eqnarray}
\\
We now turn to the second coordinate patch. One could guess that, since the potential describing the model in this patch was obtained by considering the mirror transform of the potential in the first patch, the analysis will yield similar results with the roles of the two pure spinors exchanged. We will show that this is indeed the case. The starting point is again the potential given by (\ref{Vsemilog2}) and the coordinates by (\ref{cos2}). The pure spinors are still of the form (\ref{pspss3}) with the 2-forms now given by
\begin{eqnarray}
 i\, \Omega^+ + \Xi^+  = \frac{k}{8 \pi} \left( -2 dl\wedge d \bar l - dl \wedge d\bar r - d \bar l \wedge 
d r + \frac{2}{1+ e^{r+ \bar r}} dr \wedge d\bar r  \right),
\end{eqnarray}
 and
 \begin{eqnarray}
i\,  \Omega^- + \Xi^-  = \frac{k}{8 \pi} \left(  2 dl\wedge d \bar l -  dl \wedge d\bar r + 3d \bar l \wedge 
d r + \frac{2}{1+ e^{r+ \bar r}} dr \wedge d\bar r  \right).
 \end{eqnarray}
Calculating the Mukai pairings results in
\begin{eqnarray}
\big( \phi_+,\bar \phi_+\big)&=&- \frac{k^2}{4 \pi^2}\, \frac{1}{1+e^{r+\bar r}}\,=-
\frac{k^2}{4 \pi^2}\,\cos^2\psi\, \nonumber\\
\big( \phi_-,\bar \phi_-\big)&=&- \frac{k^2}{4 \pi^2}\, \frac{e^{r+\bar r}}{1+e^{r+\bar r}}\,=-
\frac{k^2}{4 \pi^2}\,\sin^2\psi\,.\label{Muksc2}
\end{eqnarray}
As expected this Mukai pairing indicates one-loop superconformal invariance but becomes degenerate at the locus $\psi=0$. To make the type changing occurring there manifest we introduce the coordinates 

\begin{eqnarray}
w=e^l,\qquad z=e^{\bar l-\bar r}.
\end{eqnarray}
This choice of coordinates diagonalises $J_+$, while $J_-$ is the negative of (\ref{jminchange}) as can be checked by explicit calculation.\\
\\
By performing the necessary transformation we can once again put the pure spinors in their canonical form. We find that
\begin{eqnarray}
\phi_+&=&\bar w\, e^{\Lambda_-}, \nonumber\\
\phi_-&=&e^{\Lambda_+}.
\end{eqnarray}
We see that $\phi_+$ and $\phi_-$ are again of symplectic type if we restrict ourselves to regular points and that they correspond to switching $\phi_+ \leftrightarrow \phi_-$, which is once again a sign of mirror symmetry at work. At the locus $\psi =0$, or $w=0$, the type changes from (0,0) to (2,0) and $\phi_+$ becomes complex. This is also reflected on the level of the complex structures since $J_+ = J_-$ when we are at the type-changing locus. Recalculating the Mukai pairings shows that they are everywhere non-degenerate, however $\phi_+$ is no longer $H$-closed, nor are the generalised Calabi-Yau conditions satisfied.\\
\\
In conclusion, we have found explicit expressions for the different possible choices of the complex structures on the group manifold $SU(2) \times U(1)$. We found that there were essentially two distinct possibilities parametrising the target space, one consisting of a chiral and twisted-chiral superfield and one consisting of a semi-chiral superfield. Both descriptions are T-dual to each other, as is made manifest by considering the affine isometries whose transformation parameters take values in the maximal torus. As expected, no type changing occurred in the chiral/twisted-chiral parametrisation, which remained of type $(1,1)$. To fully describe the manifold we required two different generalised K\"{a}hler potentials, each defined on a certain coordinate patch. Both descriptions are related to each other via a generalised K\"{a}hler transformation.\\
\\
When considering the semi-chiral parametrisation we observed type-changing occurring where one of the pure spinors corresponding to the generalised complex structures changed type. We found two generalised K\"{a}hler potentials describing both coordinate patches, which crucially are not only related via a generalised K\"{a}hler transformation but via a Legendre transform as well. As should be expected for a parametrisation consisting of semi-chiral superfields, type changing occurred. We found that in both coordinate patches one of the pure spinors changed from type 0 to type 2. Furthermore, the role of the type-changing spinors was exchanged in the two coordinate patches. These results are summarised in table 7.1. We also found that in most cases the generalised Calabi-Yau conditions were not satisfied. This is to be expected from the discussion in \cite{hullampere} since we have disregarded contributions from the dilaton. Even though the one-loop contribution to the $\beta$-function vanishes, as should be expected since any $\mathcal{N}=(2,2)$ model formulated on a group manifold is superconformally invariant, in general these models will not constitute valid supergravity solutions when not coupled to a dilaton in the manner described in \cite{hullampere}.

\begin{table}
\begin{center}
\begin{tabular}{c@{\qquad}c}
\hline $S^3\times S^1$ & locus \\
\hline \hline \\
patch 1 (\ref{Vsemilog}) & $\psi = \pi/2$: $(0,0) \rightarrow (0,2)$
\\
&\\
\hline \hline \\
patch 2 (\ref{Vsemilog2}) & $\psi = 0 $: $(0,0) \rightarrow (2,0)$\\
& \\
\hline
\end{tabular}\begin{picture}(0,0) \put(-235,-30){\begin{tikzpicture}  \draw[<->] (0,0) arc (270:90:0.8cm);\end{tikzpicture}} \put(-280, -5){mirror} \put(-290,-15){symmetry}  \end{picture}
\caption{Summary of type changing for $SU(2)\times U(1)$. \label{capTab1}}
\end{center}
\end{table}

\newpage
\section{$SU(2) \times SU(2)$}
Our next example will be the $\mathcal{N}=(2,2)$ $\sigma$-model with the six-dimensional manifold $SU(2) \times SU(2)$ as its target space. It was previously briefly considered in \cite{sevrintroost}. We will perform the same analysis as the $SU(2) \times U(1)$ case : starting from the possible complex structures on the Lie algebra we will calculate the complex structures on the group manifold in real coordinates and determine the superfield content. We will then derive the generalised K\"{a}hler potentials in the different cases and use them to investigate the generalised K\"{a}hler geometry by calculating the pure spinors and analysing where type changing occurs.

\subsection{Parametrisation and topology}
We begin by choosing a parametrisation of the group element in terms of real coordinates on the manifold
\begin{eqnarray}
g= \left(
\begin{array}{cccc} \cos \psi_1 \, e^{i\varphi _{11}} & \sin \psi_1 \,e^{i\varphi _{12}}&0&0\\
-\sin\psi_1\, e^{-i\varphi _{12}} & \cos\psi_1\, e^{-i\varphi _{11}}&0&0\\
 0&0&\cos \psi_2 \, e^{i\varphi _{21}} & \sin \psi_2 \,e^{i\varphi _{22}}\\
0&0&-\sin\psi_2\, e^{-i\varphi _{22}} & \cos\psi_2\, e^{-i\varphi _{21}}
\end{array}\right),
\label{su2su2coor}
\end{eqnarray}
where we have denoted the phases by $\varphi _{11},\, \varphi_{12},\, \varphi _{21},\, \varphi_{22} \in \mathbb{R}\,\mbox{mod}\,2 \pi $ and the angles $\psi_1,\psi_2\in[0, \pi /2]$. The topology of the group is locally that of a product of 3-spheres $S^3 \times S^3$, with the first 3-sphere parametrised by $\varphi_{11}$, $\varphi_{12}$ and $\psi_1$, and the second three-sphere by $\varphi_{21}$, $\varphi_{22}$ and $\psi_2$. \\
\\
For this choice of parametrisation we get the line element
\begin{eqnarray}
ds^2 = &\frac{k}{2 \pi} \left(  d \psi_1^2 + \cos^2 \psi_1 \, d\varphi_{11}^2 + \sin^2 \psi_1 \, d \varphi_{12}^2 \right. \nonumber \\
&+\left. d \psi_2^2 + \cos^2 \psi_2 \, d\varphi_{21}^2 + \sin^2 \psi_2 \, d \varphi_{22}^2 \right), \label{S3S3met}
\end{eqnarray}
and the torsion 3-form
\begin{eqnarray}
H = \frac{k}{2 \pi} \left( \sin 2 \psi_1 \, d\varphi_{11} \wedge d\varphi_{12} \wedge d \psi_1 + \sin 2 \psi_2 \, d\varphi_{21} \wedge d\varphi_{22} \wedge d \psi_2  \right). \label{S3S3tor}
\end{eqnarray}
We see that at the endpoints $\psi_{1,2}=0$ and $\psi_{1,2}=\pi/2$ the group manifold is pinched down to $S^1 \times S^3$ and $S^3 \times S^1$, respectively, which is consistent with the line-element and torsion 3-form for $SU(2) \times U(1).$

\subsection{The complex structures}
The first step is to write down the different possible complex structures on the complexified Lie algebra. We chose to write the Lie algebra generators as
\begin{eqnarray}
h =\frac 1 2 \,  \left(\begin{array}{cccc} 1&0&0&0\\
0&-1&0&0\\0&0&i&0\\0&0&0&-i
\end{array} \right), \quad
e_1 = \left(\begin{array}{cccc} 0&1&0&0\\
0&0&0&0\\0&0&0&0\\0&0&0&0
\end{array} \right),\quad
e_2 = \left(\begin{array}{cccc} 0&0&0&0\\
0&0&0&0\\0&0&0&1\\0&0&0&0
\end{array} \right)
\end{eqnarray}
and their complex conjugates $\bar h=h^\dagger$, $\bar e_1=e_1^\dagger$ and  $\bar e_2=e_2^\dagger$. \\
\\
The Lie algebra $\mathfrak{su}(2) \oplus \mathfrak{su}(2)$ is of rank 2, and as such has a two-dimensional Cartan subalgebra spanned by $h$ and $\bar h$. This leaves us with two possible choices for complex structures on the Lie algebra

\begin{eqnarray}
\mathbb{J}_1&=&\mbox{diag} \big(+i,+i,+i,-i,-i,-i\big) \nonumber \\
\mathbb{J}_2&=&\mbox{diag} \big(-i,+i,+i,+i,-i,-i\big),
\end{eqnarray}
where we have labelled the rows and columns as $h,e_1,e_2,\bar h,\bar e_1, \bar e_2$. As before we will use the right- and left-invariant vielbeins to determine the complex structures on the target manifold, again allowing for the left- and right choice to be made independently. This leaves the following possibilities
\begin{enumerate}
  \item $J_+^\mu{}_\nu=L^\mu_C\,\mathbb{J}_1^C{}_D\,L^D_\nu$,\quad
  $J_-^\mu{}_\nu=R^\mu_C\,\mathbb{J}_1^C{}_D\,R^D_\nu\,$.
   \item $J_+^\mu{}_\nu=L^\mu_C\,\mathbb{J}_1^C{}_D\,L^D_\nu$,\quad
  $J_-^\mu{}_\nu=R^\mu_C\,\mathbb{J}_2^C{}_D\,R^D_\nu\,$.
  \item $J_+^\mu{}_\nu=L^\mu_C\,\mathbb{J}_2^C{}_D\,L^D_\nu$,\quad
  $J_-^\mu{}_\nu=R^\mu_C\,\mathbb{J}_1^C{}_D\,R^D_\nu\,$.
  \item $J_+^\mu{}_\nu=L^\mu_C\,\mathbb{J}_2^C{}_D\,L^D_\nu$,\quad
  $J_-^\mu{}_\nu=R^\mu_C\,\mathbb{J}_2^C{}_D\,R^D_\nu\,$.
\end{enumerate}
Having chosen the complex structures on the group manifold the next step is to determine the superfield content by analysing the relevant kernels.
\begin{itemize}
\item The first and fourth choice yield the following eigenvalues.\\
$(J_+ + J_-) \rightarrow$ $\pm 2i$ and $\pm2i \cos \psi_1 \cos \psi_2$, two-fold degenerate.\\
$(J_+ - J_-) \rightarrow$ 0 and $\pm i \sqrt{3-2\cos2\psi_1 \cos^2 \psi_2-\cos 2\psi_2}$, both two-fold degenerate. 
\item The second and third choice yield the eigenvalues\\
$(J_+ + J_-) \rightarrow$ 0 and $\pm i \sqrt{3+2\cos2\psi_1 \cos^2 \psi_2+\cos 2\psi_2}$, both two-fold degenerate.\\
$(J_+ - J_-) \rightarrow$ $\pm 2i$ and $\pm2i \sin \psi_1 \sin \psi_2$, two-fold degenerate.\\
\end{itemize}
The structure of the relevant kernels implies that the first and fourth choice will lead to a parametrisation in terms of a chiral superfield and a semi-chiral superfield, while the second and third choice will lead to a parametrisation in terms of a twisted-chiral superfield and a semi-chiral superfield. In what follows we will work with the first and second choices as we can expect that, just like in the case of $S^3 \times S^1$, choices with the same kernel structure will lead to similar results after a suitable coordinate transformation. Furthermore, the values of the eigenvalues seem to indicate that we can expect type-changing to occur at the loci $\psi_1=\pi/2$, $\psi_2=\pi/2$,$\psi_1=\psi_2=0$ and $\psi_1=0$, $\psi_2=0$,$\psi_1=\psi_2=\pi/2$ respectively.\\
\\
We now turn to the first choice for the complex structures, which when explicitly calculated in real coordinates read
	\begin{eqnarray}
	 J_+=\left(\begin{array}{cccccc}
	 0 & -\tan \psi_1 & 0 & -\cos^2 \psi_2  & 0 & \sin^2 \psi_2 \\
	 \frac{\sin 2\psi_1}{2} & 0 &  \frac{\sin 2\psi_1}{2}  & 0 & 0 & 0 \\
	 0 &- \cot \psi_1 & 0 & \cos^2 \psi_2 & 0 & -\sin^2 \psi_2 \\
	 \cos^2 \psi_1 & 0 & -\sin^2 \psi_1 & 0 &  -\tan \psi_2 & 0 \\
	 0 & 0 & 0 &  \frac{\sin 2\psi_2}{2}  & 0 &  \frac{\sin 2\psi_2}{2}  \\
	 -\cos^2 \psi_1 & 0 & \sin^2 \psi_1 & 0 & -\cot \psi_2 & 0
 	\end{array}   \right)
	\end{eqnarray}
and
	\begin{eqnarray}
	 J_-=\left(\begin{array}{cccccc}
	 0 & \tan \psi_1 & 0 & -\cos^2 \psi_2  & 0 & -\sin^2 \psi_2 \\
	 -\frac{\sin 2\psi_1}{2} & 0 &\frac{\sin 2\psi_1}{2}  & 0 & 0 & 0 \\
	 0 & -\cot \psi_1 & 0 & -\cos^2 \psi_2 & 0 & -\sin^2 \psi_2 \\
	 \cos^2 \psi_1 & 0 & \sin^2 \psi_1 & 0 &  \tan \psi_2 & 0 \\
	 0 & 0 & 0 & -\frac{\sin 2\psi_2}{2} & 0 & \frac{\sin 2\psi_2}{2} \\
	 \cos^2 \psi_1 & 0 & \sin^2 \psi_1 & 0 & -\cot \psi_2 & 0
 	\end{array}   \right).  
	\end{eqnarray}
This results in the potential
\begin{eqnarray}
 V(l,\bar l,r,\bar r,z,\bar z)&=&\frac{k}{4\pi }\Big(-(l-r)( \bar{l}- \bar{r})+\int^{r+\bar r}
dq\,
 \ln\big(1+e^q\big)+ \nonumber\\
&&\qquad \int^{z+ \bar{z}+i(l-\bar l)}dq\,
 \ln\big(1+e^q\big)\Big),\label{s3s3V}
\end{eqnarray}
with the following choice of complex coordinates
\begin{eqnarray}
&&l=\ln\sin \psi_2+i\,\ln\cos \psi_1 + \varphi_{11}+i\, \varphi_{22}\,,  \nonumber\\
&&r=\ln\tan\psi_2+i(\varphi_{22}-\varphi_{21})\,,\nonumber\\
&&z=\ln\sin \psi_1-i\ln\sin \psi_2+\varphi_{22}+i \,\varphi_{12}\,.
\end{eqnarray}
This potential is well defined on any patch where $ \psi_1,\,\psi_2\neq\pi/2$.\\
\\
For the second choice of the complex structures, with $J_+$ as before and the second complex structure now given by
	\begin{eqnarray}
	 J_-=\left(\begin{array}{cccccc}
	 0 & \tan \psi_1 & 0 & \cos^2 \psi_2  & 0 & \sin^2 \psi_2 \\
	 -\frac{\sin 2\psi_1}{2} & 0 &\frac{\sin 2\psi_1}{2}  & 0 & 0 & 0 \\
	 0 & -\cot \psi_1 & 0 & \cos^2 \psi_2 & 0 & \sin^2 \psi_2 \\
	 -\cos^2 \psi_1 & 0 & -\sin^2 \psi_1 & 0 &  \tan \psi_2 & 0 \\
	 0 & 0 & 0 & -\frac{\sin 2\psi_2}{2} & 0 & \frac{\sin 2\psi_2}{2} \\
	 -\cos^2 \psi_1 & 0 & -\sin^2 \psi_1 & 0 & -\cot \psi_2 & 0
 	\end{array}   \right),  
	\end{eqnarray}
 we find the ``mirror'' potential
\begin{eqnarray}
 V(l,\bar l,r,\bar r,w,\bar w)&=&\frac{k}{4\pi }\Big((l-\bar r)( \bar{l}- r)-\int^{r+\bar r}
dq\,
 \ln\big(1+e^q\big)- \nonumber\\
&&\qquad \int^{w+ \bar{w}+i(l-\bar l)}dq\,
 \ln\big(1+e^q\big)\Big),\label{s3s3V2}
\end{eqnarray}
using the complex coordinates
\begin{eqnarray}
&&l=\ln\cos \psi_2+i\,\ln\sin \psi_1 - \varphi_{12}-i\, \varphi_{21}\,,  \nonumber\\
&&r=\ln\cot\psi_2+i(\varphi_{21}-\varphi_{22})\,,\nonumber\\
&&w=\ln\cos \psi_1-i\ln\cos \psi_2-\varphi_{21}-i\, \varphi_{11}\,,
\end{eqnarray}
and the potential is well defined as long as $ \psi_1,\,\psi_2\neq 0$.\\
\\
The expressions derived above are related to those we found in the previous section for $SU(2) \times U(1) \subset SU(2) \times SU(2)$ in a natural way. Looking at the potential (\ref{s3s3V}), we see that letting $\psi_1 \rightarrow 0$ we obtain the potential 
\begin{eqnarray}
 V_2(l',\bar l',r',\bar r')=\frac{k}{4\pi }\Big(-\ln \frac{l'}{ r'}\,\ln \frac{\bar l'}{\bar r'}\,+\int^{r'\bar r'}\frac{dq}{q}\,
 \ln\big(1+q\big)\Big),\label{piano1}
\end{eqnarray}
which is related to the potential (\ref{Vsemi}) via a Legendre transform with respect to $l$ and $\bar l$ and where we have introduced the new coordinates
\begin{eqnarray}
l'= \frac{\bar l}{r}\,, & \bar l'= \frac{l}{\bar r} \\
r'=\frac 1 r \,,& \bar r ' = \frac{1}{\bar r}\,.
\end{eqnarray}
This further establishes the relationship between $SU(2) \times SU(2)$ and $SU(2) \times U(1)$. We will find that this is apparent on the level of the pure spinors as well.\\
\\
To conclude this subsection we will consider isometries of the form $g \rightarrow h_-gh_+$, where we will denote the elements of the maximal torus as 
\begin{eqnarray}
h_\pm= \left(
\begin{array}{cc}
e^{i \varepsilon_\pm^1 \sigma_3}&0\\
0& e^{i \varepsilon_\pm^2 \sigma_3}
\end{array}\right).
\end{eqnarray}
Under isometries of this form the coordinates transform as 
\begin{eqnarray}
l &\rightarrow&  l+ \phi_- + \bar{ \phi }_+ \nonumber \\
 r &\rightarrow& r+ \phi_+-\bar \phi_+ \nonumber \\
z &\rightarrow&  z+ i\bar \phi_--i\bar \phi_+,
\end{eqnarray}
with
\begin{eqnarray}
\phi_\pm\equiv \varepsilon_\pm^1+i\, \varepsilon^2_\pm\,.
\end{eqnarray}
The model is indeed invariant under these transformations, provided that the superconstraints remain valid. This is the case when
\begin{eqnarray}
\mathbb{D}_+ \phi_-=\bar{\mathbb{D}}_+ \phi_-=\mathbb{D}_- \phi_-=0 \nonumber \\
\mathbb{D}_- \phi_+=\bar{\mathbb{D}}_- \phi_+=\mathbb{D}_+ \phi_+=0.
\end{eqnarray}

\subsection{Generalised K\"{a}hler geometry and type changing}
Having calculated the different complex structures on the manifold and the resulting generalised K\"{a}hler potential we turn to the associated pure spinors for both parametrisations. By calculating the Mukai pairings we will identify the loci where type changing occurs. As before we will investigate these cases by putting the pure spinors in their canonical form an re-calculating the Mukai pairings in the new coordinate system.

\subsubsection{Chiral and semi-chiral parametrisation}

Starting from the generalised K\"ahler potential (\ref{s3s3V}) we find that the pure spinors are given by

\begin{eqnarray}
\phi_+ &=& d \bar z \wedge e^{i\, \Omega^+ + \Xi^+},\\
\phi_- &=& e^{i\, \Omega^- + \Xi^-},
\end{eqnarray}
with the 2-forms $i\, \Omega^+ + \Xi^+$
\begin{eqnarray}
i\, \Omega^+ + \Xi^+ = \frac{k}{8 \pi} \left( -\frac{2}{1+ e^{i\, (l - \bar l ) + z + \bar z}} dl \wedge d \bar l - dl\wedge d\bar r -\frac{2 i}{1+ e^{-i\, (l - \bar l) - z - \bar z}} dl\wedge d \bar z\right. \nonumber \\  - d\bar l \wedge d r + \frac{2 i}{ 1+ e^{-i\,(l - \bar l) - z -\bar z}} d\bar l \wedge d z + \frac{4i}{1+e^{-i\,(l - \bar l) - z -\bar z}} d \bar l \wedge d \bar z \nonumber\\
\left. + \frac{2}{1+e^{r+\bar r}} dr\wedge d\bar r - \frac{2}{1+ e^{-i\,(l - \bar l) - z -\bar z}} dz\wedge d\bar z  \right)
\end{eqnarray}
and 
\begin{eqnarray}
i\, \Omega^- + \Xi^- = \frac{k}{8 \pi} \left( \frac{2}{1+ e^{i\, (l-\bar l) + z + \bar z}} dl\wedge d\bar l - dl\wedge d\bar r  - \frac{2i}{1+ e^{-i(l-\bar l) - z - \bar z}} dl \wedge d\bar z\right. \nonumber \\ + 3 d\bar l \wedge d r - \frac{2i}{1+e^{-i(l-\bar l) - z - \bar z}} d\bar l \wedge d z \nonumber \\ \left. + \frac{2}{1+ e^{r+ \bar r}}  dr \wedge d \bar r  - \frac{2}{1+e^{-i\, (l-\bar l) - z - \bar z}} dz \wedge d \bar z \right).
\end{eqnarray}
Using these expressions we can calculate the Mukai pairings in this coordinate system
\begin{eqnarray}
(\phi_+, \bar \phi_+) &=& \frac{k^2}{4 \pi^2} \frac{1}{(1+e^{r+\bar r}) (1+ e^{i\, l - i\, \bar l + z +\bar z})} =  \frac{k^2}{4 \pi^2} \cos^2 \psi_1 \, \cos^2 \psi_2 , \nonumber\\
(\phi_-, \bar \phi_-) &=& - \frac{k^3}{8 \pi^3} \frac{e^{i\, l - i\, \bar l + r+ \bar r + z + \bar z}}{(1+e^{r+\bar r}) (1+ e^{i\, l - i\, \bar l + z +\bar z})} =  - \frac{k^3}{8 \pi^3} \sin^2 \psi_1 \, \sin^2 \psi_2 .
\end{eqnarray}
While superconformally invariant, the generalised complex structures do not satisfy the generalised Calabi-Yau conditions. As expected we can explicitly identify the loci where type changing will occur, one at $\psi_1=0$, one at $\psi_2=0$ and a third at $\psi_1=\psi_2=\pi/2$. We will now investigate these loci in turn by finding appropriate coordinates where this becomes explicit.\\
\\
\noindent \underline{Type changing locus 1} : $\psi_1 = \psi_2=0$ \label{sect:5211}\\ 
We introduce the following complex coordinates,
\begin{eqnarray}
z^1 = e^l,\qquad z^2 = e^{-i\, \bar l + i\, r - \ln \big(1+ e^{i\, l - i\, \bar l + z + \bar z} \big)}, \qquad z^3 = e^{z + i\, l},
\end{eqnarray}
This choice of coordinates diagonalises $J_+=$ diag$\{+i,+i,+i,-i,-i-i\}$. We see that the type changing locus corresponds to $z_1=z_3=0$, where $J_-$ is diagonalised as well and is equal to $J_+$. As a consequence of this the type changing at the locus takes us from type (1,0) to type (3,0). As we will see this is reflected on the level of the pure spinors as well once appropriate coordinates are introduced.\\
\\
Explicitly calculating the pure spinors in the new coordinates yields, after rescaling and a $b$-transform
\begin{eqnarray}
\phi_+ &=& i \sqrt{z^1 \bar z^1 z^2  \bar z^2 z^3  \bar z^3} \left( \frac{d\bar z^3}{\bar z^3} + i\, \frac{d \bar z^1}{\bar z^1} \right) \wedge e^{-b\wedge}\, e^{\Lambda_+} \nonumber \\
\phi_- &=& i \sqrt{z^1 \bar z^1 z^2 \bar z^2 z^3  \bar z^3}\, e^{-b\wedge}\, e^{\Lambda_-}.
\end{eqnarray} 
Both pure spinors are $H$-closed when omitting the $b$-transform dependent part, as can be seen from the fact that $d\Lambda_+=d\Lambda_-=db=H$. The first spinor can be rescaled to the form
\begin{eqnarray}
\phi_+ = i\, \frac{2 \pi}{k} \bar z^1 \bar z^2 \bar z^3 \left( \frac{d\bar z^3}{\bar z^3} + i\, \frac{d \bar z^1}{\bar z^1} \right) \wedge e^{\Lambda_+}, \label{p1l1psp}
\end{eqnarray}
which at the type changing locus $z_1=z_3=0$ becomes a complex type pure spinor in its canonical form
\begin{eqnarray}
\phi_+ = d\bar z_1 \wedge d\bar z_2 \wedge d\bar z_3,
\end{eqnarray}
signifying that the type changes from $(1,0)$ to $(3,0)$. Conversely, at the type changing locus the other pure spinor is of symplectic type
\begin{eqnarray}
\phi_- = e^{\Lambda_-}.
\end{eqnarray}
The second pure spinor $\phi_-$ is still $H$-closed, while the pure spinor undergoing the type changing $\phi_+$ is not. Finally, we calculate the Mukai pairing at the type changing locus
\begin{eqnarray}
(\phi_+, \bar \phi_+) &=& \frac{1}{1+ z^1 \bar z^1 (z^2)^{-i} (\bar z^2)^i }\, \frac{1}{1+ z^3 \bar z^3}\\
(\phi_-, \bar \phi_-) &=& -\frac{k^3}{8 \pi^3}\, \frac{1}{ (z^2)^i (\bar z^2)^{-i}+ z^1 \bar z^1 }\, \frac{1}{1+ z^3 \bar z^3} \,\frac{1}{z^2 \bar z^2}.
\end{eqnarray} 
Both pairings are now non-vanishing at $z_1=z_3=0$. The generalised Calabi-Yau conditions are not satisfied, which implies that this configuration is not a valid supergravity background.\\
\\
\noindent \underline{Type changing locus 2} : $\psi_1=\pi/2$ \label{sect:5211}\\ 
We now switch to coordinates appropriate for describing the second locus
\begin{eqnarray}
w^1 = e^{-i\, l}, \qquad w^2 = e^{-\bar r - i\, \bar z}, \qquad z^1 = e^{i\, z}.
\end{eqnarray}
Neither complex structure is diagonalised by this choice of coordinates. However, when investigating the locus corresponding to $w^1=0$ we find that both complex structures become diagonal
\begin{eqnarray}
J_+|_{w^1=0} &=& {\rm diag} (+i, -i, +i, -i, +i, -i) \nonumber \\\
J_- |_{w^1=0} &=& {\rm diag}(-i, +i, -i, +i, +i, -i).
\end{eqnarray}
We immediately  see that, at the type changing locus, $\ker(J_+ + J_-)_{w^1=0}$ is of complex dimension two, while $\ker(J_+ + J_-)_{w^1=0}$ is of complex dimension one. We find that at the locus $w^1=0$, or equivalently $\psi=\pi/2$, the model is described by one chiral and two twisted-chiral superfields, which corresponds to the type changing from (1,0) to (1,2).\\
\\
Writing down the pure spinors using the new coordinates yields
\begin{eqnarray}
\phi_+ &=& -\, \sqrt{w^1 \bar w^1 w^2 \bar w^2 z^1 \bar z^1}\, \frac{d\bar z^1}{\bar z^1} \wedge e^{-b\wedge}\, e^{\Lambda_+} ,\\
\phi_- &=&i\, \sqrt{w^1 \bar w^1 w^2 \bar w^2 z^1 \bar z^1}\, e^{-b\wedge}\, e^{\Lambda_-},
\end{eqnarray}
where the fact that $d\Lambda_+=d\Lambda_-=db=H$ again ensures the $H$-closure of both pure spinors after the relevant $b$-transform. Next we put the pure spinors in their canonical forms through a rescaling
\begin{eqnarray}
\phi_+ &=& d\bar z^1 \wedge e^{\Lambda_+}  \\
\phi_- &=& -i\, \frac{2 \pi}{k} \bar w^1 \bar w^2 e^{\Lambda_-}
\end{eqnarray}
and investigate the behaviour at the locus $w^1=0$. $\phi_+$ remains unchanged and we recover the pure spinor already found for $S^3 \times S^1$ (\ref{Lampsimctcp2}). This is consistent with the observation that when looking at the metric (\ref{S3S3met}) and the torsion (\ref{S3S3tor}) we see that at $\psi_1=\pi/2$ they coincide with the metric and torsion found for  $S^3 \times S^1$. The resulting manifold is then parametrised by the chiral superfield $z^1$ and the twisted chiral superfield $w^2$. The pure spinor $\phi_-$ changes from a symplectic to a complex type spinor
\begin{eqnarray}
\phi_- \big|_{w^1=0} = d \bar w^1 \wedge d \bar w^2 \wedge e^{\Lambda_- |_{w^1=0}},
\end{eqnarray}
where $\Lambda |_{w^1=0}$ denotes the piece of $\bar w^1 \bar w^2 \Lambda_-$ that does not vanish when $w^1=0$. We can again make contact with the previously found description for $S^3 \times S^1$ by observing that $d \bar w^2 \wedge e^{\Lambda |_{w^1=0}}$ is exactly the pure spinor found in (\ref{Lammsimctcp2}).\\
\\
Finally, we compute the Mukai pairings and check that they are non-vanishing at the locus $w^1=0$
\begin{eqnarray}
(\phi_+, \bar \phi_+) &=& \frac{k^2}{4 \pi^2}\, \frac{1}{w^1 \bar w^1+(z^1)^{-i} (\bar z^1)^{i} }\, \frac{1}{w^2 \bar w^2+ z^1 \bar z^1}\, \frac{1}{z^1 \bar z^1},\\
(\phi_-, \bar \phi_-) &=& -\frac{k}{2 \pi} \, \frac{1}{1+ w^1 \bar w^1 (z^1)^{i} (\bar z^1)^{-i} }\, \frac{1}{w^2 \bar w^2+ z^1 \bar z^1}.
\end{eqnarray} 
This is indeed the case. The generalised Calabi-Yau conditions are once again not satisfied.\\
\\
\noindent \underline{Type changing locus 3} : $\psi_2=\pi/2$ \label{sect:5211}\\ 
We now turn to the third locus under consideration by introducing the coordinates
\begin{eqnarray}
w^1 = e^{-i\, l}, \qquad w^2 = e^{\bar l - \bar r - i\, \ln\big( 1+ e^{i\, l -i\, \bar l + z + \bar z} \big)}, \qquad z^1 = e^z,
\end{eqnarray}
which diagonalises $J_+$ to its canonical form. For this choice of coordinates the type changing locus corresponds to $w^2=0$, where the second complex structure is diagonalised as well. The resulting complex structures 
\begin{eqnarray}
J_+ |_{w^2=0} &=& {\rm diag} (+i, +i, +i, -i, -i, -i) \nonumber \\\
J_- |_{w^2=0} &=& {\rm diag}(-i, +i, -i,+ i,+ i, -i).
\end{eqnarray}
make it clear that $\ker (J_+ + J_-) |_{w^2=0} $ is of complex dimension two and that $\ker (J_+ - J_-) |_{w^2=0} $ is of complex dimension one, signalling the presence of two twisted-chiral superfields and one chiral superfield at the type changing locus, corresponding to the types of the generalised complex structures changing from (1,0) to (1,2).\\
\\
For our choice of coordinates we find the pure spinors in their simplified form after the proper $b$-transform (with once again $d\Lambda_+=d\Lambda_-=db=H$)
\begin{eqnarray}
\phi_+ &=& i\, \sqrt{w^1 \bar w^1 w^2 \bar w^2 z^1   \bar z^1}\, \frac{d\bar z^1}{\bar z^1} \wedge e^{-b \wedge}\, e^{\Lambda_+}\nonumber\\
\phi_- &=&  i\, \sqrt{w^1 \bar w^1 w^2  \bar w^2  z^1 \bar z^1}\, e^{-b \wedge}\, e^{\Lambda_-},
\end{eqnarray}
which can be rescaled after omitting the $b$-dependent part to yield 
\begin{eqnarray}
\phi_+ &=& d\bar z^1 \wedge e^{\Lambda_+} \nonumber\\
\phi_- &=& -i\, \frac{2 \pi}{k} \bar w^1 \bar w^2 e^{\Lambda_-}.
\end{eqnarray}
By taking $w^2=0$ we see that $\phi_+$, modulo the $b$-transform, contains the expression found in (\ref{Lampsimctcp2}). The corresponding geometry at the locus is described by a $S^3 \times S^1$ parametrised by the twisted-chiral superfield $w^1$ and the chiral superfield $z^1$. One can verify that this is consistent with the calculated forms of the metric and the torsion 3-form when $\psi_2=\pi/2$. Turning to the other pure spinor, we see that it indeed changes from a symplectic to a more complex type
\begin{eqnarray}
\phi_-\big|_{w^2 = 0} = d\bar w^1 \wedge d\bar  w^2 \wedge e^{\Lambda_-|_{w^2=0} }.
\end{eqnarray}
We recover the expression found in (\ref{Lammsimctcp2}) from the factor $d\bar w^1\wedge e^{\Lambda_-|_{w^2=0}}$, again confirming that at the locus $w^2=0$ the resulting geometry is that of the $S^3 \times S^1$ found before. We check that the Mukai pairings for the rescaled pure spinors are well-behaved at the type changing locus
\begin{eqnarray}
(\phi_+, \bar \phi_+) &=& \frac{k^2}{4 \pi^2}\, \frac{1}{w^2 \bar w^2+(w^1)^{i} (\bar w^1)^{-i} }\, \frac{1}{w^1 \bar w^1+ z^1 \bar z^1}\, \frac{1}{z^1 \bar z^1}\\
(\phi_-, \bar \phi_-) &=& -\frac{k}{2 \pi} \, \frac{1}{1+(w^1)^{-i} (\bar w^1)^{i}  w^2 \bar w^2}\, \frac{1}{w^1 \bar w^1+ z^1 \bar z^1},
\end{eqnarray}
which shows that the generalised Calabi-Yau conditions are not satisfied, consistent with the observation that one of the pure spinors is no longer ($H$-)closed.

\subsubsection{The twisted-chiral and semi-chiral parametrisation}
The second parametrisation of the manifold $S^3 \times S^3$, corresponding to the second choice of complex structures, led us to a description on the level of the $\sigma$-model in terms of a generalised K\"{a}per potential (\ref{s3s3V2}) that was related to the first potential (\ref{s3s3V}) by a mirror transform. As such, one could expect the results to be similar to those of the previous subsection but with the role of the two pure spinors exchanged. We will calculate this explicitly and show that this is indeed the case. We begin by calculating the pure spinors starting from the potential (\ref{s3s3V2})
 \begin{eqnarray}
\phi_+ &=& e^{i\, \Omega^+ + \Xi^+}\\
\phi_- &=& d\bar w \wedge e^{i\, \Omega^- + \Xi^-},
\end{eqnarray}
with,
\begin{eqnarray}
i\, \Omega^+ + \Xi^+ = \frac{k}{8 \pi} \left( \frac{2}{1+ e^{i\,(l - \bar l)+w + \bar w}} dl\wedge d\bar l - dl \wedge dr + \frac{i}{1+e^{-i\,(l-\bar l) - w - \bar w}} dl \wedge dw   \right. \nonumber \\  - \frac{i}{1+e^{-i\,(l-\bar l) - w - \bar w}} dl \wedge d\bar w + 3 d\bar l \wedge d\bar r - \frac{3i}{1+e^{-i\,(l-\bar l) - w - \bar w}} d\bar l \wedge d w \nonumber \\  - \frac{i}{1+ e^{-i\,(l-\bar l) - w - \bar w}} d\bar l \wedge d \bar w - \frac{2}{1+e^{r + \bar r}} dr \wedge d\bar r \nonumber \\ \left. - \frac{2i}{1+ e^{-i\,(l-\bar l ) - w - \bar w}} dw \wedge d \bar w  \right), \nonumber \\
\end{eqnarray}
and
\begin{eqnarray}
i\, \Omega^- + \Xi^- = \frac{k}{8 \pi} \left( -\frac{2}{1+e^{i\,(l-\bar l) + w + \bar w }} dl \wedge d\bar l - dl\wedge dr + \frac{i}{1+e^{-i\,(l-\bar l) - w - \bar w}} dl\wedge dw \right. \nonumber\\  + \frac{3}{1+ e^{-i\,(l-\bar l) - w - \bar w}} d l \wedge d\bar w - d\bar l \wedge d \bar r + \frac{i}{1+e^{-i\,(l-\bar l) - w - \bar w}} d\bar l \wedge dw \nonumber  \\   - \frac{i}{1+e^{-i\,(l- \bar l ) - w - \bar w}} d\bar l \wedge d \bar w - \frac{2}{1+e^{r + \bar r}} d r\wedge d\bar r \nonumber \\ \left. + \frac{2}{1+e^{-i\,(l-\bar l ) - w -\bar w}} dw \wedge d \bar w  \right) . \nonumber \\
\end{eqnarray}
By calculating the Mukai pairings one should be able to discern the type changing loci directly
\begin{eqnarray}
(\phi_+, \bar \phi_+) &=& \frac{k^3}{8 \pi^3} \frac{e^{i\, l - i \, \bar l + r + \bar r+ w + \bar w}}{(1+e^{r+\bar r})(1+e^{i\, l -i\, \bar l + w + \bar w})} = \frac{k^3}{8 \pi^3} \cos^2 \psi_1 \, \cos^2\psi_2\nonumber \\ 
\\
(\phi_-, \bar \phi_-) &=& -\frac{k^2}{4 \pi^2} \frac{1}{(1+e^{r+\bar r})(1+e^{i\, l -i\, \bar l + w + \bar w})} = - \frac{k^2}{4 \pi^2} \sin^2 \psi_1 \, \sin^2 \psi_2  . \nonumber \\
\end{eqnarray}
While the generalised Calabi-Yau condition is not satisfied, this description yields a superconformally invariant theory at one loop. Again we see that there are three type changing loci where one of the Mukai pairings vanishes~: $\psi_1 = \psi_2= \pi/2 $, $\psi_1 = 0$ and $\psi_2 = 0$. By introducing new coordinates adapted to the type changing loch we will investigate the resulting geometry more thoroughly.\\
\\
\underline{Type changing locus 1} :  $\psi_1 = \psi_2= \pi/2 $ \label{sect:5211}\\ 
In order to investigate the first locus we switch to the complex coordinates 
\begin{eqnarray}
w^1 = e^l, \qquad w^2 = e^{-i\, \bar l + i\, r - \ln\big(1+ e^{i\, l - i\, \bar l + w +\bar w} \big)}, \qquad w^3 = e^{w+i\, l}.
\end{eqnarray}
By employing these coordinates $J_+$ is put in its canonical form. At the type changing locus, now given by $w^1=w^3=0$, $J_-$ is put in diagonal form as well and both complex structures are opposite to each other. This signifies the type changing from (0,1) to (0,3) and leaves us with a description in terms of three twisted-chiral superfields when restricted to the locus.\\
\\
Let us make this more manifest by looking at the pure spinors
\begin{eqnarray}
\phi_+ &=& \sqrt{w^1 \bar w^1 w^2 \bar w^2 w^3 \bar w^3 } \, e^{-b\wedge}\, e^{\Lambda_+}\\
\phi_- &=& \sqrt{w^1 \bar w^1 w^2 \bar w^2 w^3 \bar w^3 } \, \left( \frac{d \bar w^3}{\bar w^3} + i\, \frac{d \bar w^1}{\bar w^1} \right)\wedge e^{-b\wedge}\, e^{\Lambda_-}
\end{eqnarray}
By undoing the $b$-transform, where we have $d\Lambda_+ = d \Lambda_- = db = H$, we are left with simplified forms for the spinors that are $H$-closed. Further rescaling these expressions yields
\begin{eqnarray}
\phi_+ &=& e^{\Lambda_+} \\
\phi_- &=& i\, \frac{2 \pi}{k} \bar w^1 \bar w^2 \bar w^3 \left( \frac{d \bar w^3}{\bar w^3} + i\, \frac{d \bar w^1}{\bar w^1} \right) \wedge  e^{\Lambda_-}.
\end{eqnarray}
We now clearly see that when we restrict ourselves to the type changing locus $z^1=z^3=0$ we are left with the pure spinors in their canonical form, with $\phi_+$ of symplectic type and $\phi_-$ of complex type
\begin{eqnarray}
\phi_+ &=& e^{\Lambda_+} \\
\phi_- &=& d \bar w^1 \wedge d \bar w^2 \wedge d \bar w^3,
\end{eqnarray}
with $\phi_-$ no longer being $H$-closed. Note that this is consistent with the fact that for a purely twisted-chiral parametrisation we have a target space geometry that is K\"{a}hler with the generalised complex structures $\mathcal{J}_1$ and $\mathcal{J}_2$ corresponding to the usual symplectic and complex structures, respectively. The generalised Calabi-Yau conditions are not satisfied as can be seen from computing the Mukai pairings
\begin{eqnarray}
(\phi_+, \bar \phi_+) &=& - \frac{k^3}{8 \pi^3}\, \frac{1}{(w^2)^i (\bar w^2)^{-i} + w^1 \bar w^1} \, \frac{1}{1+ w^3 \bar w^3}\, \frac{1}{w^2 \bar w^2} \\
(\phi_-, \bar \phi_-) &=& \frac{1}{1+ w^1 \bar w^1 (w^2)^{-i} (\bar w^2)^{i}}\, \frac{1}{1+w^3 \bar w^3},
\end{eqnarray}
who are now both non-vanishing at the type changing locus.\\
\\
\underline{Type changing locus 2} : $\psi_1=0$\\
In order to clarify the second type changing locus we employ the coordinates
\begin{eqnarray}
z^1 = e^{-i\, l}, \qquad z^2 = e^{-r - i\, \bar w}, \qquad w^1 = e^{i\, w}.
\end{eqnarray}
By performing this coordinate transformation we find that neither complex structure is diagonalised. However, when restricting ourselves to the type-changing locus $z^1=0$ we find that they both become diagonal and are given by
\begin{eqnarray}
J_+ |_{z^1=0} &=& {\rm diag}(+i, -i, +i, -i, +i, -i) \nonumber \\
J_- |_{z^1=0} &=& {\rm diag}(+i, -i, +i, -i, -i, +i).
\end{eqnarray}
As can be seen from considering the kernel structure, we are left with two complex directions along $\ker(J_+ - J_-)|_{z^1=0} $ and one complex direction along $\ker(J_+ + J_-)|_{z^1=0} $. This leaves us with the locus described by two chiral superfields and a twisted chiral superfield, indicating that at $z^1=0$ the type changes from (0,1) to (2,1).\\
\\
Explicitly calculating the pure spinors yields
\begin{eqnarray}
\phi_+ &=& \sqrt{w^1 \bar w^1 z^1 \bar z^1 z^2 \bar z^2} \wedge e^{-b\wedge}\, e^{\Lambda_+}\, \\
\phi_- &=& i\, \sqrt{w^1 \bar w^1 z^1 \bar z^1 z^2 \bar z^2}\, \frac{d\bar w^1}{\bar w^1} \wedge e^{-b \wedge}\, e^{\Lambda_-},
\end{eqnarray}
where obtains $H$-closed expressions by once again undoing the $b$-transform with $d\Lambda_+ = d \Lambda_- = db = H$.\\
\\
Rescaling the resulting expressions leaves us with the forms
\begin{eqnarray}
\phi_+ &=& -i\, \frac{2 \pi}{k} \bar z^1 \bar z^2\, e^{\Lambda_+} \nonumber \\ 
\phi_- &=& d\bar w^1 \wedge e^{\Lambda_-}.
\end{eqnarray}
At the type changing locus $z^1=0$, $\phi_-$ does not change type and contains the previously obtained expression (\ref{Lammsimctcp1}). This is consistent with a local geometry describing a $S^3 \times S^1$ parametrised by a chiral superfield $z^2$ and a twisted chiral superfield $w^1$, as could be expected from considering the $S^3 \times S^3$ metric (\ref{S3S3met}) and the torsion 3-form (\ref{S3S3tor}) when restricted to $\psi_1=0$.\\
\\
The second complex spinor $\phi_+$ changes type when considering the locus $z^1=0$ and is given by
\begin{eqnarray}
\phi_+ \big|_{z^1 = 0} = d \bar z^1 \wedge d \bar z^2 \wedge e^{\Lambda_- |_{z^1=0}},
\end{eqnarray} 
where we can make contact with the $S^3 \times S^1$ description by remarking that $d \bar z^2 \wedge e^{\Lambda_- |_{z^1=0}}$ exactly corresponds to the pure spinor as given in eq.~(\ref{Lampsimctcp1}), where the local geometry is described by the chiral superfield $z^2$ and the twisted-chiral superfield $w^1$. As can be expected, $\phi_+$ is no longer $H$-closed.\\
\\
\underline{Type changing locus 3} : $\psi_2=0$\\
We now turn to our last type changing locus. The relevant complex coordinates are 
\begin{eqnarray}
z^1 = e^{-i\, l}, \qquad z^2 = e^{\bar l - r - i\, \ln \big( 1 + e^{i\, l - i\, \bar l + w + \bar w } \big)}, \qquad w^1 = e^w.
\end{eqnarray}
This choice of coordinates puts $J_+$ in its canonical form. At the type changing locus $z^2=0$, $J_-$ is also diagonalised and we obtain the complex structures
\begin{eqnarray}
J_+ |_{z^2=0} &=& {\rm diag}(+i, +i, +i, -i, -i, -i) \nonumber \\
J_- |_{z^2=0} &=& {\rm diag}(+i, -i, +i, -i, -i, +i).
\end{eqnarray}
It is clear that this results in $\ker(J_+ - J_-)|_{z^2=0} $ being of complex dimension two, while $\ker(J_+ + J_-)|_{z^2=0} $ is of complex dimension one. As such, the type changes from (0,1) to (2,1) and the locus $z^2=0$ is parametrised by two chiral superfields and one twisted chiral superfield. After the coordinate transformation we are left with the pure spinors
\begin{eqnarray}
\phi_+ &=& \sqrt{z^1 \bar z^1 z^2 \bar z^2 w^1 \bar w^1}\, e^{-b\wedge} \, e^{\Lambda_+} \\
\phi_- &=& \sqrt{z^1 \bar z^1 z^2 \bar z^2 w^1 \bar w^1}\, \frac{d \bar w^1}{\bar w^1} \wedge e^{-b \wedge} \, e^{\Lambda_-}, 
\end{eqnarray}
which can be untwisted to yield $H$-closed expressions since as before we have that $d\Lambda_+ = d \Lambda_- = db = H$.\\
\\
To make the type changing manifest we rescale the pure spinors to 
\begin{eqnarray}
\phi_+ &=& - i\, \frac{2 \pi}{k} \bar z^1 \bar z^2  e^{\Lambda_+}\\
\phi_- &=& d \bar w^1 \wedge e^{\Lambda_-}.
\end{eqnarray} 
and look at their behaviour at the locus $z^2=0$. The second spinor $\phi_-$ does not change type and can be shown to contain the previously derived expression (\ref{Lammsimctcp1}). The first spinor $\phi_+$ changes type as can be seen from its new form
\begin{eqnarray}
\phi_+\big|_{z^2=0} &=& d \bar z^1 \wedge d\bar z^2\wedge e^{\Lambda_+ |_{z^2=0} }, 
\end{eqnarray}
with $d\bar z^1 \wedge e^{\Lambda_+ |_{z^2=0}}$ corresponding to the pure spinor in (\ref{Lampsimctcp1}). Locally, the geometry is that of a $S^1 \times S^3$ parametrised by a chiral and a twisted chiral superfield. The spinor undergoing type changing $\phi_+$ is again no longer $H$-closed, which is also reflected in the Mukai pairings
\begin{eqnarray}
(\phi_+, \bar \phi_+) &=& - \frac{k}{2 \pi} \frac{1}{1+ (z^1)^{-i} (\bar z^1)^{i} z^2 \bar z^2 }\, \frac{1}{z^1 \bar z^1 + w^1 \bar w^1} \\
(\phi_-, \bar \phi_-) &=&\frac{k^2}{4 \pi^2} \frac{1}{ (z^1)^{i} (\bar z^1)^{-i}+ z^2 \bar z^2}\, \frac{1}{z^1 \bar z^1 + w^1 \bar w^1}\, \frac{1}{w^1 \bar w^1}.
\end{eqnarray}
The pairings are now well-behaved at the type changing locus, and do not satisfy the generalised Calabi-Yau conditions.\\
\\
Concluding this section, we have found that by considering the various possible pairs of complex structures on $SU(3) \times SU(3)$ there are two distinct parametrisations, one consisting of a chiral and semi-chiral superfield and one consisting of a twisted-chiral and semi-chiral superfield. The resulting geometry on the space $S^3 \times S^3$ possesses loci where one of the $S^3$ factors is pinched down to an $S^1$, effectively reducing the model to the case with $S^3 \times S^1$ considered before. By explicitly calculating the pure spinors corresponding to the generalised complex structures we were able to make type changing manifest. The resulting behaviour is summarised in Table 7.2. Again we find that the role of the pure spinors is exchanged when moving from one patch to there other, signalling that mirror symmetry could be at work. Here as well we find that although the models considered are UV-finite, they do not constitute valid supergravity backgrounds.

\begin{table}
\begin{center}
\begin{tabular}{c@{\qquad}c@{\qquad}c@{\qquad}c}
\hline $S^3\times S^3$ & locus 1 & locus 2 & locus 3 \\
\hline \hline \\
patch 1 (\ref{s3s3V}) & $\psi_1 = 0 = \psi_2 $:  & $\psi_1= \pi/2$:  & $\psi_2= \pi/2$: \\
&$ (1,0)\rightarrow (3,0)$ & $(1,0)\rightarrow (1,2)$& $(1,0)\rightarrow (1,2)$\\
& & &\\
\hline \hline \\
patch 2 (\ref{s3s3V2}) & $\psi_1 = \pi/2 = \psi_2 $: & $\psi_1 = 0$:    &  $\psi_2= 0$:  \\
&  $(0,1)\rightarrow (0,3)$  & $(0,1)\rightarrow (2,1)$& $(0,1)\rightarrow (2,1)$\\
& & & \\
\hline
\end{tabular}\begin{picture}(0,0) \put(-375,-32){\begin{tikzpicture}  \draw[<->] (0,0) arc (270:90:1.1cm);\end{tikzpicture}} \put(-391, 0){mirror} \put(-400,-10){symmetry}  \end{picture}
\caption{Summary of type changing for $SU(2)\times SU(2)$ \label{capTab2}}
\end{center}
\end{table}

\newpage
\section{$SU(3)$}
In this final section we present some preliminary results on the eight dimensional group manifold $SU(3)$. Due to the sheer computational obstacles of working with eight-dimensional operators whose expressions quickly become rather involved no conclusive results can be presented, however it should give some indication towards future work.

\subsection{Parametrisation and topology}
We first choose the Gell-Mann matrices as generators for the Lie algebra $\mathfrak{su}(3)$
\begin{eqnarray}
T_1 =\left( \begin{matrix} 0 & 1 & 0 \\ 1 & 0 & 0 \\ 0 & 0 & 0 \end{matrix} \right)
& T_2= \left( \begin{matrix} 0 & -i & 0 \\ i & 0 & 0 \\ 0 & 0 & 0 \end{matrix} \right)
& T_3= \left( \begin{matrix} 1 & 0 & 0 \\ 0 & -1 & 0 \\ 0 & 0 & 0 \end{matrix} \right) \nonumber \\
T_4 = \left( \begin{matrix} 0 & 0 & 1 \\ 0 & 0 & 0 \\ 1 & 0 & 0 \end{matrix} \right)
& T_5=  \left( \begin{matrix} 0 & 0 & -i \\ 0 & 0 & 0 \\ i & 0 & 0 \end{matrix} \right)
& T_6 = \left( \begin{matrix} 0 & 0 & 0 \\ 0 & 1 & 0 \\ 0 & 1 & 0 \end{matrix} \right) \nonumber \\
T_7= \left( \begin{matrix} 0 & 0 & 0 \\ 0 & 0 & -i \\ 0 & i & 0 \end{matrix} \right)
& T_8=\frac{1}{\sqrt{3}}\left( \begin{matrix} 1 & 0 & 0 \\ 0 & 1 & 0 \\ 0 & 0 & -2 \end{matrix} \right) &.
\end{eqnarray}
An arbitrary group element $g \in SU(3)$ can be parametrised\footnote{This parametrisation is consistent with the factorisation of the Lie algebra described in chapter 6.} as
\begin{eqnarray}
g=U_1 (\varphi_1,\psi_1,\epsilon_1) \exp(i T_5 \chi) U_2(\varphi_2,\psi_2,\epsilon_2) \exp(i T_8 \tau),
\end{eqnarray}
where $U_1$ and $U_2$ correspond to two inequivalent $SU(2)$ subgroups and are given by
\begin{eqnarray}
U_{1}  (\varphi_1,\psi_1,\epsilon_1)&=& \left( \begin{matrix} e^{i(\epsilon_1 + \varphi_1)}\cos \psi_1 & e^{-i(\epsilon_1- \varphi_1)} \sin \psi_1 & 0 \\ -e^{i(\epsilon_1- \varphi_1)} \sin \psi_1 & e^{-i(\epsilon_1 + \varphi_1)}\cos \psi_1 & 0 \\ 0 & 0 & 1 \end{matrix} \right) \nonumber \\
U_{2} (\varphi_2,\psi_2,\epsilon_2) &=& \left( \begin{matrix} e^{i(\epsilon_2 + \varphi_2)}\cos \psi_2 & e^{-i(\epsilon_2- \varphi_2)} \sin \psi_2 & 0 \\ -e^{i(\epsilon_2- \varphi_2)} \sin \psi_2 & e^{-i(\epsilon_2 + \varphi_2)}\cos \psi_2 & 0 \\ 0 & 0 & 1 \end{matrix} \right).
\end{eqnarray}

\subsection{The complex structures}
In what follows we will work with the complexified Lie algebra by defining the Cartan subalgebra
\begin{eqnarray}
h=\frac{1}{2} \big( T_3 +i T_8 \big), \qquad \bar h =\frac{1}{2}\big( T_3 -i T_8 \big)
\end{eqnarray}
and the positive and negative roots
\begin{eqnarray}
e_1=\frac{1}{2} \left(T_1 + i T_2 \right), \quad e_2=\frac{1}{2} \left(T_4 + i T_5 \right), \quad e_3=\frac{1}{2} \left(T_6 + i T_7 \right), \\
\bar e_1=\frac{1}{2} \left(T_1 - i T_2 \right), \quad \bar e_2=\frac{1}{2} \left(T_4 - i T_5 \right), \quad \bar e_3=\frac{1}{2} \left(T_6 - i T_7 \right).
\end{eqnarray}
Labelling rows and columns as $(h, e_1,e_2,e_3, \bar h, \bar e_1, \bar e_2, \bar e_3)$ this leaves us with two choices for the complex structure on the Lie algebra
\begin{eqnarray}
\mathbb{J}_1&=&\mbox{diag} \big(+i,+i,+i,+i,-i,-i,-i,-i\big) \nonumber \\
\mathbb{J}_2&=&\mbox{diag} \big(-i,+i,+i,+i,+i,-i,-i,-i\big).
\end{eqnarray}
As a result, four choices for a pair of complex structures on the group manifold are possible
\begin{enumerate}
  \item $J_+^\mu{}_\nu=L^\mu_C\,\mathbb{J}_1^C{}_D\,L^D_\nu$,\quad
  $J_-^\mu{}_\nu=R^\mu_C\,\mathbb{J}_1^C{}_D\,R^D_\nu\,$.
   \item $J_+^\mu{}_\nu=L^\mu_C\,\mathbb{J}_1^C{}_D\,L^D_\nu$,\quad
  $J_-^\mu{}_\nu=R^\mu_C\,\mathbb{J}_2^C{}_D\,R^D_\nu\,$.
  \item $J_+^\mu{}_\nu=L^\mu_C\,\mathbb{J}_2^C{}_D\,L^D_\nu$,\quad
  $J_-^\mu{}_\nu=R^\mu_C\,\mathbb{J}_1^C{}_D\,R^D_\nu\,$.
  \item $J_+^\mu{}_\nu=L^\mu_C\,\mathbb{J}_2^C{}_D\,L^D_\nu$,\quad
  $J_-^\mu{}_\nu=R^\mu_C\,\mathbb{J}_2^C{}_D\,R^D_\nu\,$.
\end{enumerate}
At this point we should note that, just like $SU(2) \times U(1)$, $SU(3)$ allows for an enhanced $\mathcal{N}=(4,4)$ supersymmetry, implying that every choice will give rise to a 2-sphere worth of complex structures. We will not consider this here.\\
\\
In order to determine the field content, we should investigate the various kernels. Due to the computational complexity involved this is a rather daunting task. We will take simpler approach and compute the determinants of $(J_+-J_-)$ and $(J_+ +J_-)$. Doing so we find that for the first and fourth choice the determinants are non-zero, while those of the second and third vanish. These values seem to indicate that the first and fourth choice, and the second and third choice, are related to each other. For the first and fourth choice the determinants do not vanish for arbitrary points on the manifolds, implying that these choices correspond to a parametrisation in terms of two semi-chiral superfields. Specifically, we find that
\begin{eqnarray}
\det  (J_+ - J_-) = 2^8 \cos^4 \psi_1 \cos^4 \psi_2 \sin^8 \chi,
\end{eqnarray}
while the expression for $\det ( J_+ + J_- )$ is more involved. We see that for certain values of the real coordinates $\psi_1, \psi_2$ and $\chi$ the determinants do vanish, signalling the occurrence of type-changing at these loci. We expect that the second and third choices will require one chiral and one twisted-chiral superfield in addition to a semi-chiral superfield in order to parametrise the relevant directions of the kernels.

\subsection{The generalised K\"{a}hler potential}
In order to derive the relevant expressions for the pure spinors corresponding with the generalised K\"{a}hler structures we require an explicit form for the generalised K\"{a}hler potential. In the case of $SU(3)$ this potential is not yet known for the various superfield parametrisations. Deriving said potential is again wrought with computational difficulties. For completeness' sake we will provide partial results here.\\
\\
When considering the semi-chiral parametrisation corresponding to the choice of complex structures
\begin{eqnarray}
J_+^\mu{}_\nu&=&L^\mu_C\,\mathbb{J}_1^C{}_D\,L^D_\nu \nonumber \\
J_-^\nu{}_\nu&=&R^\mu_C\,\mathbb{J}_1^C{}_D\,R^D_\nu\,,
\end{eqnarray}
we must first attempt to diagonalise these complex structures. Doing so will provide a way to determine the generalised K\"{a}hler potential. Since we are considering a semi-chiral parametrisation where the complex structures do not commute we will have to derive two sets of coordinates, one diagonalising $J_+$ and one diagonalising $J_-$. This in itself is a fairly non-trivial problem.\\
\\
One way of proceeding is considering the coordinate transformation from the real coordinates on the manifold $x^\mu$ to the complex coordinates $z^a(x^\mu)$ diagonalising the complex structure $J_+$ and noting that
\begin{eqnarray}
\partial_\nu z^a J^\nu_{\pm \mu} &=& \partial_\mu z^b J^a_{+ b} \nonumber \\
&=&i \partial_\mu z^a.
\end{eqnarray}
Conversely, by construction we have that
\begin{eqnarray}
L^A_{\phantom{a}\nu} J^\nu_{+ \mu} &=& \nonumber L^B_{\phantom{b}\mu} J^a_{+ b} \nonumber \\
&=& i L^A_{\phantom{a}\mu}.
\end{eqnarray}
Putting both together yields the system of partial differential equations
\begin{eqnarray}
\partial_\mu z^a = f^a_{\phantom{a}B}(z,\bar z) L^B_{\phantom{b}\mu}.
\end{eqnarray}
The vanishing of the Nijenhuis tensor ensures that this system is integrable and that the required coordinates exist. Eliminating the auxiliary functions $f^a_{\phantom{a}B}(z,\bar z)$ will give us a system of coupled partial differential equations which can be solved to yield the required coordinates $z^a(x^\mu)$. To obtain the coordinates diagonalising $J_-$ on the other hand we need to consider the system
\begin{eqnarray}
\partial_\mu z^a = g^a_{\phantom{a}B}(z,\bar z) R^B_{\phantom{b}\mu}.
\end{eqnarray}
Proceeding in this manner we find that we can diagonalise $J_+$ by choosing the coordinates
\begin{eqnarray}
l_1 &=& e^{-i(\epsilon_1 + \varphi_1 )} \cos \psi_1 \tan \chi \nonumber \\
l_2 &=& e^{-i(\epsilon_1 - \varphi_1)} \sin \psi_1 \tan \chi \nonumber \\
r_1 &=& e^{-\tau +i(-\epsilon_1 + \epsilon_2 - \varphi_1 - \varphi_2)} (\cos \chi)^{-1/2+i \sqrt{3}/2} \nonumber \\
&& \left( e^{2i(\epsilon_1 + \varphi_2)} \cos \chi \sin \psi_1 \cos \psi_2 +  \cos \psi_1 \sin \psi_2 \right) \nonumber \\
r_2 &=& e^{-\tau +i(-\epsilon_1 + \epsilon_2 - \varphi_1 + \varphi_2)} (\cos \chi)^{-1/2+i \sqrt{3}/2} \nonumber \\
&& \left( e^{2i(\epsilon_1 + \varphi_2)} \cos \chi \cos \psi_1 \cos \psi_2 - \sin \psi_1 \sin \psi_2 \right),
\end{eqnarray}
and their complex conjugates $\bar l_1, \bar l_2, \bar r_1$ and $\bar r_2$. On the other hand, we can diagonalise $J_-$ by using the complex coordinates
\begin{eqnarray}
\tilde l_1 &=& e^{i(\epsilon_2 + \varphi_2 + \sqrt{3}\tau )} \cos \psi_2 \tan \chi \nonumber \\
\tilde l_2 &=& e^{i(-\epsilon_2 + \varphi_2+ \sqrt{3}\tau)} \sin \psi_2 \tan \chi \nonumber \\
\tilde r_1 &=& e^{\tau +i(\epsilon_1 + \epsilon_2 - \varphi_1 + \varphi_2)} (\cos \chi)^{-1/2+i \sqrt{3}/2} \nonumber \\
&& \left( e^{-2i(\epsilon_1 + \varphi_2)} \cos \chi \cos \psi_1 \sin \psi_2 + \sin \psi_1 \cos \psi_1\right) \nonumber \\
\tilde r_2 &=& e^{\tau +i(\epsilon_1 - \epsilon_2 - \varphi_1 + \varphi_2)} (\cos \chi)^{-1/2+i \sqrt{3}/2} \nonumber \\
&& \left( e^{-2i(\epsilon_1 + \varphi_2)} \cos \chi \cos \psi_1 \cos \psi_2 - \sin \psi_1 \sin \psi_2 \right),
\end{eqnarray}
and their complex conjugates. Other solutions yielding useable coordinates obviously exist. Unfortunately, the coordinates found do not put the 2-form $\Omega=2g\left[J_+, J_- \right]^{-1}$ in its canonical form for either choice. In principle it should be possible to find holomorphic (with respect to the diagonalised complex structure) coordinate transformations which accomplish this. Doing so would yield the two sets of coordinates
\begin{eqnarray}
q^{\tilde \alpha}=l^{\tilde \alpha},& \qquad p^{\tilde \alpha}=V_{\tilde \alpha} \nonumber \\
q^{\tilde \mu}=V_{\tilde \mu},& \qquad p^{\tilde \mu}=r^{\tilde \mu},
\end{eqnarray}
providing us with expressions which can be integrated to the explicit form of the potential $V$ as described in \cite{sevrintroost}. Doing so would make it possible to calculate the pure spinors, where hopefully one could express them in forms that make the expected type changing manifest.\\
\\
It should be possible to relate the occurrence of type changing to specific points on the manifold where it is pinched down to a lower-dimensional one, similar to the relation between $S^3 \times S^3$ and $S^3 \times S^1$. Locally, $SU(3)$ is isomorphic to $S^3 \times S^5$, although the global fibre structure is non-trivial\footnote{In fact, it can be shown that $SU(3)$ is the only consistent non-trivial $S^3$ fibration over $S^5$.}, see for example \cite{aguilar}. Presumably, the type changing loci correspond to the choices of coordinates were $SU(3)$ degenerates to a lower-dimensional manifold. A related issue is that any expression for the pure spinors will be coordinate dependent, so that when moving to different coordinate patches covering the manifold these expressions will vary. We would like to stress that this is mere conjecture at this point and that without a coordinate-free form for expressing the pure spinors, or an explicit form for them in terms of some set of suitable complex coordinates this cannot be verified. It should be interesting to see if future work can shed light on these issues.

\chapter{Discussion and outlook}
\thispagestyle{empty}
Having provided a thorough overview of how the bihermitian geometry arising in the study $\mathcal{N}=(2,2)$ non-linear $\sigma$-models can be described by generalised K\"{a}hler geometry we provided a number of concrete examples in chapter 7 by considering the generalised K\"{a}hler geometry of Wess-Zumino-Witten models. By virtue of the Lie algebra structure one can consistently categorise the different possible complex structure pairs that make up the bihermitian geometry, which in turn determines the superfield content of the model.\\
\\
The $\sigma$-model with $SU(2) \times U(1)$ as its target space was already known to allow for a description in terms of a chiral and twisted-chiral superfield pair, as well as a description in terms of a semi-chiral superfield. Furthermore, it was known that these descriptions were related through T-duality provided a proper isometry was present. $SU(2) \times U(1)$ is special in the sense that it is the only non-trivial WZW-model which can be described without semi-chiral superfields. When considering this parametrisation we found that we require two different expressions for the generalised K\"{a}hler potential, which are related to each other by a generalised K\"{a}hler transformation. By explicitly calculating the pure spinors characterising the generalised complex geometry we found that this description remained of type $(1,1)$ in both coordinate patches.\\
\\
The description of $SU(2) \times U(1)$ in terms of a semi-chiral superfield provided a richer structure. Here as well we found that we needed two different expressions for the generalised K\"{a}hler potential to describe the full manifold, however these two expression were not simply related by a generalised K\"{a}hler transformation but required a Legendre transform as well. This is a further indication that a description in terms of semi-chiral superfields allows for, and indeed requires, a broader class of reformulations of the generalised K\"{a}hler potential.\\
\\
Another feature which should be generic once semi-chiral superfields are involved is the occurrence of type changing. We explicitly showed this in the case of the semi-chiral parametrisation of $SU(2) \times U(1)$ by considering the pure spinors and their Mukai pairings. We found specific loci where the Mukai pairings became degenerate, which could be rectified by rescaling the pure spinors after performing a $b$-transform. The price to be paid for this however was that the resulting spinors were not $H$-closed anymore, reflecting the ambiguity in the (local) definition of the pure spinors. By proceeding in this manner we found that one of the two pure spinors changed from a symplectic type to a complex type, with the role of the spinors undergoing the type changing exchanged when moving from one coordinate patch to the other, itself a consequence of mirror symmetry.\\
\\
$S^3 \times S^3$ was considered as an example of a WZW-model which does not allow for an additional $\mathcal{N}=(4,4)$ supersymmetry. By considering the various complex structure pairs on the manifold we confirmed the existence of two possible descriptions. Both descriptions now require a semi-chiral superfield, with the other superfield respectively being a chiral superfield and a twisted-chiral superfield. Both descriptions are valid on a different coordinate patch of the target manifold and allow for a description in terms of a generalised K\"{a}hler potential, with both potentials related through a generalised K\"{a}hler transformation and a Legendre transform.\\
\\
By explicitly considering the pure spinors and the Mukai pairings we found that we could investigate the type changing loci through rescalings of the pure spinors. The type changing loci were related to points on the manifold were locally it is pinched down to a $S^3 \times S^1$, which we were able to show explicitly by making contact with the previous description for $SU(2) \times U(1)$. We found that at the type changing loci the symplectic-like pure spinor took on a more complex-like form, with the pure spinors exhibiting this behaviour being swapped when moving from one coordinate patch to the other.\\
\\
Unfortunately, the example of $SU(3)$ remains unresolved for now. Since it allows for a description purely in terms of semi-chiral superfields on the one hand, and a description in terms of a semi-chiral multiplet alongside chiral or twisted-chiral superfields on the other hand it should exhibit the various kinds of behaviour described above. Specifically, by considering the kernel of the commutators of the possible complex structures we would expect type changing to occur at various loci. It would be interesting to relate this fact to points on the manifold where $SU(3)$ is pinched down to a lower dimensional manifold that we have describe before to see if the various expressions arising are consistent with the lower dimensional cases. An important first step would be to obtain a relatively simple expression for the generalised K\"{a}hler potential for the various parametrisations. This certainly seems to be doable, however we were not able to provide a satisfactory answer here. The fact that these various potentials should be related through Legendre transforms as well as generalised K\"{a}hler transformations should also be kept in mind, as they will almost certainly be required to formulate the different descriptions in a consistent manner. \\
\\
Our discussion so far makes it clear that while a local $\sigma$-model description in terms of generalised K\"{a}hler structures of WZW-models sheds light on the various properties of these models, a global description would be preferable to make their behaviour more transparent. In particular, the description at the type changing loci and the ambiguities arising through the definition of the pure spinors should be dealt with in a more covariant (in the context of generalised complex geometry) manner. Treating examples such as the ones considered in this work in a systematic way may provide some guidance in this matter.

\appendix 
\chapter{Notation and Conventions}
\thispagestyle{empty}
\section{Coordinates}
We denote the worldsheet coordinates by $ \tau,\sigma \in\mathds{R}$, and the worldsheet light-cone coordinates are
defined by,
\begin{eqnarray}
\sigma ^\pp= \tau + \sigma ,\qquad \sigma ^== \tau - \sigma .\label{App1}
\end{eqnarray}
Collectively, indices of coordinates on the worldsheet are denoted by Greek letters from the beginning alphabet $\{\alpha, \beta, \dots\}$.\\
\\
Indices of real coordinates corresponding to a Riemannian manifold are denoted by Greek letters from the middle of the alphabet $\{\mu,\nu,\rho,\,\sigma, \dots \}$. \\
\\
Indices of complex coordinates corresponding to a complex manifold are denoted by Latin indices from the beginning of the alphabet $\{a,b,c,d, \dots\}$.\\ The components of the complex conjugates with respect to a given complex structure are defined as $z^{\bar a}= \bar{z}^a$.\\
\\
Indices corresponding to Lie algebra elements are taken to be capitalised Latin letters from the beginning of the alphabet $\{A,B,C,D,\dots\}$.\\
\\
The indices of the various superfields are chosen so that they consistently denote which kind of constrained superfield is under consideration (it should be clear within the context that these indices stand for superfields and not for the generic coordinates considered above)
\begin{itemize}
\item Chiral : $(z^\alpha,z^{\bar{\alpha}})$ with $\alpha,\bar{\alpha}\in \{1, \dots, n_c\}$
\item Twisted-chiral : $(w^\mu,w^{\bar{\mu}})$ with $\mu,\bar{\mu}\in \{1, \dots, n_t\}$
\item Left semi-chiral $(l^{\tilde{\alpha}},l^{\bar{\tilde{\alpha}}})$ with $\tilde{\alpha},\bar{\tilde{\alpha}}\in \{1, \dots, n_s\}$
\item Right semi-chiral $(r^{\tilde{\mu}},r^{\bar{\tilde{\mu}}})$ with $\tilde{\mu},\bar{\tilde{\mu}}\in \{1, \dots, n_s\}$.
\end{itemize}
Often, we will use collective coordinates representing the various superfields for constructing objects built of the derivates of a function $V$.
\begin{eqnarray}
	V_{XY}=\left(
	\begin{matrix}
	V_{xy} & V_{x \bar{y}} \\
	V_{\bar{x} y} & V_{\bar{x} \bar{y}} 
	\end{matrix} \right),
\end{eqnarray}
where $X,Y \in \{z,w,l,r \}$ and $x,y \in \{\alpha, \mu, \tilde{\alpha}, \tilde{\mu}$\} (and complex conjugates) and where the size of the matrix $V_{XY}$ depends on the amount of superfields of the considered type.
\section{Superspace}
The $N=(1,1)$ (real) fermionic coordinates are denoted by $ \theta ^+$ and $ \theta ^-$ and the
corresponding derivatives satisfy,
\begin{eqnarray}
D_+^2= - \frac{i}{2}\, \partial _\pp \,,\qquad D_-^2=- \frac{i}{2}\, \partial _= \,,
\qquad \{D_+,D_-\}=0.\label{App2}
\end{eqnarray}
The $N=(1,1)$ integration measure is explicitly given by,
\begin{eqnarray}
\int d^ 2 \sigma \,d^2 \theta =\int d\tau \,d \sigma \,D_+D_-.
\end{eqnarray}
Passing from $N=(1,1)$ to $ N=(2,2)$ superspace requires
the introduction of two more real fermionic coordinates $ \hat \theta ^+$ and $ \hat \theta ^-$
where the corresponding fermionic derivatives satisfy,
\begin{eqnarray}
\hat D_+^2= - \frac{i}{2} \,\partial _\pp \,,\qquad \hat D_-^2=- \frac{i}{2} \,\partial _= \,,
\end{eqnarray}
and again all other -- except for (\ref{App2}) -- (anti-)commutators do vanish.
The $N=(2,2)$ integration measure is,
\begin{eqnarray}
\int d^2 \sigma \,d^2 \theta \, d^2 \hat \theta =
\int d \tau\, d \sigma \,D_+D_-\, \hat D_+ \hat D_-.
\end{eqnarray}
We can use a complex basis for the superderivatives,
\begin{eqnarray}
\mathds{D}_\pm\equiv \hat D_\pm+i\, D_\pm,\qquad
\bar{\mathds{D}}_\pm\equiv\hat D_\pm-i\,D_\pm,
\end{eqnarray}
where they satisfy,
\begin{eqnarray}
\{\mathds{D}_+,\bar{\mathds{D}}_+\}= -2i\, \partial _\pp\,,\qquad
\{\mathds{D}_-,\bar{\mathds{D}}_-\}= -2i\, \partial _=,
\end{eqnarray}
and all other anti-commutators do vanish.

\section{The Nijenhuis tensor and related objects}
Given a linear map $A^\mu_{\pa \nu}$, one can define a (1,2)-tensor called the Nijenhuis tensor as follows
	\begin{eqnarray}
	N(A)^\mu_{\pa \nu \rho} \equiv A^\sigma_{\pa [\nu} A^\mu_{\pa \rho],\sigma} + A^\mu_{\pa \sigma} A^\sigma_{\pa 
	[\nu,\rho]}
	\end{eqnarray}
One can generalise this expression somewhat by introducing a (1,2)-tensor built out of two linear maps $A^\mu_{\pa \nu}$ and $B^\mu_{\pa \nu}$
	\begin{eqnarray}
	N(A,B)^\mu_{\pa \nu \rho} \equiv A^\sigma_{\pa [\nu} B^\mu_{\pa \rho],\sigma} + A^\mu_{\pa \sigma} B^\sigma_{\pa 
	[\nu,\rho]} + B^\sigma_{\pa [\nu} A^\mu_{\pa \rho],\sigma} + B^\mu_{\pa \sigma} A^\sigma_{\pa 
	[\nu,\rho]},
	\end{eqnarray}
which reduces to the Nijenhuis tensor $N(A)$ when $A=B$. Additionally, it is useful to introduce the quantity\footnote{This object is not a tensor, unless $A$ and $B$ commute.}
	\begin{eqnarray}
	M(A,B)^\mu_{\pa \nu \rho} \equiv \frac 1 2 \left( A^\mu_{\pa \sigma} B^\sigma_{\pa \nu,\rho} - B^\mu_{\pa \sigma} A^\sigma_{\pa \rho,\nu} + B^\sigma_{\pa \nu} A^\mu_{\pa \rho,\sigma} - A^\sigma_{\pa \rho} B^\mu_{\pa \nu, \sigma}\right),
	\end{eqnarray}
satisfying
\begin{eqnarray}
N(A,B)^\mu_{\pa \nu \rho}&=&M(A,B)^\mu_{\pa \nu \rho}+M(B,A)^\mu_{\pa \nu \rho} \nonumber \\
M(A,B)^\mu_{\pa \nu \rho}&=&-M(B,A)^\mu_{\pa \rho \nu}.
\end{eqnarray}

\section{The Lie derivative}

\section{Lie algebras and Lie groups}
We denote the generators of a Lie algebra by $T_A$, $A\in\{1,\cdots,d\}$
and satisfying the algebra $[T_A,T_B]=i\,f_{AB}{}^C\,T_C$. The
Cartan-Killing metric on the algebra, $\eta_{AB}$, is defined by,
\begin{eqnarray}
 \eta_{AB}\equiv -\frac{1}{\tilde h}\,f_{AC}{}^Df_{BD}{}^C,
\end{eqnarray}
with $\tilde h$ the dual Coxeter number of the Lie algebra. In general we
have,
\begin{eqnarray}
 \mbox{Tr}\big(T_AT_B\big)=x\,\eta_{AB},
\end{eqnarray}
with $x$ the index of the representation.\\
\\
Denoting a group element by $G$ and a set of coordinates on the group by
$x^\mu$, $\mu \in\{1,\dots ,d\}$, we denote the left- and right-invariant
vielbeins by $L_\mu^B$ and $R_\mu^B$.\\
\\
The metric $g$ and 3-form $H$ are obtained via
\begin{eqnarray}
 g_{\mu\nu}&=&-\frac{k}{8\pi  x}\,\mbox{Tr}\,\partial _\mu g g^{-1}\partial _\nu gg^{-1} \nonumber \\
 &=&
 \frac{k}{8\pi }R_\mu^C\,R_\nu^D\,\eta_{CD}\, \nonumber \\
  H_{\mu\nu\rho}&=&\frac{k}{24\pi x}\,\mbox{Tr}\,dgg^{-1}\wedge dgg^{-1}\wedge dgg^{-1} \nonumber \\
  &=&\frac{k}{48\pi }R_\mu^DR_\nu^ER_\rho^Ff_{DEF}\,dx^\mu\wedge dx^\nu\wedge dx^\rho.
\end{eqnarray}
where $k\in \mathds{N}$. The metric on the group is,
\begin{eqnarray}
 g_{\mu \nu}=-\frac{k}{8\pi  x}\,\mbox{Tr}\,\partial _\mu g g^{-1}\partial _\nu gg^{-1}=
 \frac{k}{8\pi }R_\mu^C\,R_\nu^D\,\eta_{CD}\,.
\end{eqnarray}

\chapter{Nederlandse samenvatting}

\section*{Het verhaal tot dusver}

Een belangrijke vraag in de theoretische fysica, en meer specifiek de theoretische elementaire deeltjesfysica, is het vinden van een eengemaakt theoretisch kader dat de tot nu toe ontdekte elementaire deeltjes en de fundamentele wisselwerking volgens dewelke ze interageren kan verklaren. Enerzijds beschikt men over het zogenaamde standaardmodel dat de gekende materiedeeltjes, alsook de elektromagnetische, de sterke en de zwakke kernkracht beschrijft aan de hand van zogenaamde niet-Abelse ijktheori\"{e}n. Het standaardmodel heeft tot op heden met zeer grote precisie de test van het experiment doorstaan, met als meest recente ontwikkeling de ontdekking van een nieuw scalair deeltje aan het CERN dat vermoedelijk overeenkomt met het Brout-Englert-Higgs deeltje dat lang een ontbrekend ingredi\"{e}nt van het standaardmodel was. De algemene relativiteitstheorie van haar kant beschrijft de gravitationele wisselwerking die verantwoordelijk is voor een groot deel van de waargenomen fenomenen op kosmische schaal. In tegenstelling tot het standaardmodel is de algemene relativiteitstheorie een klassieke theorie, wat betekent dat zij niets zegt over de kwantummechanische effecten die zich op kleine schaal zullen voordoen. \\
\\
Afgezien van het feit dat het wenselijk zou zijn om beide theoretische kaders te verenigen tot eenzelfde geheel zijn er goede redenen om aan te nemen dat dit niet heel het verhaal kan zijn. Men kan zich experimentele regimes voorstellen waar beide beschrijvingen van belang zullen zijn, onder andere wanneer men het vroege universum zou willen beschrijven. De zoektocht naar een kwantummechanische zwaartekrachttheorie is al enkele decennia aan de gang, en van de verscheidene kandidaten die zijn voorgesteld is de snaartheorie tot op heden de meest belovende. Snaartheorie vindt zijn oorsprong in pogingen tot het beschrijven van de sterke kernkracht, maar al snel werd duidelijk dat deze theorie veel meer bevat dan wat oorspronkelijk werd vermoed. Niet alleen bleek dat de theorie op een natuurlijke manier zwaartekracht op een kwantummechanische manier beschreef, de theorie bevat ook een heel gamma aan nieuwe elementen waaronder de ijktheorie\"{e}n die het standaardmodel onderbouwen. Of de snaartheorie al dan niet betrekking heeft op ons universum is tot op heden een open vraag. Wat wel vaststaat is dat haar onderzoek een uitermate vruchtbare bodem is gebleken voor het ontwikkelen van nieuwe fysische en wiskundige technieken, en dat zij vaak op verrassende wijze het verband tussen beiden kan illustreren. Het onderwerp van deze verhandeling, veralgemeende complexe meetkunde en niet-lineaire $\sigma$-modellen, is een elegant voorbeeld van hoe wiskundige structuren en fysische theorie\"{e}n elkaar kunnen aanvullen en inspireren.

\section*{De kracht van de meetkunde}
Snaartheorie, en haar nuttigere supersymmetrische veralgemening supersnaartheorie, bevatten als basisingredi\"{e}nten open en gesloten snaren wiens verschillende trillingstoestanden corresponderen met wat wij ervaren als zijnde elementaire deeltjes. In het massaloze  deeltjesspectrum van de gesloten snaar vinden we naast het gravitonveld, dat aanleiding zal geven tot kromming van de ruimte-tijd en op die manier zwaartekracht, ook het zogenaamde Kalb-Ramond veld dat er voor zal zorgen dat de resulterende ruimte-tijd torsie kan bevatten. In principe zou men vanuit de wisselwerking tussen individuele snaren deze geometrische structuren moeten kunnen afleiden. Tot op heden is dit jammer genoeg niet mogelijk en moeten we ons beroepen op indirecte technieken. E\'{e}n manier om deze meetkunde te bestuderen is gebruik te maken van zogenaamde niet-lineaire $\sigma$-modellen, waarbij de co\"{o}rdinaten van de ruimte waarin de snaar zich voortbeweegt de vrijheidsgraden vormen van een veldentheorie op het wereldvlak van deze snaar. Het consistent zijn van deze veldentheorie zal sterke beperkingen opleggen aan de geometrie van omringende ruimte.\\
\\
Wanneer men op deze manier $\sigma$-modellen bestudeert die een uitgebreidere vorm van supersymmetrie bevatten, meer bepaald $\mathcal{N}=(2,2)$ supersymmetrie, dan vindt men dat de resulterende geometrie deze is van van een een zogenaamde bihermitische vari\"{e}teit. Dit betekent onder andere dat de omringende ruimte de structuur heeft van een complexe vari\"{e}teit met twee complexe structuren en dat de metriek hermitisch is met betrekking tot beiden. De algebra\"{i}sche relaties tussen deze complexe structuren zijn innig verbonden met de aanwezige soorten supervelden op het wereldvlak. Men onderscheidt in deze chirale, twisted-chirale en semi-chirale supervelden, waarvan het is aangetoond dat deze voldoende zijn om elk $\mathcal{N}=(2,2)$ niet-lineair $\sigma$-model te kunnen beschrijven.\\
\\
Deze bihermitische structuren kunnen worden beschreven aan de hand van een zogenaamde veralgemeende K\"{a}hler meetkunde. Dit is een speciaal geval van veralgemeende complexe meetkunde, dewelke in zekere zin interpoleert tussen symplectische meetkunde enerzijds, en complexe meetkunde anderzijds. Het basisidee is dat men de raakbundel en de co-raakbundel van een vari\"{e}teit beschouwt als \'{e}\'{e}zelfde object, de veralgemeende raakbundel. Een bijzonder nuttige methode voor het beschrijven van veralgemeende complexe structuren is gebruik maken van het feit dat deze kunnen worden gekarakteriseerd door een zogenaamde zuivere spinorbundel, naar analogie van de canonische lijnbundel bij complexe vari\"{e}teiten. Een interessant fenomeen is dat het type van de veralgemeende complexe structuur, waarmee we bedoelen in hoeveel richtingen op de veralgemeende raakbundel deze symplectisch of complex is, niet noodzakelijk constant is doorheen de vari\"{e}teit. In de context van $\sigma$-modellen zal deze situatie zich voordoen als het model zogenaamde semi-chirale supervelden bevat. De zuivere spinorbundel bevat voldoende informatie om te kunnen nagaan waar het type van de veralgemeende complexe structuur zal veranderen, waardoor een expliciete vorm voor de zuivere spinoren wenselijk is.\\
\\
Belangrijk voor het aanwenden van veralgemeende complexe meetkunde bij niet-lineaire $\sigma$-modellen is dat in de gebruikelijke K\"{a}hler meetkunde de metriek van deze meetkunde volledig wordt bepaald door \'{e}\'{e}n enkele functie, de K\"{a}hler-potentiaal. Een algemeen $\mathcal{N}=(2,2)$ niet-lineair $\sigma$-model kan  langs zijn kant ook volledig beschreven worden door een lagrange-dichtheid die een scalaire functie is van de aanwezige supervelden. Aangezien deze lagrange-dichtheid  het model volledig bepaalt  kan men deze identificeren met een zogenaamde veralgemeende K\"{a}hler-potentiaal die volledig de resulterende meetkunde vastlegt, naar analogie van het K\"{a}hler geval. Indien er een $b$-veld aanwezig is zal de bihermitische meetkunde in het algemeen niet meer K\"{a}hler zijn maar een een ingewikkeldere structuur vertonen. Tot op heden zijn er maar weinig expliciete voorbeelden van zulke niet-K\"{a}hlerse bihermitische vari\"{e}teiten waarbij er een expliciete uitdrukking voor de veralgemeende K\"{a}hlerpotentiaal gekend is. Door modellen te beschouwen waarbij er meer supersymmetrie aanwezig is kunnen we, gebruikmakende van het feit dat men beschikt over een continue familie van complexe structuren, deze niet-K\"{a}hlerse bihermitische meetkunde bestuderen.
\section*{Wess-Zumino-Witten modellen}

Een belangrijke subklasse van niet-lineaire $\sigma$-modellen zijn de zogenaamde Wess-Zumino-Witten modellen waarbij we een reductieve Lie-groep als doelruimte kunnen nemen. Men kan nagaan welke mogelijke complexe structuren er aanwezig kunnen zijn op de Lie-groep door de mogelijke complexe structuren te onderzoeken op de corresponderende Lie-algebra. Het blijkt dat de enige toegestane vrijheid voor het kiezen van deze complexe structuur op de Lie-algebra haar actie op de Cartan subalgebra is. Eenmaal deze keuze gemaakt liggen de complexe structuren op de Lie-groep vast. Daar deze ook de veldinhoud van het $\sigma$-model bepalen kunnen we op deze manier nagaan welk $\sigma$-model correspondeert met de mogelijke meetkundige structuren op de doelruimte. In dit werk werden de $\mathcal{N}=(2,2)$ Wess-Zumino-Witten modellen op de doelruimtes $SU(2) \times U(1)$,  $SU(2) \times SU(2)$ en $SU(3)$ onderzocht. 

\section*{$SU(2) \times U(1)$}
Dit model is het enige niet-triviale model dat een beschrijving toelaat zonder gebruik te maken van semi-chirale supervelden. Door het expliciet berekenen van de toegestane complexe structuren op de doelruimte werd het bestaan bevestigd van twee mogelijke parametrisaties, de ene in termen van een chiraal en een twisted-chiraal superveld, de andere in termen van een semi-chiraal superveld.  Deze twee parametrisaties corresponderen met de keuzevrijheid die men heeft in het kiezen van een complexe structuur op de Cartan-subalgebra. Daar de Cartan-decompositie van de Lie-algebra slechts gedefini\"{e}erd is modulo een groepsconjugatie verkrijgen we op deze manier twee families van complexe structuren, consistent met het feit dat $SU(2) \times U(1)$ uitgebreidere supersymmetrie toelaat.\\
\\
Bij het uitrekenen van de veralgemeende K\"{a}hler-potentiaal dient opgemerkt te worden dat deze een lokale uitdrukking is die slechts gedefini\"{e}erd is modulo een veralgemeende K\"{a}hler-transformatie en een legendre-transformatie. In het algemeen zal deze potentiaal een andere vorm hebben wanneer we ons in een andere co\"{o}rdinatenkaart bevinden. Op overlappende kaarten dienen deze uitdrukkingen echter overeen te komen. Bij de chirale/twisted-chirale parametrisatie is het voldoende om een veralgemeende K\"{a}hler-transformatie uit te voeren om van de ene beschrijving naar de andere te gaan. In het geval van de semi-chirale parametrisatie echter dient men ook een legende-transformatie uit te voeren om dit aan te tonen. Daar de veralgemeende K\"{a}hler-potentiaal in beide gevallen is gekend is het mogelijk expliciet de zuivere spinoren uit te rekenen die corresponderen met de veralgemeende complexe structuren op de doelruimte. Zoals verwacht toont dit aan dat in het chirale/twisted-chirale geval de types van de veralgemeende complexe structureren niet veranderen. In het semi-chirale geval daarentegen  doet zich dit wel voor. Dit is te verwachten daar men anders een symplectische 2-vorm zou kunnen neerschrijven die globaal defini\"{e}erd is, wat niet mogelijk is voor dit soort vari\"{e}teiten. De gevonden uitdrukking voor de zuivere spinoren stelt ons in staat om expliciet uit te rekenen waar deze verandering van type zich zal voordoen.\\
\\
Daar beide formuleringen van het $\sigma$-model op $SU(2) \times U(1)$ met elkaar zijn verbonden door $T$-dualiteit dient dit ook tot uiting te komen op het niveau van de veralgemeende K\"{a}hler-potentialen. Dit is inderdaad blijkt inderdaad het geval te zijn. 

\section*{$SU(2) \times SU(2)$}

Het veranderen van het type van de veralgemeende complexe structuren beperkt zich niet tot modellen waarbij er extra supersymmetrie aanwezig is. Een voorbeeld hiervan is het niet-lineaire $\sigma$-model op $SU(2) \times SU(2)$. Ook hier kunnen we door de mogelijke complexe structuren op de Lie-algebra te beschouwen nagaan welke complexe structuren er zich kunnen voordoen op de doelruimte. Er doen zich twee gevallen voor, \'{e}\'{e}n waarbij de vari\"{e}teit geparametriseerd wordt door een chiraal en een semi-chiraal superveld, en \'{e}\'{e}n parametrisatie door een twisted-chiraal en een semi-chiraal superveld. Ook hier zullen deze aanleiding geven tot verschillende veralgemeende K\"{a}hler-potentialen, die op verschillende kaarten gerelateerd zijn aan elkaar door middel van veralgemeende K\"{a}hler-transformaties en legendre-transformaties. \\
\\
Beide beschrijvingen bevatten noodzakelijk een semi-chiraal superveld, wat impliceert dat de types van de veralgemeende complexe structuren niet overal dezelfde zullen zijn. Door middel van de veralgemeende K\"{a}hler-potentiaal kan dit expliciet worden aangetoond op het niveau van de zuivere spinoren. We vinden drie loci waar het type van de veralgemeende structuren verandert, waarbij de rollen van de zuivere spinoren worden omgewisseld als we de beschrijvingen met elkaar vergelijken.

\section*{$SU(3)$}
In dit werk werd een aangezet gegeven om het niet-lineaire $\sigma$-model op $SU(3)$ volledig te beschrijven in termen van veralgemeende K\"{a}hler-meetkunde. De mogelijke complexe structuren op de Lie-algebra geven aanleiding tot twee verschillende parametrisaties,  \'{e}\'{e}n in termen van een chiraal, twisted-chiraal en semi-chiraal superveld en een andere in termen van twee semi-chirale supervelden.\\
\\
Jammer genoeg is de veralgemeende K\"{a}hler-potentiaal voor deze $\sigma$-modellen niet expliciet bekend. In dit werk wordt een aanzet gegeven om deze te berekenen in het geval van een semi-chirale parametrisatie. Door beide complexe structuren afzonderlijk te diagonaliseren is het mogelijk om een stel co\"{o}rdinaten te vinden dewelke verbonden zullen zijn door een canonieke transformatie. Dit is het gevolg van de interpretatie dat de veralgemeende K\"{a}hler-potentiaal deze canonieke transformaties (of symplectomorfismen) genereert in het geval van een zuiver semi-chirale parametrisatie. Deze zullen aanleiding geven tot een stel gekoppelde parti\"{e}el differentiaalvergelijkingen die kunnen worden ge\"{i}ntegreerd tot de potentiaal. Omwille van de computationele moeilijkheden die hiermee gepaard gaan werd deze niet gevonden. De volledige beschrijving van $SU(3)$ in termen van veralgemeende K\"{a}hler meetkunde blijft op dit moment een open vraag.

\bibliographystyle{plain}
\bibliography{Referenties.bib}

\end{document}